\documentclass[fleqn,usenatbib]{mnras}

\usepackage{mathptmx}
\usepackage[T1]{fontenc}
\usepackage{graphicx}	% Including figure files
\usepackage{amsmath}	% Advanced maths commands
\usepackage{amssymb}	% Extra maths symbols
\usepackage{colortbl}    % Ajout
\usepackage{mathrsfs}

  \title[The Of?p star HD~108]{Diving into the magnetosphere of the Of?p star HD~108\thanks{Based on data obtained with the {\it Chandra} X-ray Observatory, the {\it TESS} mission, whose funding is provided by the NASA Explorer Program, and the {\it XMM-Newton} satellite. Also based on spectra collected at the Observatoire de Haute Provence, as well as with the TIGRE telescope.}}

\author[G.\,Rauw et al.]{Gregor Rauw$^1$\thanks{E-mail: g.rauw@uliege.be}, Ya\"el Naz\'e$^1$\thanks{Senior Research Associate FRS-FNRS (Belgium)}, Asif ud-Doula$^2$, Coralie Neiner$^3$\\
  $^1$ Space sciences, Technologies and Astrophysics Research (STAR) Institute, Universit\'e de Li\`ege, All\'ee du 6 Ao\^ut, 19c, B\^at B5c, 4000 Li\`ege, Belgium \\
  $^2$ Penn State Scranton, 120 Ridge View Drive, Dunmore, PA 18512, USA \\
  $^3$ LESIA, Paris Observatory, PSL University, CNRS, Sorbonne University, Universit\'e de Paris Cit\'e, 5 Place Jules Janssen, 92195 Meudon, France}

\date{Accepted XXX. Received YYY; in original form ZZZ}

\pubyear{2023}

\begin{document}
\label{firstpage}
\pagerange{\pageref{firstpage}--\pageref{lastpage}}
\maketitle
\begin{abstract}
We analyse optical and X-ray spectroscopy of the Of?p star HD~108, known for its strong dipolar magnetic field and its optical line profile variability with a timescale of $54 \pm 3$\,yrs, interpreted as the stellar rotation period. Optical emission lines have now recovered from their minimum emission state reached in 2007--2008. The variations of the equivalent width of the H$\alpha$ emission provide constraints on the inclination of the rotation axis ($i$) and the obliquity of the magnetic axis ($\beta$). The best agreement between model and observations is found for ($i$, $\beta$) pairs with $i + \beta \simeq 85^{\circ}$ and $i \in [30^{\circ},55^{\circ}]$. The Balmer emission lines display stochastic variability at the $\sim 5$\% level on timescales of a few days. {\it TESS} photometry unveils transient modulations on similar timescales in addition to prominent red noise variations. A {\it Chandra} X-ray observation of December 2021, when the star was at a higher emission level, indicates a slight increase of the flux and a spectral hardening compared to the August 2002 {\it XMM-Newton} observation, taken near minimum emission state. Magnetohydrodynamic simulations are used to compute synthetic X-ray spectra. With our current best estimate of the $\dot{M}_{B=0}$ mass-loss rate, the simulated X-ray luminosity and spectral energy distribution agree very well with the observations. Finally, the radial velocities vary on a period of 8.5 years with a peak-to-peak amplitude of 10 -- 11\,km\,s$^{-1}$, suggesting orbital motion with an unseen companion of at least 4\,M$_{\odot}$.   
\end{abstract}
\section{Introduction}
Ever since the discovery of HD~108's emission-line spectrum, nearly a century ago \citep{Mer25}, the properties and variability of this star have challenged astrophysicists \citep[e.g.][]{Man55,Bru71,And73,JMV,Und94}. Extending the pioneering compilation of literature data of \citet{And73}, \citet{Naz01} presented a long-term spectroscopic monitoring complemented by literature data from over eight decades. They showed that HD~108 undergoes recurrent transitions of its H\,{\sc i} and He\,{\sc i} lines from absorption to (apparent) P-Cygni profiles and in some cases pure emissions. These authors tentatively suggested a timescale of variability of 56 years. They further showed that the apparent P-Cygni like profiles are not genuine P-Cygni profiles, but result instead from a narrow variable emission on top of a photospheric absorption. As a result of the variable emission infilling the He\,{\sc i} $\lambda$\,4471 classification line, the apparent spectral type was found to vary between O\,4 in August 1987 and O\,7.5 in September 2000 \citep{Naz01}. Beside the H\,{\sc i} Balmer lines and the He\,{\sc i} lines, strong variations were also seen in the Si\,{\sc iii} $\lambda\lambda$\,4552, 4568 and C\,{\sc iii} $\lambda\lambda$\,4647-52 emissions. N\,{\sc iii} $\lambda\lambda$\,4634-41 and He\,{\sc ii} $\lambda$\,4686 displayed lower amplitude variations, whilst O\,{\sc ii} emissions and the S\,{\sc iv} $\lambda\lambda$\,4486, 4504 emissions were only marginally variable.

In parallel, \citet{Naz04} presented an {\it XMM-Newton} observation of HD~108 taken in August 2002. The X-ray spectrum unveiled the signature of two optically thin thermal plasma components with $kT$ near 0.2 and 1.4\,keV, and the star was found to be X-ray over-luminous compared to normal O-type stars.

In fact, HD~108 belongs to the Of?p category that was originally defined by \citet{Wal72,Wal73} as Of stars displaying certain peculiarities, among which the presence of strong C\,{\sc iii} $\lambda\lambda$\,4647--52 emission lines. Over the last two decades, it was found that recurrent line profile variations and X-ray overluminosity are common features of Of?p stars \citep{Naz08}. The three historical members of this class display a wide range of timescales for their spectral variations: 7.031\,d for HD~148\,937 \citep{Naz08a}, 537.6\,d for HD~191\,612 \citep{Wal04} and $\sim 55$\,yr for HD~108. The explanation of these intriguing properties came from the discovery that these stars feature relatively strong dipolar magnetic fields \citep{Don06,Mar10,Wad12a}. \citet{Wal10} added two more stars to the Of?p category: NGC~1624-2 and CPD$-28^{\circ}$\,2561 which were subsequently found to also feature strong dipolar magnetic fields, and to display cyclic spectral variations with periods of respectively 158\,d and 73.4\,d \citep{Wad12b,Wad15}.

The properties of Of?p stars are nowadays interpreted in the framework of the oblique magnetic rotator model that was originally proposed for $\sigma$~Ori\,E \citep{Sho90} and $\theta^1$~Ori\,C \citep{Sta96}. The magnetic field deflects the stellar winds from both magnetic hemispheres towards the magnetic equator where they collide \citep{Bab97,udD02,udD22}. The ensuing strong shock produces a hot plasma in the post-shock region, which generates the observed X-ray emission \citep[for a review, see][]{udD16}. The shocked plasma subsequently cools, forming a disk-like structure roughly located in the plane of the magnetic equator. As the magnetic field axis is inclined with respect to the rotation axis, the strength of the observed Balmer emission lines is modulated on the stellar rotation period \citep{Sun12}: the strongest emission is observed when the magnetically-confined wind region is seen face-on and the weakest emission when the confined wind is seen edge-on. The material trapped by the strong dipolar magnetic field also brakes the stellar rotation \citep{udD09}. In this context, HD~108 is the most extreme case known to date as its rotation has been braked to near 0\,km\,s$^{-1}$ \citep{Mar10}.

In this paper, we present a new set of optical and X-ray observations of HD~108 (Sect.\,\ref{obs}). In Sect.\,\ref{optvar}, we use these data to investigate optical variability on various timescales and to assess the multiplicity of HD~108. The X-ray data of HD~108 are analysed in Sect.\,\ref{Xdata}. Section\,\ref{MHDADM} describes our efforts to model the magnetosphere of HD~108, and Sect.\,\ref{discuss} summarizes our findings and conclusions. 

\section{Observations and data processing} \label{obs}
\subsection{Spectroscopic monitoring}
Since the last report on HD~108's spectral variations by \citet{Naz10}, we continued monitoring the spectral changes of the star through yearly snapshot observations with the Aur\'elie spectrograph \citep{Gillet} at the 1.52\,m telescope of the Observatoire de Haute Provence (OHP) in France. Our observations used different gratings and covered different wavelength domains. Most of the time we used a 600\,l\,mm$^{-1}$ grating to observe a 430\,\AA-wide wavelength domain centered at 4670\,\AA\ or 6630\,\AA. Some campaigns rather used a 1200\,l\,mm$^{-1}$ grating to observe a 210\,\AA-wide domain centered at 4565\,\AA. Until 2018, the detector was an EEV CCD with $2048 \times 1024$ pixels. From 2019 on, it was replaced by an Andor CCD camera with $2048 \times 512$ pixels. Both detectors had pixel sizes of 13.5\,$\mu$m squared. Typical integration times of individual exposures were 15 -- 30\,min (depending on the weather conditions). The OHP spectra were reduced using version 17FEBpl\,1.2 of the {\sc midas} software developed at ESO. 

From 2013 on, we also monitored HD~108 with the 1.2~m TIGRE telescope \citep{Schmitt,Gon22} at La Luz Observatory near Guanajuato (Mexico). TIGRE is operated in a fully robotic way and uses the refurbished HEROS echelle spectrograph \citep{Kaufer2,Schmitt} which offers a spectral resolving power of 20\,000 over the optical range from 3760 -- 8700\,\AA. Most of the time, a single observation was taken per year, but in 2013, 2014 and 2015, we performed more intensive campaigns to investigate the variations on time scales of several days, up to a few weeks. The TIGRE spectra were reduced with the HEROS reduction pipeline \citep{Mittag,Schmitt}.

Finally, we retrieved archival echelle spectra obtained with the NARVAL instrument at the Telescope Bernard Lyot (TBL) and with the ESPaDOnS instrument at the Canada-France-Hawaii Telescope (CFHT). These data are described by \citet{Mar10} and \citet{Shu17}.

For all the spectra covering the regions around the He\,{\sc i} $\lambda$\,5876 and H$\alpha$ lines, we used the {\tt telluric} tool within {\sc iraf} in combination with the atlas of telluric lines of \citet{Hinkle} to remove the absorptions due to the Earth's atmosphere. All spectra were continuum-normalized using the MIDAS software adopting best-fit spline functions adjusted to the same set of continuum windows for all spectra. We further re-analysed all OHP data taken from 1987 on and which were previously discussed by \citet{Naz01,Naz04,Naz06} and \citet{Naz10}. The journal of our observations including the results of some measurements is given in Table\,\ref{Journal}.

\subsection{Photometry}
High-cadence space-borne photometry of HD~108 was obtained with the Transiting Exoplanet Survey Satellite \citep[{\it TESS},][]{TESS}. {\it TESS} gathers high-precision photometry in the 6000\,\AA\, to 1\,$\mu$m bandpass. The sky is divided into partially overlapping sectors covering an area of $24^{\circ} \times 96^{\circ}$. Each sector is observed for about 27\,days, i.e.\ two consecutive spacecraft orbits separated by a gap around perigee passage. HD~108 was observed with {\it TESS} camera 3 during sectors 17 and 18 (i.e.\ between 7 October and 27 November 2019), with camera 3 during sector 57 (30 September to 29 October 2022) and with camera 2 during sector 58 (29 October to 26 November 2022). For sectors 17 and 18, we retrieved the 2-minute high-cadence photometric light curves from the Mikulski Archive for Space Telescopes (MAST) portal\footnote{http://mast.stsci.edu/}. The light curves produced by the {\it TESS} pipeline \citep{Jen16} provide simple background-corrected aperture photometry and so-called PDC photometry obtained after removing trends correlated with systematic spacecraft or instrument effects. For our analysis, we focused on the PDC fluxes converted into magnitudes, and we retained only those data points with a quality flag equal to 0. These PDC magnitudes have formal photometric accuracies of 0.26\,mmag. For targets with observations over several sectors, the mean PDC magnitudes can differ between consecutive sectors. For our combined analysis, we therefore subtracted the mean magnitude from each sector. For sectors 57 and 58, we extracted aperture photometry light curves at a 200\,s cadence using the {\tt Lightkurve}\footnote{https://docs.lightkurve.org} software. The background was evaluated from pixels with fluxes below the median flux. We tested background correction using either the median of the background pixels or performing a principal component analysis (PCA) with five components. Both cases yielded similar light curves, and only those obtained via the PCA technique are presented in the forthcoming sections. As before, the background-corrected fluxes were converted into magnitudes and the mean magnitude of the corresponding sector was subtracted.   

The {\it TESS} CCD detectors have pixels of size 15\,$\mu$m squared corresponding to (21\arcsec )$^2$ on the sky. Because the flux of a given source is extracted over apertures of several pixels, source crowding can become an issue. To check whether this is the case here, we searched the {\it GAIA} early data release 3 catalog \citep[EDR3,][]{EDR3} for the magnitudes of sources within a 1\arcmin\ radius of HD~108. Whilst a large number (99) of objects are located inside this area, the brightest neighbouring source is actually 6.2\,mag fainter than HD~108. Therefore, no significant contamination of the {\it TESS} photometry of HD~108 is expected. 

\subsection{X-ray observations}
The {\it Chandra} X-ray Observatory \citep{Wei00} observed HD~108 on 4 December 2021 (HJD\,2459553.178 at mid-exposure) for 15\,ks using the ACIS-S instrument (ObsID 25107). To keep photon pile-up as low as possible, only a single ACIS chip (S3) with a 1/8 subarray was used. This set-up yielded a frame time of 0.4\,s. The observed count rate of the source was $7.7\,10^{-2}$\,ct\,s$^{-1}$, which should result in a low pile-up fraction of 1.2\%. The corresponding level 2 event file provided by the {\it Chandra} pipeline was further processed using CIAO v4.14 and CALDB v4.9.6. The source spectrum was extracted with the CIAO task {\sc specextract} over a circular area of 3.5\,arcsec radius centered on the source. The background spectrum was estimated from a surrounding annulus with inner and outer radii of 3.5 and 15.0\,arcsec. A weighted response matrix and ancillary response file were generated. The source spectrum was finally binned to reach a minimum of 10 counts per bin.

Previously, HD~108 was observed for 35\,ks with {\it XMM-Newton} \citep{Jansen} on 21 August 2002 (HJD\,2452507.864 at mid exposure). A detailed description of this observation is given by \citet{Naz04}. To account for the most up-to-date status of the calibration of the instruments of {\it XMM-Newton}, we reprocessed these data with the Science Analysis System (SAS) software version 18.0.0 and using the current calibration files available in April 2022.

\section{Optical variability} \label{optvar}
\subsection{TESS photometry}\label{TESSanalysis}
The {\it TESS} photometric time series were analysed with the modified Fourier periodogram algorithm of \citet{Hec85} and \citet{Gos01}, which explicitly accounts for uneven sampling. With a nominal time step of 2\,min, the sector 17 and 18 time series have a Nyquist frequency of 360\,d$^{-1}$, whilst it amounts to 216\,d$^{-1}$ for sectors 57 and 58. The periodograms display the highest power at low frequencies, whilst the periodogram is essentially empty at frequencies above 4\,d$^{-1}$. In the individual periodograms of sectors 17 and 18, the strongest peaks are found respectively at 0.162\,d$^{-1}$ and 0.210\,d$^{-1}$ with amplitudes of respectively 1.6\,mmag and 1.2\,mmag. If we combine the data from the two sectors, we find the highest peak at 0.163\,d$^{-1}$ with an amplitude of 1.0\,mmag. This agrees well with the 6.16\,d signal found independently in the same {\it TESS} data by \citet{Tri21}. On the other hand, no significant individual peak shows up in the periodograms of sectors 57 and 58.

\begin{figure}
  \begin{center}
    \resizebox{8.5cm}{!}{\includegraphics{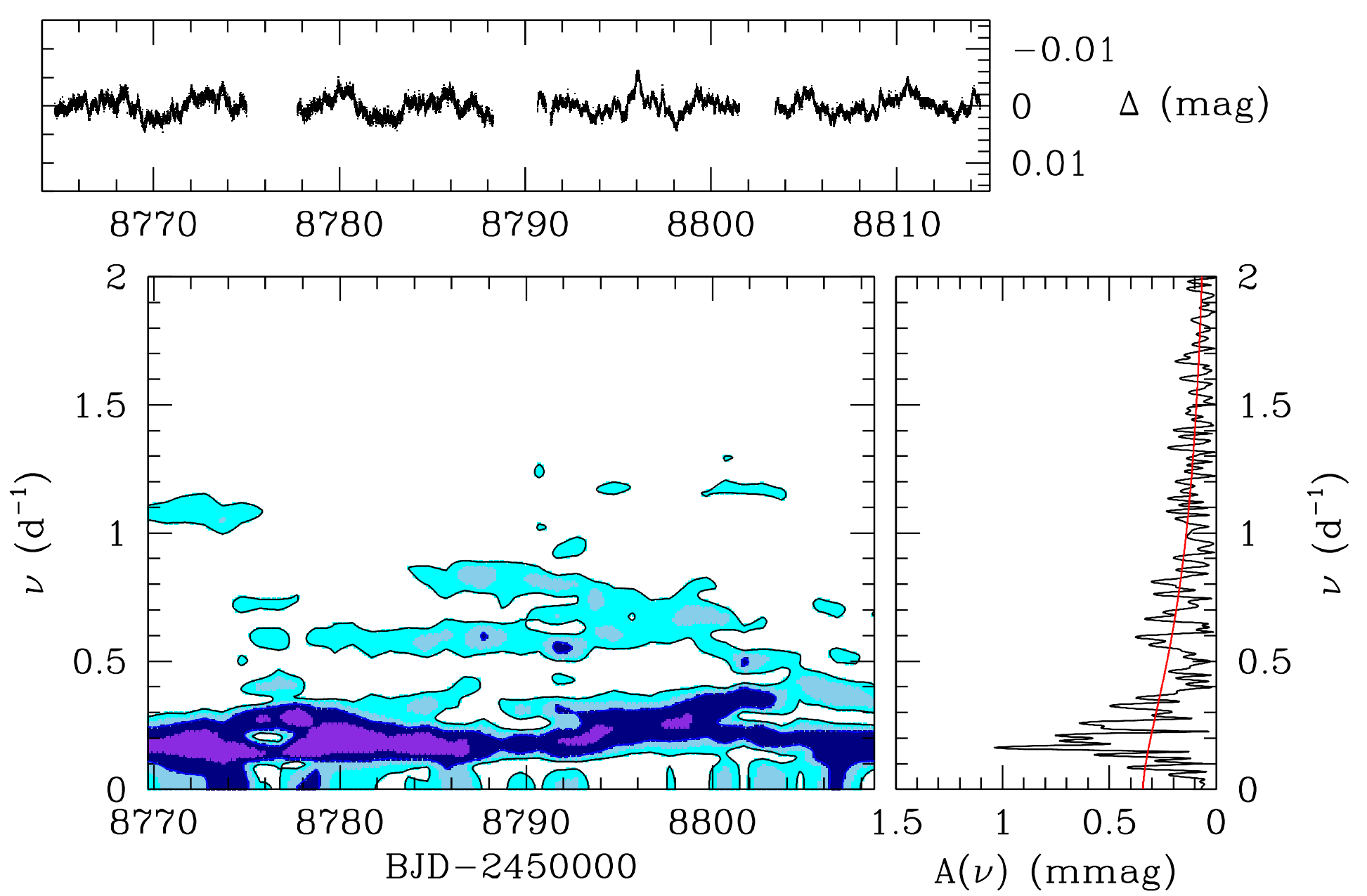}}
    \resizebox{8.5cm}{!}{\includegraphics{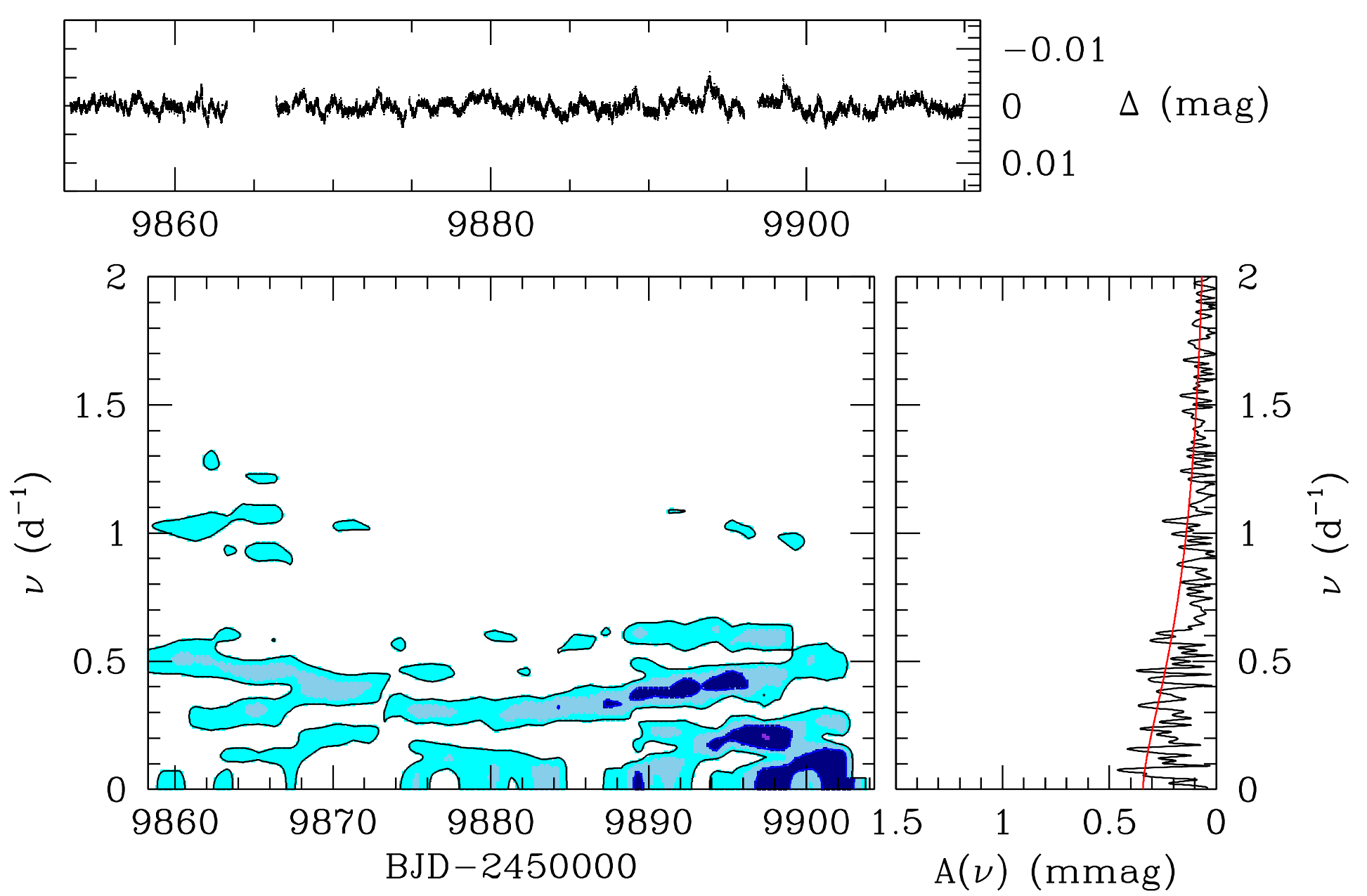}}
  \end{center}
  \caption{Time-frequency diagrams of the {\it TESS} photometry of HD~108. The top figure refers to the 2-minute PDC photometry of sectors 17 and 18, whilst the bottom figure stands for the 200\,s PCA aperture photometry from sectors 57 and 58. For each figure, the observed light curve is reproduced in the top panel. The bottom left panel provides the evolution of the Fourier periodogram with the date in the x-axis corresponding to the middle of the 10-day sliding window. The violet, dark blue, light blue, and cyan colours indicate respectively amplitudes $\geq 1.5$\,mmag, $\geq 1.0$\,mmag, $\geq 0.75$\,mmag and $\geq 0.5$\,mmag. The right vertical panel illustrates the Fourier periodogram evaluated from the full dataset of sectors 17 and 18 (top figure) and sectors 57 and 58 (bottom figure). The red curve yields our best-fit red noise relation computed from Equation\,\ref{eq1} for sectors 17 and 18.\label{spevol}}
\end{figure}

To further investigate the temporal evolution of the frequency content, we computed time-frequency diagrams (Fig.\,\ref{spevol}) by performing a Fourier analysis of the photometric time series cut into windows of ten days duration and sliding with a step of 1\,day. We see that the frequency content of the periodogram changes with time in an apparently stochastic way. This suggests that a large part of the low-frequency signal is due to a red noise component. The fact that the frequency and morphology of the highest peak change with time during sectors 17 and 18 and that it had disappeared in the sectors 57 and 58 indicates that we are dealing with a transient feature, which could be a fluctuation of the red noise component.

To describe the properties of these stochastic variations, we adopt the formalism of \citet{Stanishev} to fit the red noise part of the periodogram: 
\begin{equation}
  A(\nu) = \frac{A_0}{1 + (2\,\pi\,\tau\,\nu)^{\gamma}} + C_{\rm white\,noise}
  \label{eq1}
\end{equation}
where $A(\nu)$ is the amplitude (in mmag) at frequency $\nu$ in the periodogram built from the observations. The scaling factor $A_0$, the slope $\gamma$, the mean lifetime $\tau$ (in days), as well as the level of white noise $C_{\rm white\,noise}$ (in mmag) were determined from a fit to the power spectrum by means of a Levenberg-Marquardt routine. The corresponding best-fit parameters are given in Table\,\ref{tab:periodogram}.

\begin{table}
  \caption{Red noise properties of the periodogram of the sectors 17 and 18 {\it TESS} photometry of HD~108 \label{tab:periodogram}}
  \begin{center}
  \begin{tabular}{c c c c c c c c c}
    \hline
    $A_0$ (mmag) & $0.338 \pm 0.004$ \\
    $\tau$ (d) & $0.206 \pm 0.004$ \\
    $\gamma$ & $1.58 \pm 0.03$ \\ 
    $C_{\rm white\,noise}$ (mmag) & $0.0042 \pm 0.0004$ \\
    \hline
  \end{tabular}
  \end{center}
\end{table}

Red noise is commonly found in the analyses of high-quality space-borne photometry of all kinds of massive stars, including main-sequence OB stars, Be stars, as well as evolved Luminous Blue Variables and Wolf-Rayet stars \citep[e.g.][]{Blomme,Aerts,Tahina,Bow19,Bow20,Naze20,Naze21,Rau21}. The red noise variability is thought to arise either from convection in a subsurface convection layer \citep{Cant,Gra15,Lecoanet,Cant21} or from gravity waves generated in the convective core of the  massive star \citep{Rog13,AR15,Bow19,Bow20}. The same mechanism is thought to be responsible for the velocity fields that cause macroturbulent broadening of the spectral lines of many massive stars \citep[][and references therein]{Cant21}. 

The detection of a significant red noise variability in the photometry of HD~108, is in line with the value of the macroturbulence ($64.4 \pm 0.4$\,km\,s$^{-1}$) inferred by \citet{Sun13}. Such a high value of macroturbulence is required to explain the line widths in the absence of significant rotational broadening. Whilst strong magnetic fields could in principle prevent such turbulent motion, \citet{Sun13} showed that the magnetic fields of Of?p stars, with the exception of NGC~1624-2, are not sufficient to inhibit convective motion inside the subsurface convective layer \citep[see also][]{Mac19}.

\begin{figure*}
\begin{center}
\begin{minipage}{8.5cm}
  \resizebox{8.5cm}{!}{\includegraphics{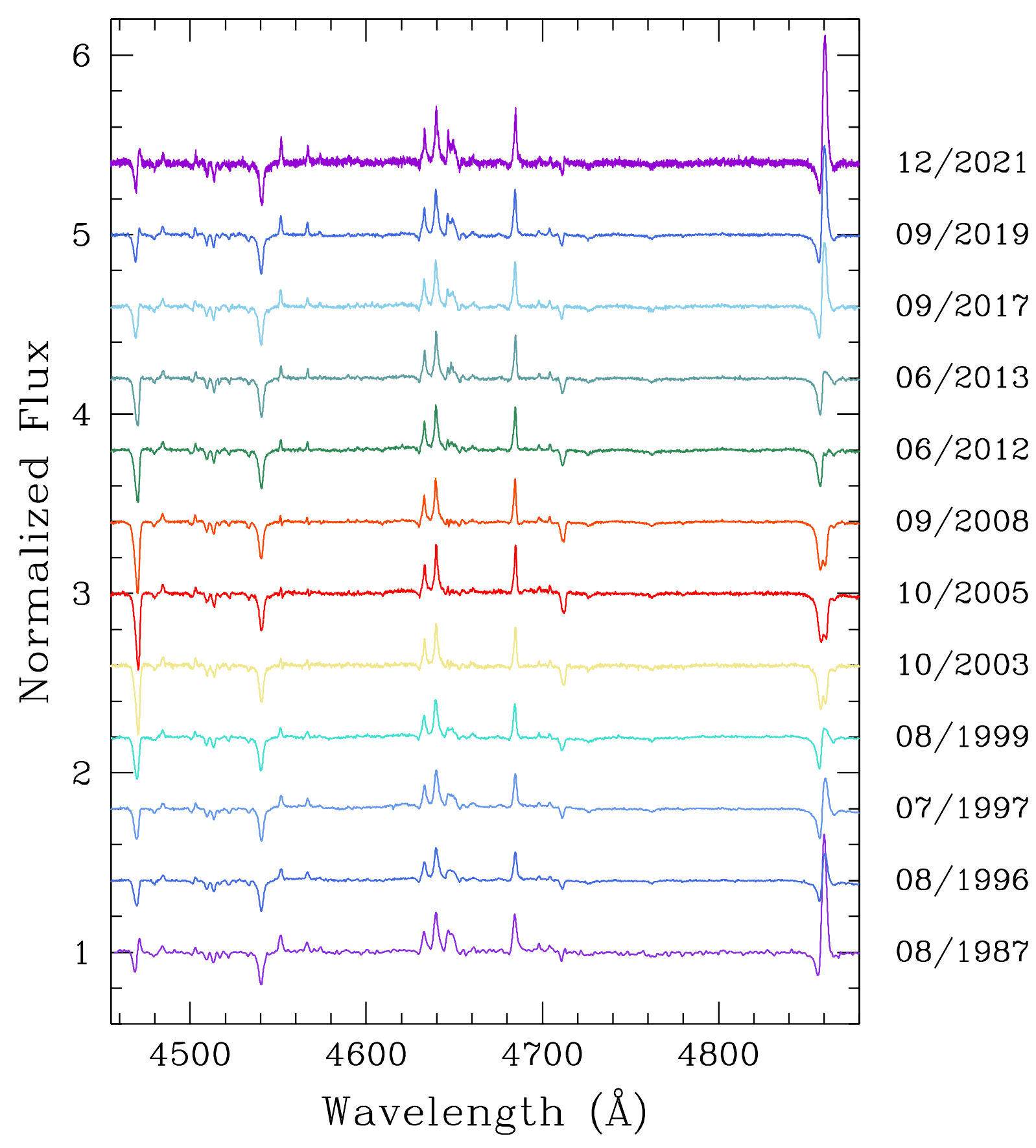}}
  \end{minipage}
  \hfill
  \begin{minipage}{8.3cm}
  \resizebox{8.3cm}{!}{\includegraphics{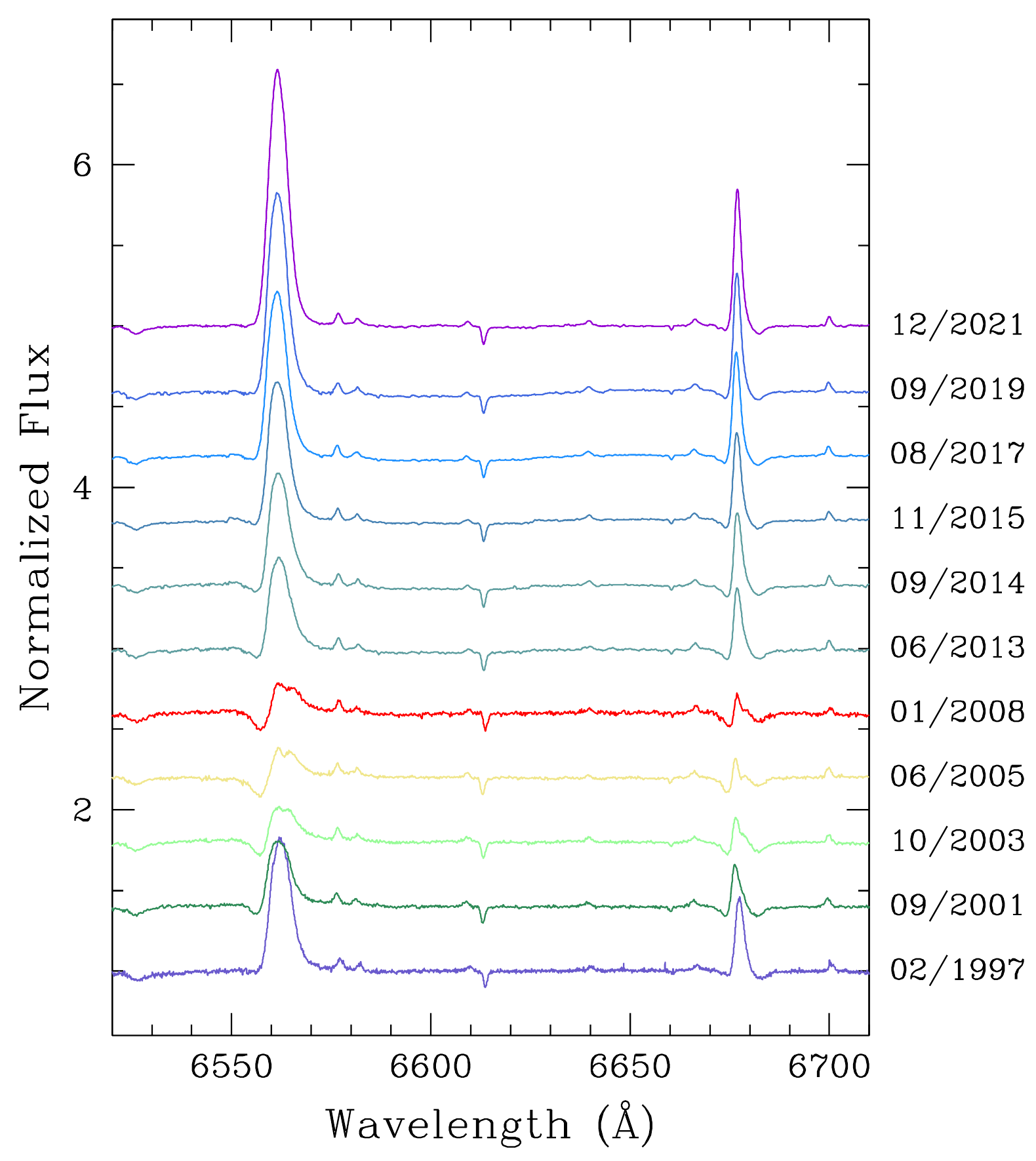}}
  \end{minipage}
\end{center}  
\caption{Long-term evolution of the blue (left) and red (right) spectrum of HD~108.\label{montage}}
\begin{center}
\begin{minipage}{8.5cm}
  \resizebox{8.5cm}{!}{\includegraphics{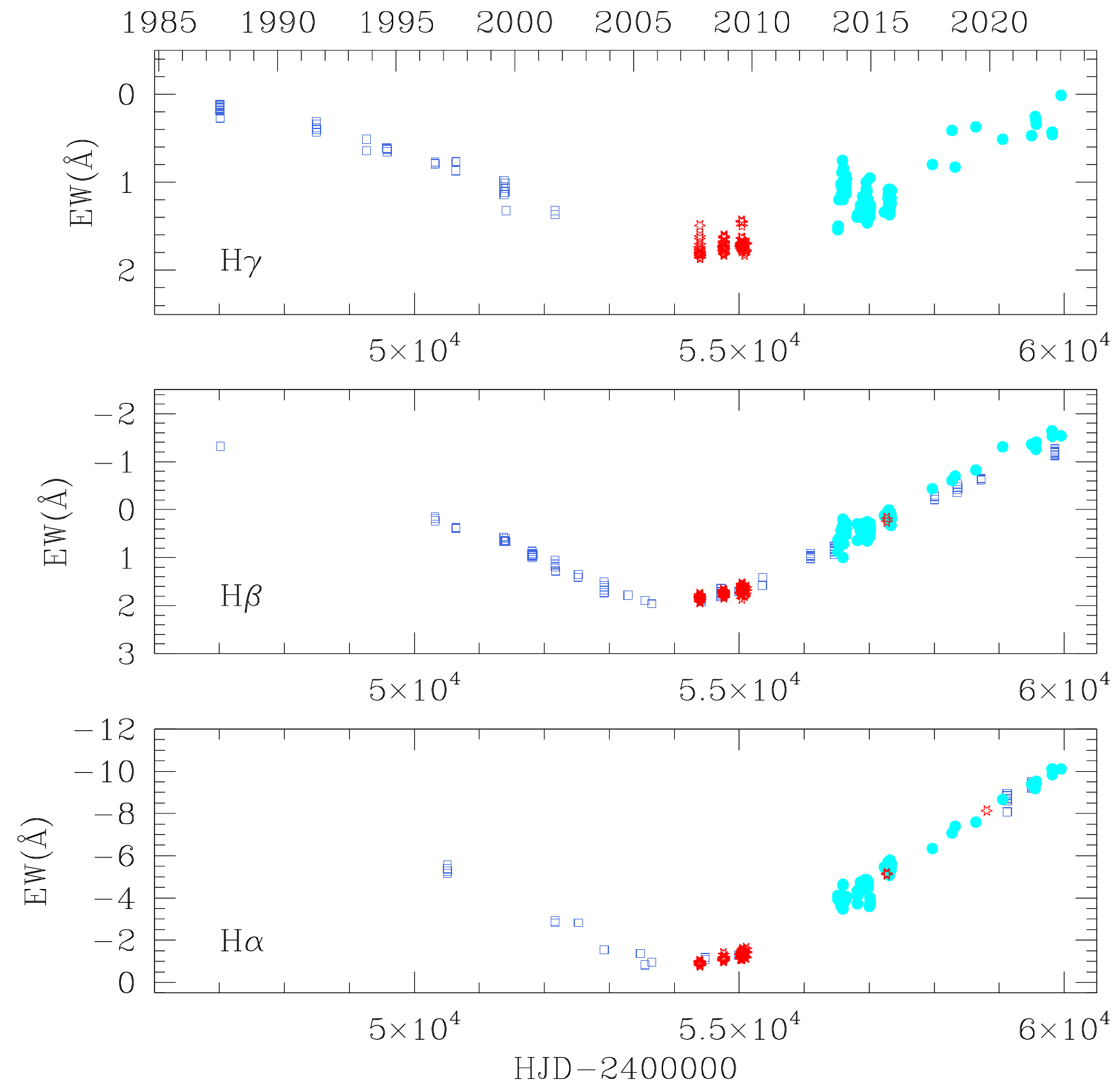}}
  \end{minipage}
  \hfill
  \begin{minipage}{8.3cm}
  \resizebox{8.3cm}{!}{\includegraphics{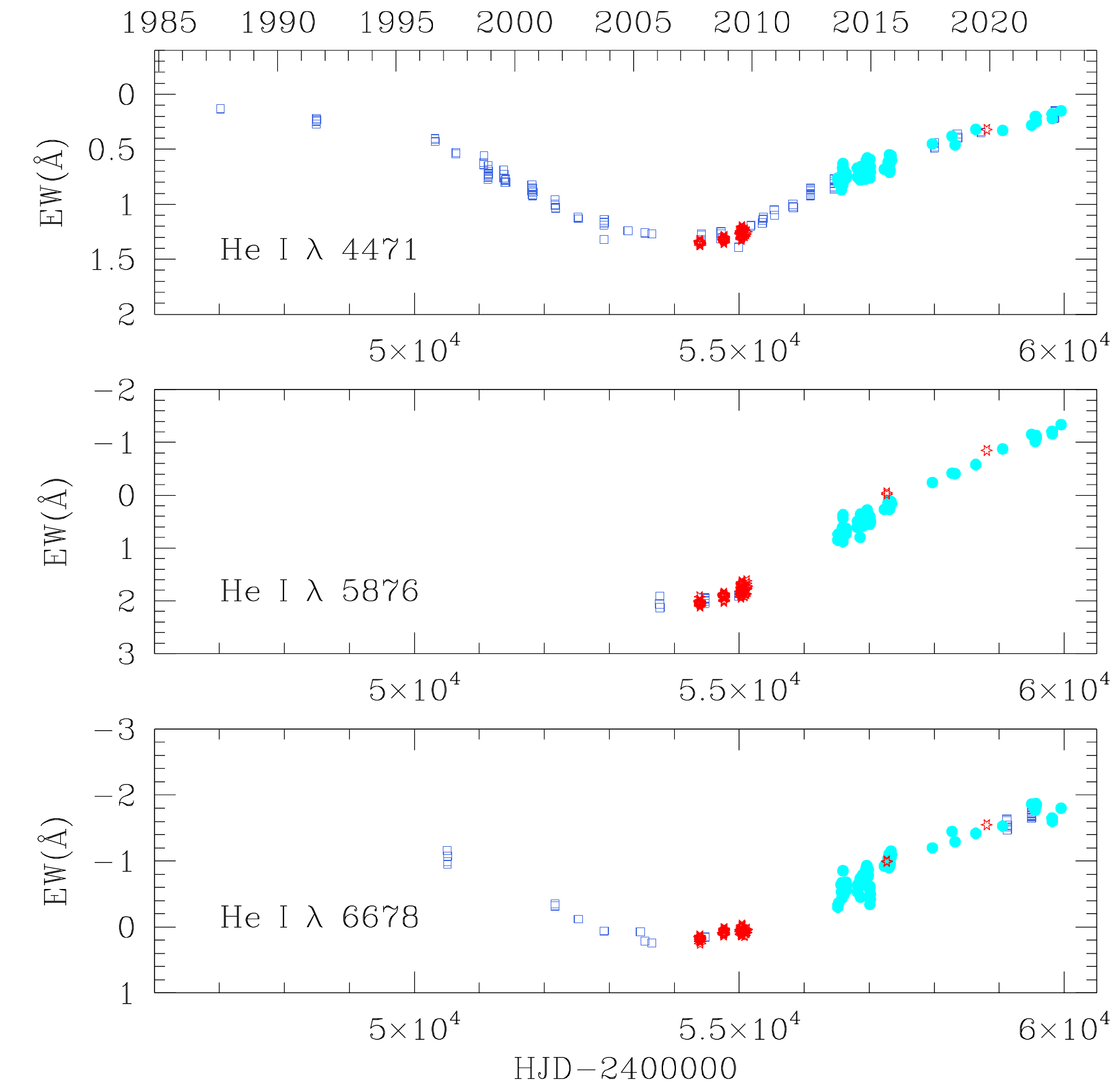}}
  \end{minipage}
\end{center}  
\caption{Long-term variations of the EWs of H\,{\sc i} (left) and He\,{\sc i} lines (right) in the spectrum of HD~108. Different symbols stand for data from different observatories: blue open squares indicate OHP data, red stars are data from the NARVAL (TBL) or ESPaDOnS (CFHT) instruments, whilst the cyan circles stand for TIGRE data. The horizontal axes are in heliocentric Julian dates, except for the top horizontal axes on each side which is expressed in calendar years.\label{EWs}}
\end{figure*}

\subsection{Long-term spectroscopic variations}\label{longtermspec}
The strength of the emission lines in the spectrum of HD~108 reached its minimum around 2007 \citep{Naz10}. Since then, the intensity of the emissions has progressively increased, reaching in the years 2021 and 2022 the level previously observed in the early 1990s (see Figs.\,\ref{montage} and \ref{EWs}). 
\begin{figure}
\begin{center}
  \resizebox{8.5cm}{!}{\includegraphics{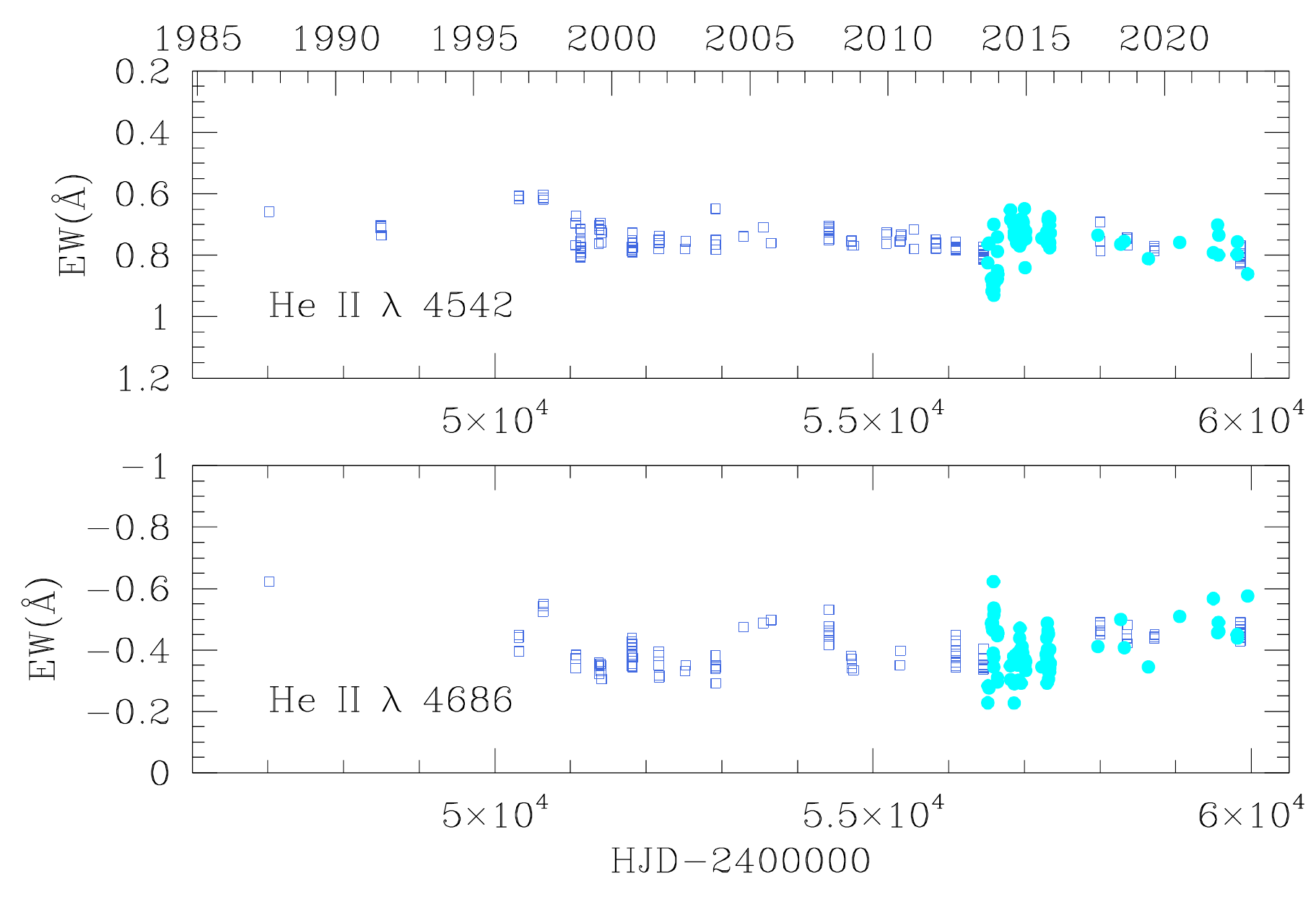}}
\end{center}  
\caption{EW of the He\,{\sc ii} $\lambda$\,4542 absorption (top panel) and the He\,{\sc ii} $\lambda$\,4686 emission (bottom panel) lines in the spectrum of HD~108 as a function of time. \label{EWHeII}}
\end{figure}

Over the part of the cycle covered by our data, the spectrum of HD~108 between 4450 and 4900\,\AA\ displays strong variations mostly in H$\beta$, He\,{\sc i} $\lambda\lambda$\,4471, 4713 and the C\,{\sc iii} emission around 4650\,\AA\ (see the left panel of Fig.\,\ref{montage} and Fig.\,\ref{EWs}). Lower level variability is also present in the emission of the Si\,{\sc iii} $\lambda\lambda$\,4552, 4568, 4575 triplet. Quite remarkably, the Si\,{\sc iii} $\lambda\lambda$\,4552, 4568, 4575 lines display an inverse P-Cygni-like profile in the years between 2003 and 2008, i.e.\ near minimum emission state. The variations of the H$\beta$ line and of the He\,{\sc i} lines mostly stem from changes of the narrow emission component superimposed on the photospheric absorption \citep{Naz08}. On the contrary, the He\,{\sc ii} $\lambda$\,4542 and N\,{\sc iii} $\lambda\lambda$\,4511 -- 4535 absorptions as well as the He\,{\sc ii} $\lambda$\,4686 and N\,{\sc iii} $\lambda\lambda$\,4634-42 emissions display essentially constant line profiles (see Figs.\,\ref{montage} and \ref{EWHeII}). The lack of strong variability of He\,{\sc ii} $\lambda$\,4686 contrasts with the case of HD~191\,612 where the strength of this emission is clearly variable \citep{Wal04}. Comparing HD~108 and HD~191\,612, we further note that the He\,{\sc i} $\lambda$\,4471 line in the spectrum of HD~191\,612 does not display an emission component above the continuum even at maximum emission strength, whilst such an emission component is seen in HD~108 in 1987 as well as in the 2021, 2022 and 2023 spectra. Also, the H$\beta$ emission of HD~108 is significantly stronger at maximum than in the case of HD~191\,612, and at the time of minimum emission strength, some residual emission remains present in the H$\beta$ line of HD~108, whilst none is seen in HD~191\,612.

Very similar remarks apply to the comparison between the variations of the H$\alpha$ and He\,{\sc i} $\lambda$\,6678 lines in HD~108 (see the right panel of Fig.\,\ref{montage} and Fig.\,\ref{EWs}) and HD~191\,612. For example, whilst the H$\alpha$ line goes into absorption during minimum state in HD~191\,612, the line remains in emission in the minimum state spectrum of HD~108.

\begin{figure}
\begin{center}
  \resizebox{8.5cm}{!}{\includegraphics{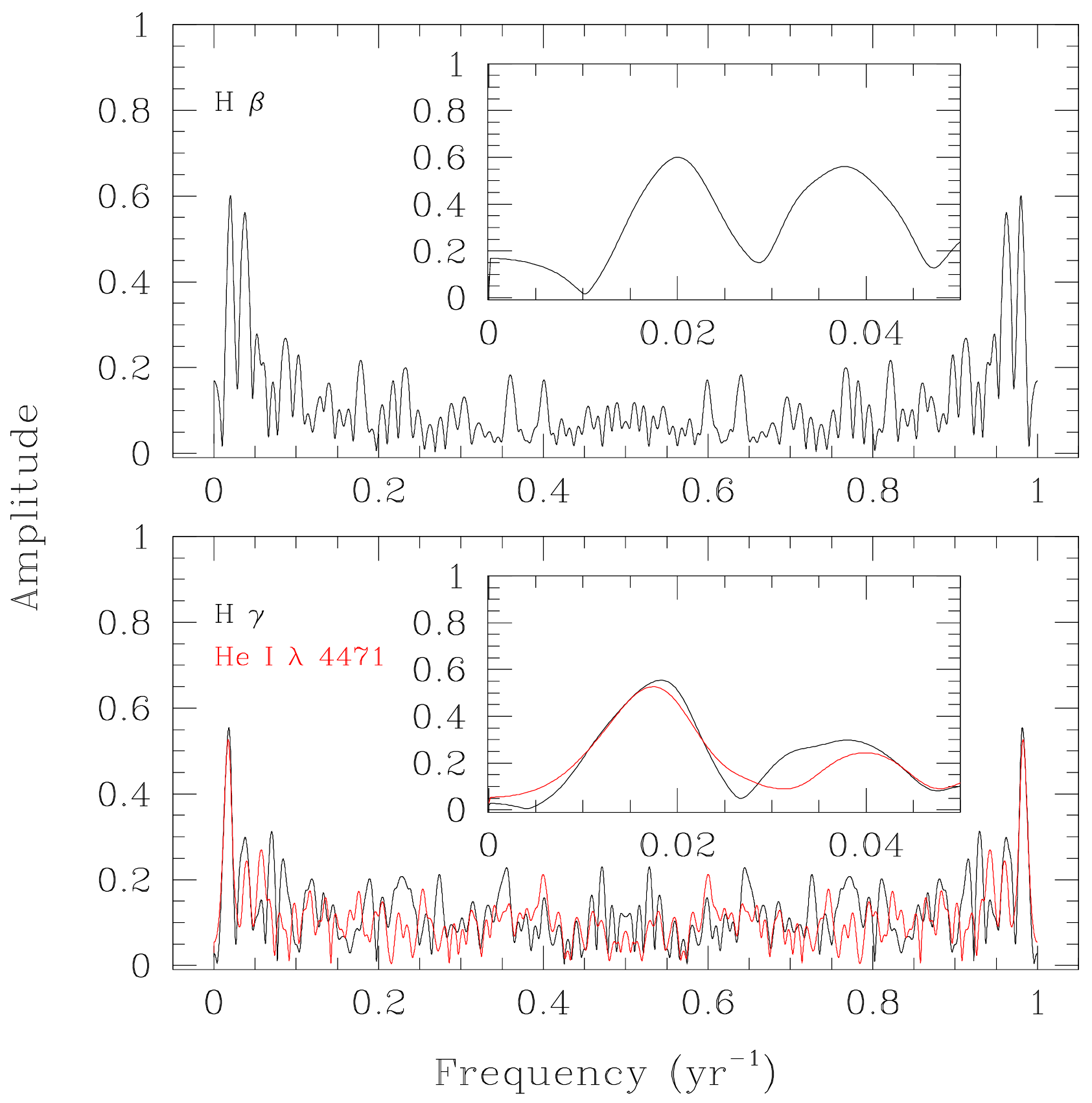}}
\end{center}  
\caption{Fourier spectra of the qualitative line profile nature of the H$\beta$ (top panel), H$\gamma$ (bottom panel, black) and He\,{\sc i} $\lambda$\,4471 (bottom panel, red) lines in the spectrum of HD~108. The inserts show a zoom on the low-frequency part of the Fourier spectra.\label{Fourlongcycle}}
\end{figure}
\begin{figure}
\begin{center}
  \resizebox{8.5cm}{!}{\includegraphics{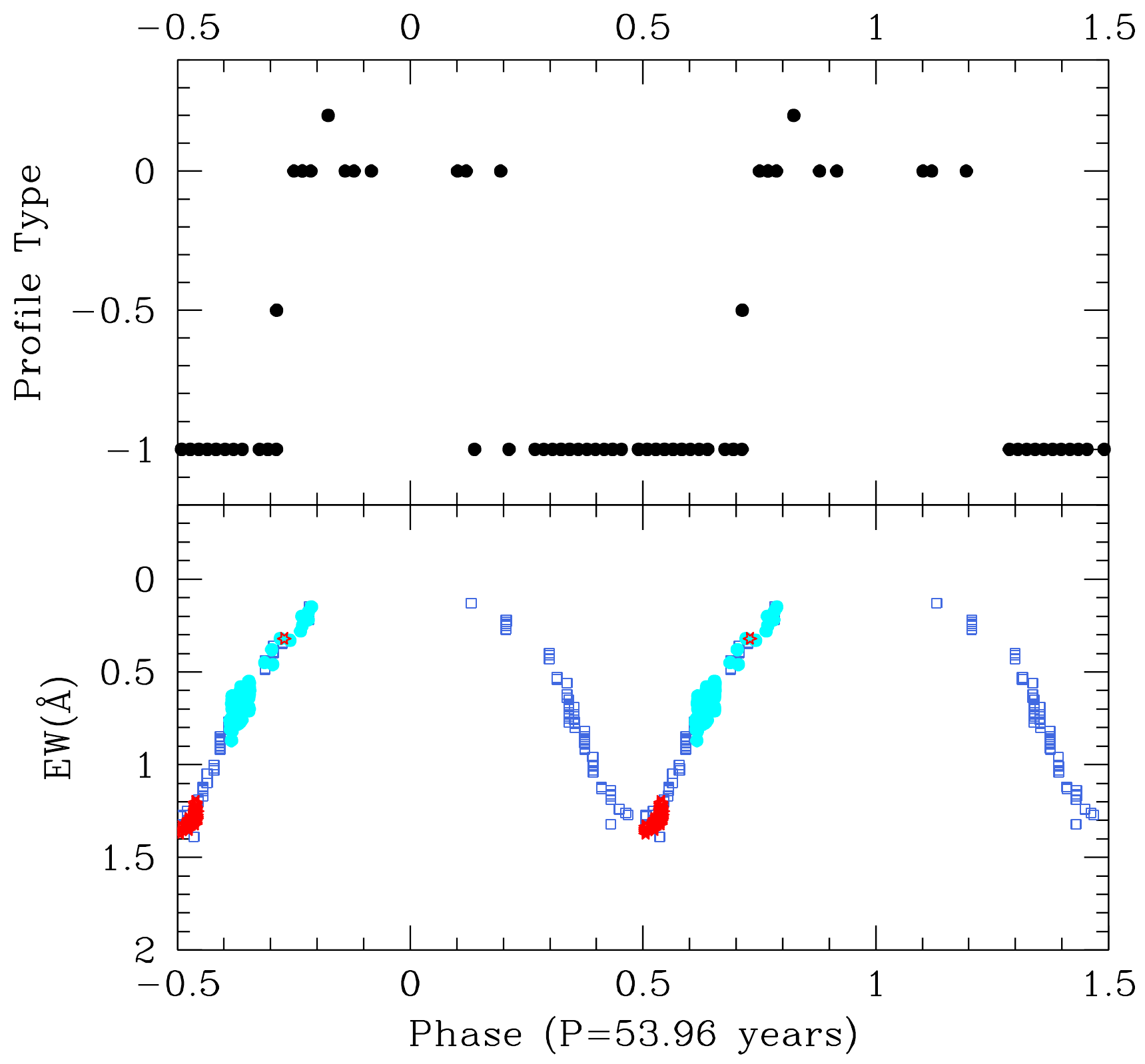}}
\end{center}  
\caption{Phase-folded variations of the He\,{\sc i} $\lambda$\,4471 line in the spectrum of HD~108. The top panel illustrates the variations of the qualitative line profile nature as estimated from literature data. The bottom panel displays our own measurements of the EW of the line. Phase $\phi = 0.5$ (i.e.\ minimum emission) corresponds to HJD\,2454284.5.\label{cycle4471}}
\end{figure}

The variable emission lines grow in intensity, but are rather narrow features. For instance, the FWHM of the H$\alpha$ emission component in the 2022 spectra amounts to 5.3\,\AA, corresponding to 245\,km\,s$^{-1}$. We measured the equivalent widths (EWs) of the most frequently observed variable lines by integrating the continuum-normalised spectra between specific boundaries. For the H$\gamma$, H$\beta$ and H$\alpha$ lines, we respectively used the wavelength intervals [4320\,\AA, 4350\,\AA], [4845\,\AA, 4875\,\AA] and [6550\,\AA, 6575\,\AA]. For the He\,{\sc i} lines at $\lambda$\,4471, $\lambda$\,5876 and $\lambda$\,6678, we respectively used [4465\,\AA, 4475\,\AA], [5866\,\AA, 5881\,\AA] and [6664\,\AA, 6694\,\AA]. We further measured the EWs of the less variable He\,{\sc ii} $\lambda$\,4542 absorption and He\,{\sc ii} $\lambda$\,4686 emission, by integrating over the wavelength intervals [4535.6\,\AA, 4547.9\,\AA] and [4680.2\,\AA, 4690.3\,\AA]. The results are illustrated in Figs.\,\ref{EWs} and \ref{EWHeII}. For the He\,{\sc ii} lines, we confirm the absence of a well-defined variability with EW(He\,{\sc ii} $\lambda$\,4542) = $0.75 \pm 0.05$\,\AA, and EW(He\,{\sc ii} $\lambda$\,4686) = $0.40 \pm 0.07$\,\AA, with the quoted standard deviation corresponding to the scatter of our full sets of data. For the variable emission lines, if we consider measurements taken over the same night or consecutive nights, we estimate that the typical standard deviation on the EWs amounts to about 5\% of their value. This value is of the same order as the observed dispersion of the He\,{\sc ii} lines. It accounts not only for the measurement uncertainties (typically about 2\% of the EW value as estimated from observations taken within less than 1\,hr of each other), but also includes the contribution of genuine short-term variations (see Sect.\,\ref{short}).

To re-assess the duration of the long emission cycle of HD~108, we combined the compilations of archival data of \citet{Bru71}, \citet{And73}, \citet{Naz01}, and \citet{Naz06} with our new data. This yields a time series spanning from 1919 until 2022. We used the following scheme to convert qualitative information on the nature of the line profiles from the oldest references into quantitative data: epochs when a line was observed in strong absorption were assigned a numerical value of $-1$, dates when the line was seen as an apparent P-Cygni profile were given a value of $0$, and epochs when the line was in net emission were assigned a value $+1$. Some intermediate values were assigned based on the details of the line description in the original reference. We then applied the Fourier method for uneven sampling of \citet{Hec85} and \cite{Gos01} to this time series. The most extensive time series are those for the H$\gamma$, H$\beta$ and He\,{\sc i} $\lambda$\,4471 lines. The resulting periodograms are illustrated in Fig.\,\ref{Fourlongcycle}.

The strongest peaks are found at 0.0200, 0.0182, 0.0174\,yr$^{-1}$ for the H$\beta$, H$\gamma$, and He\,{\sc i} $\lambda$\,4471 lines, respectively. Therefore, our current best value for the duration of the long cycle of HD~108 is $54 \pm 3$\,years, in good agreement with previous determinations \citep{Naz01,Naz10}. For the time of minimum emission line strength (defined as phase $\phi = 0.5$), we obtain a best estimate of HJD\,2454284.5. Folding our EW measurements with this period, it is interesting to note that this suggests that, unlike HD~191\,612, HD~108 spends more time near the emission maximum than near the minimum. Indeed, whilst the EW(H$\alpha$) variations of HD~191\,612 reveal an extended flat minimum that lasts for about one third of its cycle, and a more sharply peaked maximum \citep{How07}, the existing data on HD~108 rather suggest the opposite situation (i.e.\ a sharply peaked minimum, with a duration below five years (or $< 10\%$ of the cycle) and an extended (flat) maximum, see Fig.\,\ref{cycle4471}). In the oblique magnetic rotator model, these different behaviours most likely stem from different values of the inclination angle of the stellar rotation axes and different values of the obliquity of the magnetic axes (see Sect.\,\ref{ADMfit}).

Although we lack a continuous photometric monitoring of HD~108 over its full cycle, it seems likely that the spectral changes go along with brightness changes. Indeed, \citet{Bar07} reported a roughly constant optical brightness between 1989 and 1994. In the following years, the brightness declined until at least 2006 (i.e.\ the last epoch covered by these observations), which coincides with the epoch of the minimum emission state.
 
\begin{figure*}
  \begin{center}
    \begin{minipage}{5.5cm}
      \resizebox{5.5cm}{!}{\includegraphics{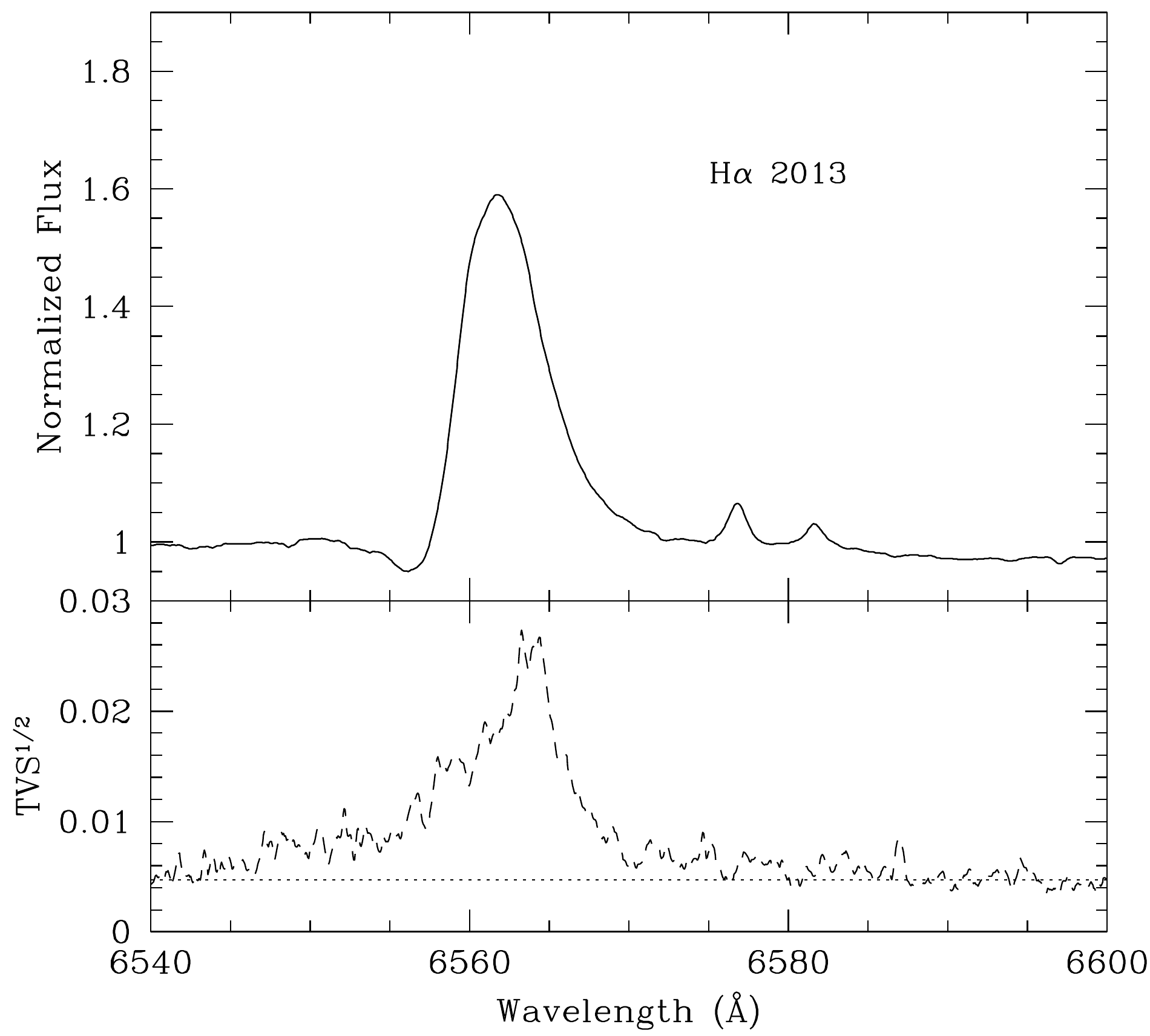}}
    \end{minipage}
    \hfill
    \begin{minipage}{5.5cm}
      \resizebox{5.5cm}{!}{\includegraphics{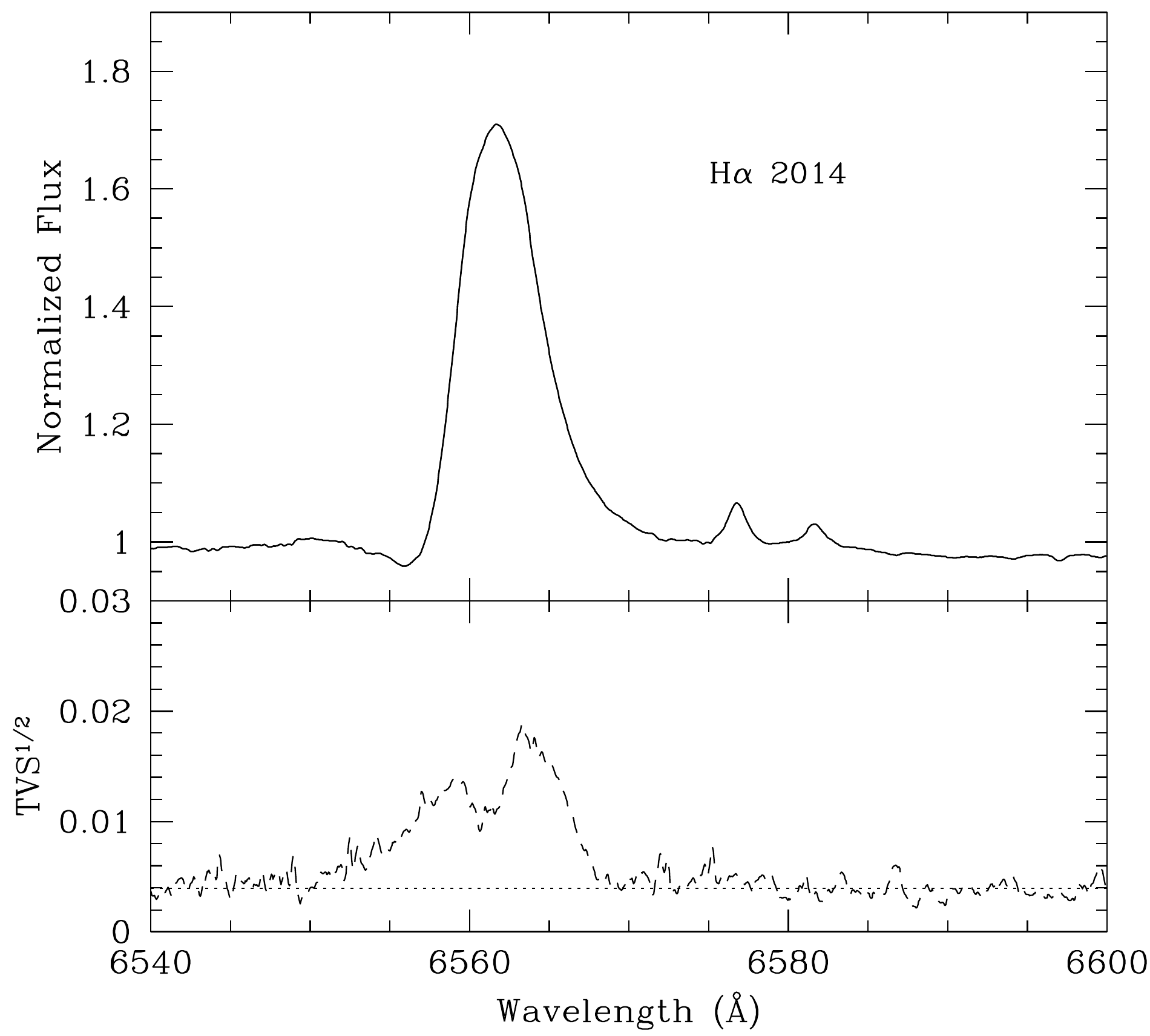}}
    \end{minipage}
    \hfill
    \begin{minipage}{5.5cm}
      \resizebox{5.5cm}{!}{\includegraphics{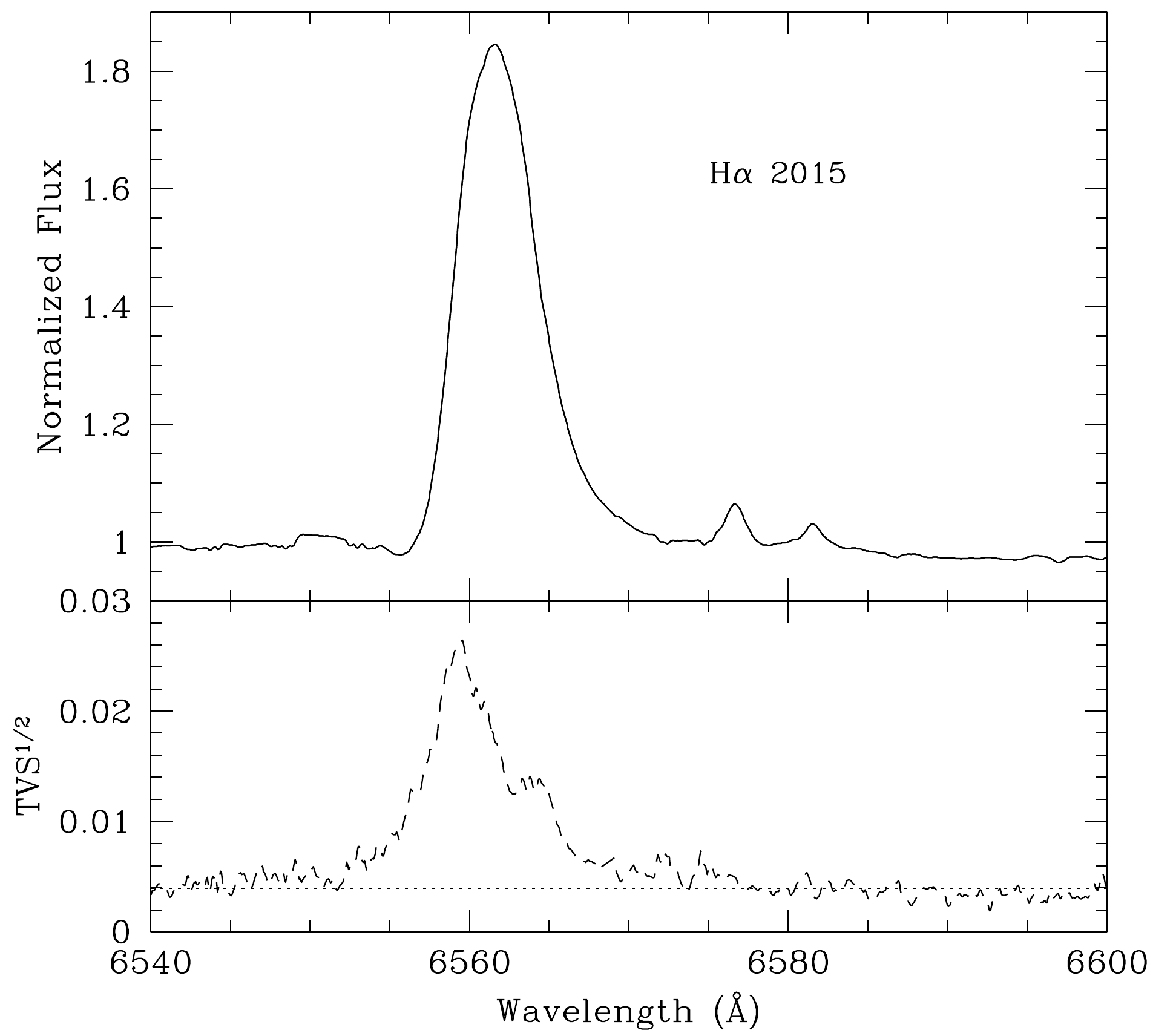}}
    \end{minipage}
  \end{center}  
\caption{Normalized mean spectrum (top panels) and TVS$^{1/2}$ (bottom panels) of the spectroscopic time series from 2013 (left column), 2014 (middle column) and 2015 (right column). The short-dashed horizontal lines in the bottom panels correspond to the 99\% significance level of variability estimated from the S/N ratios of our spectra following the method of \citet{Ful96}.\label{shortterm}}
\end{figure*}

\subsection{Short-term spectroscopic variations}\label{short}
\citet{Mar10} noted the presence of short-term spectroscopic variability mostly in the red part of the He\,{\sc i} and wind-sensitive line profiles. On the contrary, photospheric He\,{\sc ii} lines were found to be very stable. \citet{Mar10} interpreted these changes as evidence for the infall of material from the magnetosphere, a consequence of dynamical phenomena within the magnetically channeled winds \citep{udD13}.

To investigate the properties of such variations, we performed a more intensive spectroscopic monitoring with TIGRE in 2013, 2014 and 2015. Table\,\ref{Tigmon} provides an overview of the data collected during these campaigns. We focus here on the H$\alpha$ emission line as it offers an efficient diagnostic of the cool plasma in the magnetosphere. 
\begin{table}
  \caption{Main properties of the more intensive TIGRE monitoring \label{Tigmon}}
  \begin{tabular}{c c c c c c c}
    \hline
    Year & $N$ & $<S/N>$ & $\Delta\,t$ & EW(H$\alpha$) & $\nu_{\rm peak, 1}$ & $\nu_{\rm peak, 2}$ \\
    &   &  &  (days)  & (\AA)         & (d$^{-1}$)   & (d$^{-1}$) \\
    \hline
    2013 & 17 & 299 & 132 & $-3.55 \pm 0.23$ &  0.033 & 0.138 \\
    2014 & 20 & 351 & 45  & $-4.23 \pm 0.22$ &  0.178 & 0.158 \\
    2015 & 23 & 344 & 52  & $-5.09 \pm 0.19$ &  0.301 & 0.116 \\
    \hline
  \end{tabular}
  
      {\scriptsize $N$ is the number of spectra taken, $<S/N>$ their mean signal-to-noise ratio evaluated in a line free region between 6757 and 6767\,\AA, and $\Delta\,t$ the time intervall of our observations. The fifth column indicates the mean EW(H$\alpha$) evaluated between 6550 and 6575\,\AA, and its dispersion about the mean. The last two columns yield the frequencies of the highest and the second-highest peaks in the associated Fourier power spectra.}
\end{table}

Our data reveal that the level of variability of EW(H$\alpha$) is quite similar from one year to the other despite the varying overall line strength (see column 5 of Table\,\ref{Tigmon}). The scatter is typically of about 0.2\,\AA, corresponding to about 5\% of the emission EW of the H$\alpha$ line.

Figure\,\ref{shortterm} illustrates the temporal variance spectrum \citep[TVS,][]{Ful96} of the corresponding time series of spectra. The level of variability of the H$\alpha$ line (above the noise level) typically peaks at about 2\% of the continuum. Significant variability is restricted to a relatively narrow wavelength range, from 6555 to 6567\,\AA. We used the modified Fourier periodogram algorithm of \citet{Hec85} and \citet{Gos01} over this wavelength interval to identify the dominant frequencies in these short-term spectroscopic variations. In general, our Fourier periodograms lack a clearly dominant frequency, but rather consist of a number of peaks with relatively similar amplitudes. The last two columns of Table\,\ref{Tigmon} list the frequencies associated with the strongest peaks in the power spectra of the time series of the different epochs. The two highest peaks in the 2013 Fourier spectrum have a power 6.7 and 4.7 times higher than the noise level. For the 2014 and 2015 data, the power of the highest peak reaches 3.0 times the noise level, whilst the second highest peak reaches 2.5 and 2.0 times the noise level respectively in 2014 and 2015. The highest peak in the 2013 Fourier spectrum occurs at a rather low frequency which likely stems from the long-term variations of the H$\alpha$ line strength and the fact that the 2013 data are spread over nearly five months. Within the 0 -- 1\,d$^{-1}$ frequency interval, the second-highest peak of the 2013 periodogram occurs at 0.138\,d$^{-1}$. The 2014 and 2015 time series were taken over about 1.5 month making them less sensitive to the long-term variations. The strongest peaks in their Fourier spectra are found between 0.116 and 0.301\,d$^{-1}$. Whilst some of these frequencies are relatively close to the transient signal found in the sector 17 and 18 {\it TESS} photometry, none of them matches exactly with the photometric frequency. Moreover, none of the spectroscopic frequencies appears to be stable from one epoch to the other. We thus conclude that these frequencies likely reflect the recurrence of quasi-cyclic variations, occuring on timescales of a few days to about a week, rather than genuine periodicities.  

\subsection{Radial velocities}\label{radvel}
Binarity with very different periods has been suggested by a number of authors. \citet{Hut75} and \citet{Asl89} proposed rather short orbital periods of 4.6 and 5.8\,days respectively. Such short periods were ruled out by \citet{Naz01} based on intense short-term radial velocity (RV) monitorings. \citet{Bar99} claimed the detection of periodic RV variations with a period of 1627.6\,days, $K = 10.5$\,km\,s$^{-1}$ and $e = 0.43$. This result could not be confirmed though \citep{Naz01}. \citet{Naz01,Naz06} and \citet{Naz10} reported RV variations on timescales of several thousand days. Based on data collected over 18 years, \citet{Tri21} discarded binarity of HD~108. Yet, their data (e.g.\ their Fig.\,10) did show RV variations over a range of about 10 -- 15\,km\,s$^{-1}$. These authors attributed them to pulsational activity.

We decided to revisit this issue taking advantage of the large dataset at our disposal. We measured the RVs of a number of absorption and emission lines in the spectrum of HD~108. We focused on lines that do not display profile variations due to the long-term cycle. We thus measured RVs of the He\,{\sc ii} $\lambda\lambda$\,4200, 4542, N\,{\sc iii} $\lambda\lambda$\,4511 -- 4535, O\,{\sc iii} $\lambda$\,5592 absorption lines and of the He\,{\sc ii} $\lambda$\,4686 and C\,{\sc iii} $\lambda$\,5696 emission lines. For the He\,{\sc ii}, O\,{\sc iii} and C\,{\sc iii} lines, we fitted Gaussian profiles to the lower (resp.\ upper) 2/3 of the absorption (resp.\ emission) line. For the N\,{\sc iii} lines, we instead performed a cross-correlation between the observed spectra and the 4500 -- 4538\,\AA\ spectral region in a synthetic TLUSTY spectrum\footnote{For the TLUSTY template spectrum, we adopted $T_{\rm eff} = 35\,000$\,K and $\log{g} = 3.5$.} \citep{Lan03}. The RVs were then estimated by fitting a parabola to the peak of the correlation function \citep{Ver99}.

\begin{figure}
\begin{center}
  \resizebox{8.5cm}{!}{\includegraphics{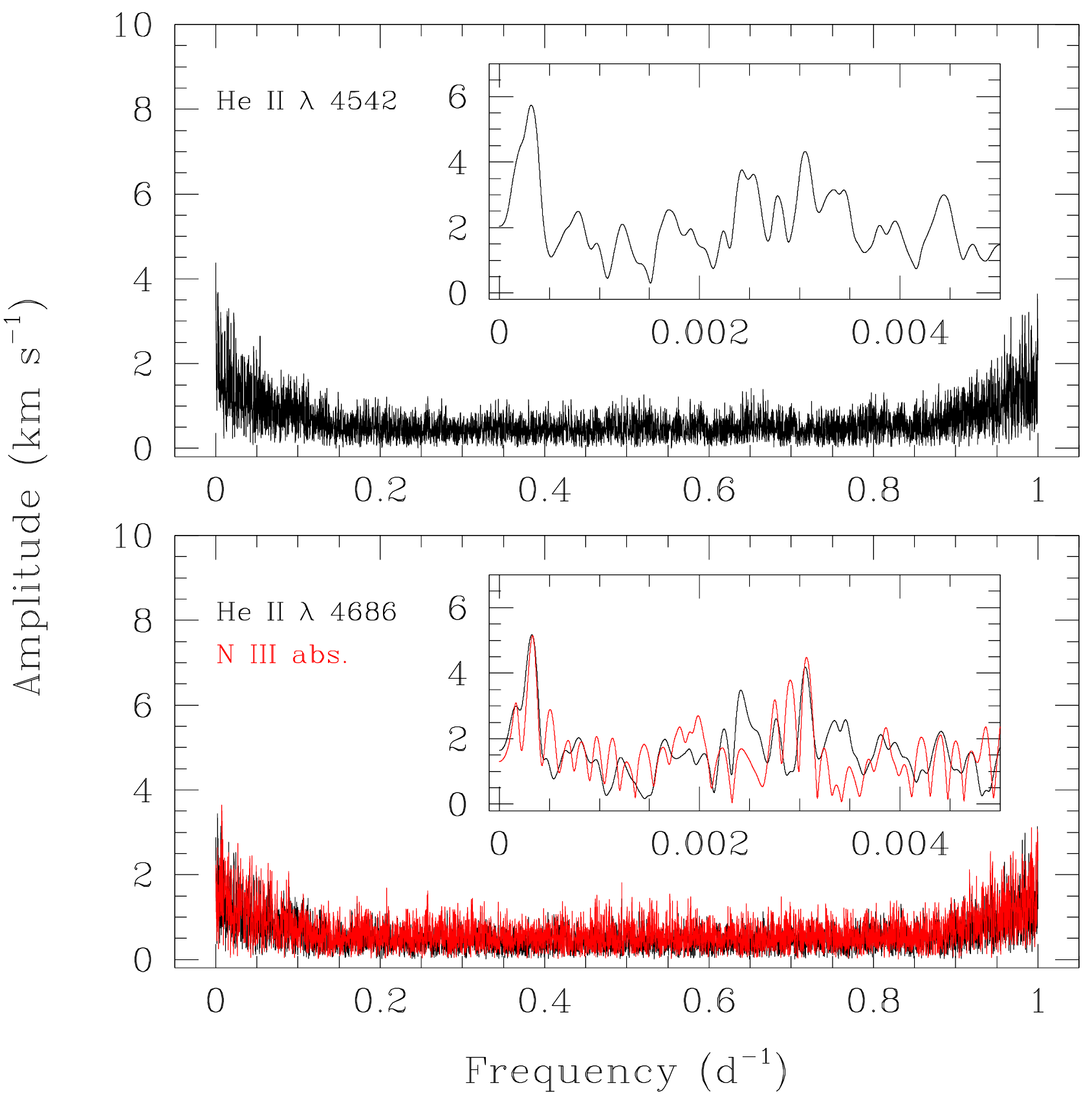}}
\end{center}  
\caption{Fourier periodograms of the RVs of the He\,{\sc ii} $\lambda$\,4542 absorption (top panel), the He\,{\sc ii} $\lambda$\,4686 emission (bottom panel, black) and the N\,{\sc iii} $\lambda\lambda$ 4511 - 4535 absorptions (bottom panel, red) in the spectrum of HD~108. The inserts zoom on the low-frequency part of the Fourier spectra, revealing in each case a dominant peak near $3.2\,10^{-4}$\,d$^{-1}$.\label{FourRVs}}
\end{figure}

\begin{table*}
  \caption{SB1 orbital solutions of HD~108. $T_0$ stands for the time of conjunction with the Of?p primary star in front, whilst $v_0$ corresponds to the apparent systemic velocity. $K_1$ and $a_1$ are the velocity amplitude and semi-major axis of the Of?p primary star. The mass function is equal to $f(m) = \frac{m_2^3\,\sin^3{i}}{(m_1 + m_2)^2}$.}
  \begin{center}
  \begin{tabular}{l c c c}
\hline
& He\,{\sc ii} $\lambda$\,4542 & He\,{\sc ii} $\lambda$\,4686 & N\,{\sc iii} $\lambda\lambda$\,4511 -- 4535 \\
\hline
$P_{\rm orb}$ (days) & 3194 & 3116 & 3072 \\
$T_0$ $(HJD - 2450000)$ & $8808.9 \pm 18.5$ & $8770.4 \pm 21.7$ & $8612.8 \pm 29.0$ \\ 
$v_0$ (km\,s$^{-1}$) & $-68.8 \pm 0.1$ & $-64.3 \pm 0.1$ & $-86.5 \pm 0.3$ \\
$K_1$ (km\,s$^{-1}$) & $5.9 \pm 0.2$ & $5.2 \pm 0.2$ & $5.8 \pm 0.5$ \\
$a_1\,\sin{i}$ (R$_{\odot}$) & $369.7 \pm 11.7$ & $321.8 \pm 10.9$ & $349.4 \pm 27.3$ \\ 
$f(m)$ (M$_{\odot}$) & $0.067 \pm 0.006$ & $0.046 \pm 0.005$ & $0.061 \pm 0.014$ \\
r.m.s.\ (km\,s$^{-1}$) & 2.17 & 2.17 & 3.35 \\
\hline
\end{tabular}
  \label{orbsol}
  \end{center}
  
%      {\scriptsize $T_0$ stands for the time of conjunction with the Of?p primary star in front, whilst $v_0$ corresponds to the apparent systemic velocity. $K_1$ and $a_1$ are the velocity amplitude and semi-major axis of the Of?p primary star. The mass function is equal to $f(m) = \frac{m_2^3\,\sin^3{i}}{(m_1 + m_2)^2}$.}
\end{table*}
\begin{figure*}
\begin{center}
\begin{minipage}{8.5cm}
  \resizebox{8.5cm}{!}{\includegraphics{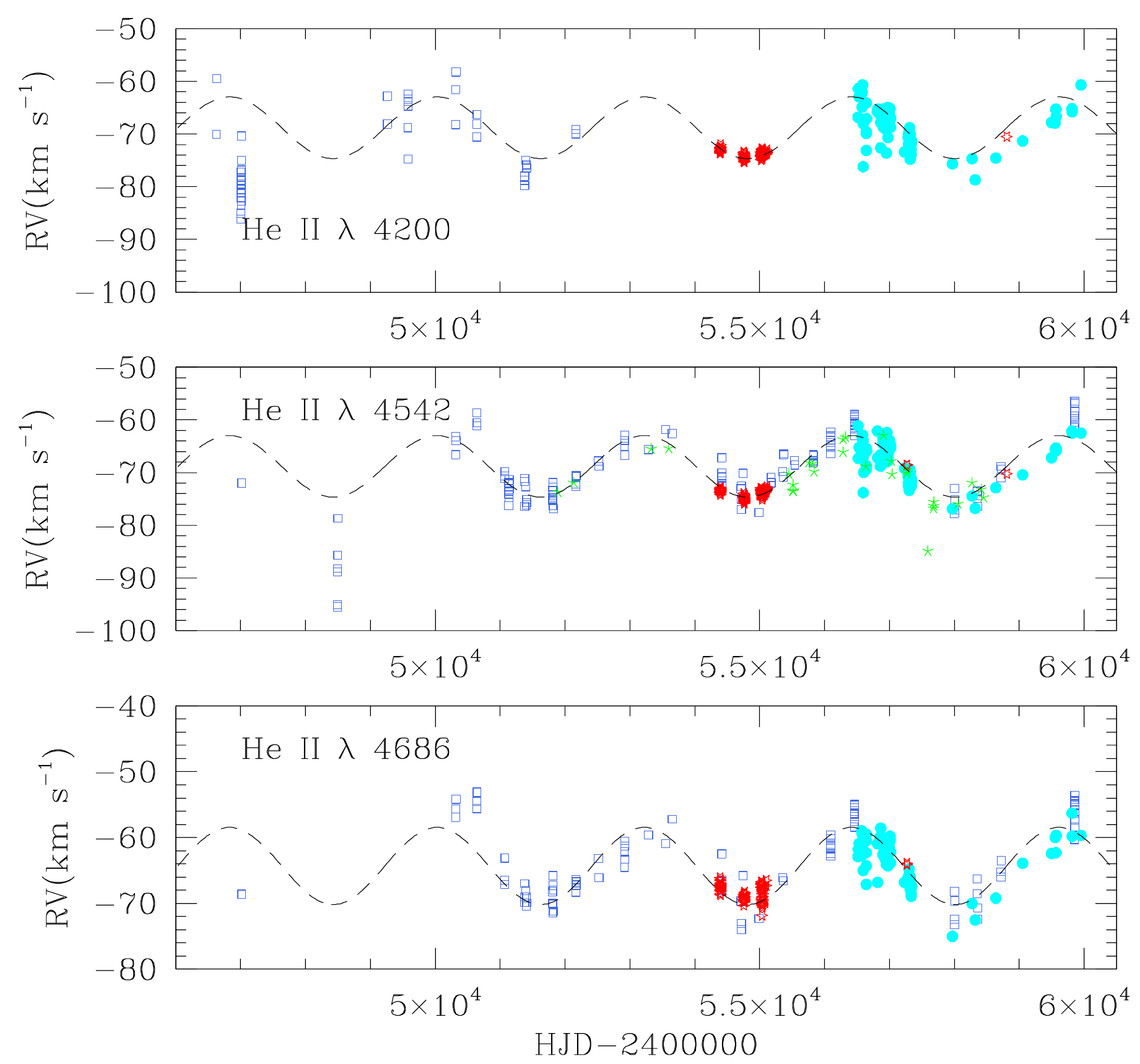}}
  \end{minipage}
  \hfill
  \begin{minipage}{8.5cm}
  \resizebox{8.5cm}{!}{\includegraphics{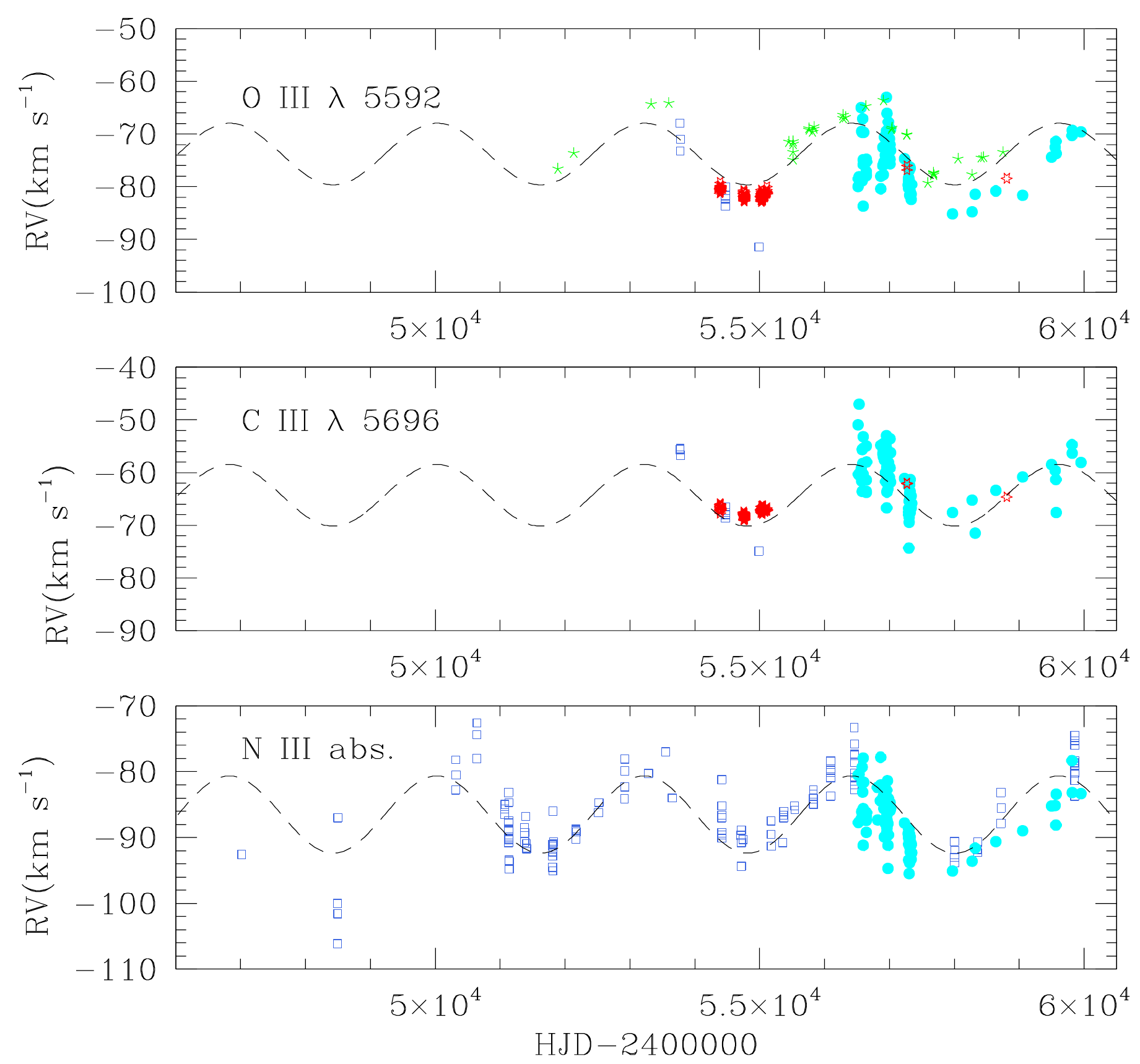}}
  \end{minipage}
\end{center}  
\caption{Left: RVs of the He\,{\sc ii} lines in the spectrum of HD~108. Different symbols refer to data from different observatories: blue open squares indicate data from OHP, red stars stand for NARVAL (TBL) or ESPaDOnS (CFHT) spectra, cyan dots correspond to TIGRE spectra, and green stars indicate RVs listed by \citet{Tri21}. The dashed line yields the best-fit SB1 orbital solution of the He\,{\sc ii} $\lambda$\,4542 line, shifted to the systemic velocity of the corresponding line. Right: same for several other absorption and emission lines. \label{RVs}}
\end{figure*}

The RVs of the He\,{\sc ii} lines reveal at least three cycles of a $\sim 10$\,km\,s$^{-1}$ modulation on a timescale close to 3000\,days. We searched for periodicities among the most densly sampled time series of RVs using the same Fourier method as above (see Fig.\,\ref{FourRVs}). In this way, we obtained periods of $3194 \pm 98$\,d for He\,{\sc ii} $\lambda$\,4542, $3114 \pm 94$\,d for He\,{\sc ii} $\lambda$\,4686, and $3076 \pm 90$\,d for N\,{\sc iii} $\lambda\lambda$\,4511 -- 4535. Within their errors, these periods agree. We note that the RV data of \citet{Tri21} agree quite well with ours (see Fig.\,\ref{RVs}).

We note that the dispersion of the RVs measured on observations taken over timescales of hours and days (i.e.\ the typical timescales of pulsations in massive stars) is significantly smaller ($\sigma_{\rm RV}$ between 0.4 and 1.7\,km\,s$^{-1}$ depending on the spectral resolution of the data) than the $\sim 10$\,km\,s$^{-1}$ amplitudes measured above. A periodicity of order 3000\,days is unlikely to be associated with pulsations in a massive near main-sequence star. Moreover, since the spectral lines selected for the RV measurements do not display line profile variations related to the long-term rotational cycle, we can safely exclude a connection between the two phenomena. Finally, we stress that lines from different ions, either in absorption or emission, display the same RV variations (both in phase and amplitude). Therefore, the most likely explanation for these periodic RV variations is binarity. We thus applied the Li\`ege Orbital Solution Package \citep[LOSP,][]{San06} to establish an SB1 orbital solution. We tested both circular and eccentric orbital solutions. For each line, we found that the eccentric solution did not improve the quality of the fits and that best-fit eccentricities were small and, for the N\,{\sc iii} lines, only marginally significant. Moreover, different lines yielded discrepant values of $e$ (e.g.\ $e = 0.12 \pm 0.03$ for  He\,{\sc ii} $\lambda$\,4542 and $e = 0.16 \pm 0.03$ for He\,{\sc ii} $\lambda$\,4686 versus $e = 0.06 \pm 0.06$ for the N\,{\sc iii} $\lambda\lambda$\,4511 -- 4535 lines) as well as of the longitude and time of periastron passage. We thus decided to focus on the solutions obtained assuming $e = 0$. The results are listed in Table\,\ref{orbsol}, and the best-fit orbital solution for the RVs of the  He\,{\sc ii} $\lambda$\,4542 line is shown in Fig.\,\ref{RVs}. Our present study thus confirms earlier reports by \citet{Naz10} and strongly suggests that HD~108 is a binary system with an orbital period near 8.5\,years, and a relatively low eccentricity.

\begin{figure}
\begin{center}
  \resizebox{8.5cm}{!}{\includegraphics{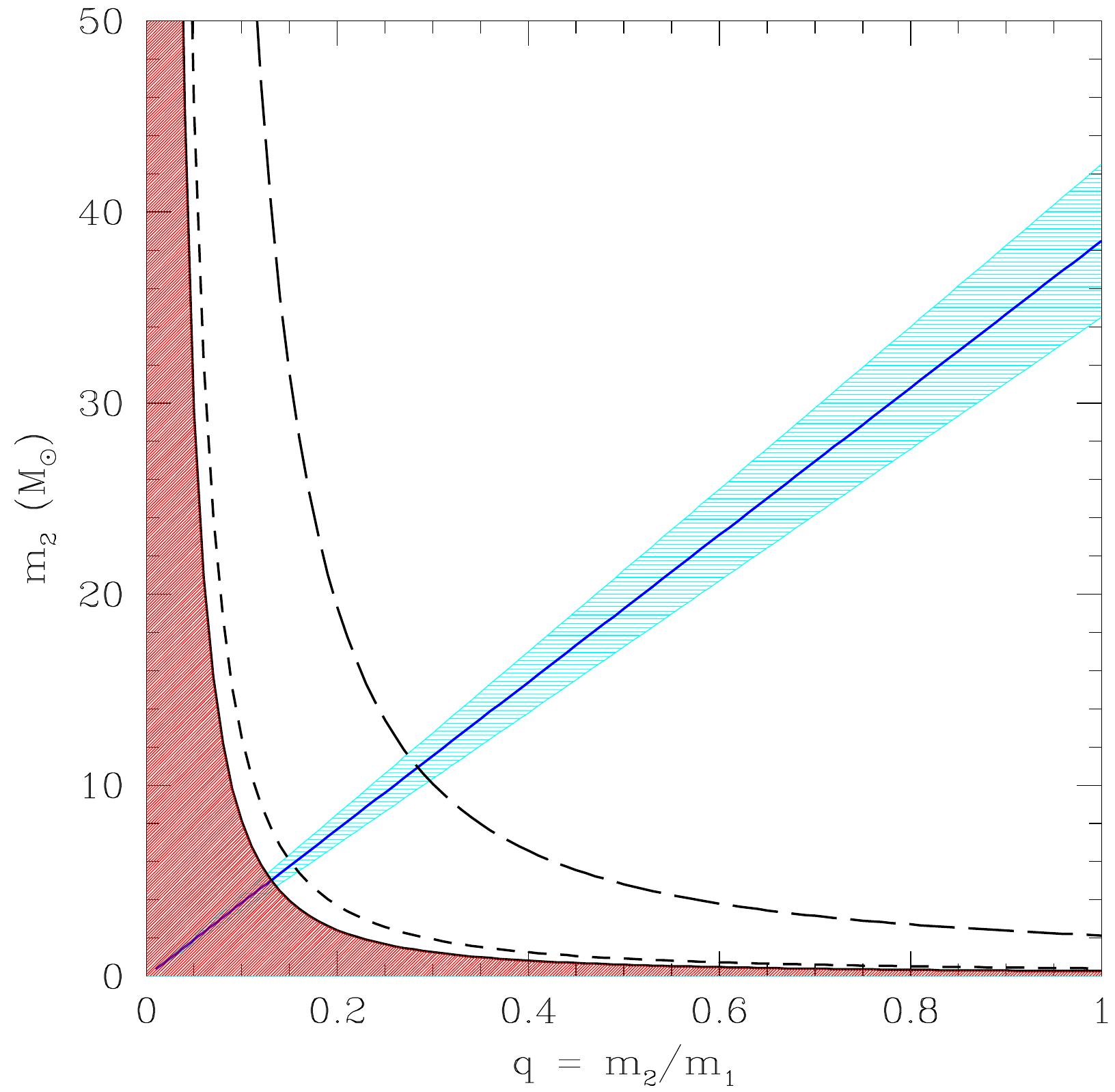}}
\end{center}  
\caption{Constraints on the mass of the companion of HD~108. The straight blue line corresponds to the companion mass computed as $m_2 = q\,m_1$ with $m_1$ the mass of the Of?p star taken to be $m_1 = 38.5 \pm 4$\,M$_{\odot}$ (see text). The light blue hatched area corresponds to the uncertainty on the mass of the Of?p star. The black curves stand for the companion mass computed from the mass function  $m_2 = \frac{f(m)\,(1 + 1/q)^2}{\sin^3{i}}$ with $f(m) = 0.067$\,M$_{\odot}$. The continuous, short-dashed and long-dashed lines correspond respectively to $i=90^{\circ}$, $i=60^{\circ}$ and $i=30^{\circ}$. The red hatched area is a forbidden region in this parameter space. \label{fm}}
\end{figure}

Figure\,\ref{fm} illustrates the observational constraints on the mass of the secondary component. The mass of the Of?p star was estimated to be $48.8^{+5.4}_{-8.6}$\,M$_{\odot}$ by \citet{Mar12} from comparison with evolutionary tracks. However, these authors adopted a distance of 2.51\,kpc, which has since then be revised to ($1.91 \pm 0.11$)\,kpc by \citet{Bai21}. This lower distance results in a downwards revision of the evolutionary mass to $38.5 \pm 4$\,M$_{\odot}$ (see also Sect.\,\ref{MHDADM}). 

From Fig.\,\ref{fm}, we then find that the companion must have a mass larger than $\sim 4$\,M$_{\odot}$ which rules out the possibility of a white dwarf or neutron star compact companion. We further see that for orbital inclinations larger than $30^{\circ}$, the mass of the companion is unlikely to exceed $\sim 12$\,M$_{\odot}$. Hence, assuming a main-sequence luminosity class, the secondary star's spectral type is most likely in the range B1 -- B5.

Spectral disentangling can in principle be used to unveil the spectrum of an unseen secondary component in an SB1 binary system \citep[e.g.][]{Mah22}. We attempted to do so for HD~108 by applying our spectral disentangling code based on the method of \citet{Gon06} to the Aur\'elie spectra between 4460 and 4700\,\AA, i.e.\ over the most frequently observed spectral domain. We tested values of $q$ between 0.1 and 0.5. Unfortunately, the variable emission features of the Of?p star generate strong artefacts in the secondary spectrum around each variable line that prevent us from unambiguously detecting any secondary spectral feature. To limit the impact of the long-term variability, we thus repeated the disentangling restricting the input dataset to 27 Aur\'elie spectra taken between 2001 and 2008, i.e. closest to the minimum emission state. Yet, whilst this attenuated the artefacts, the reconstructed secondary spectra for different values of $q$ are still dominated by residuals of the Of?p star spectrum. Hence, at this stage, we have not been able to detect a spectral signature of the secondary star. 

While few magnetic stars have been found in close ($< 30$\,d) interacting or post-interaction O-star binaries \citep[e.g.][]{Naz15b}, the presence of distant companions is not unusual. Amongst the Of?p stars, HD~191\,612 has an orbital period of about 1542\,days, whilst its rotational period is 537.6\,days. The orbit of HD~191\,612 is quite eccentric ($e = 0.44 \pm 0.04$) and the companion is likely an early B-type star \citep{How07}. As for the latter star, and contrary to HD~108, the third ``historical'' Of?p star, HD~148\,937, also has an orbital period longer than its rotation period, but the difference is even more extreme with respective values of 18--26\,yrs and 7\,d \citep{Wad19}. The orbit of HD~148\,937 is also quite eccentric ($e=0.6-0.8$) and the companion appears to be an O-type star, about 1.5 times more massive than the magnetic Of?p star. The orbit of HD~108 thus appears quite different from those of the other two stars.

\section{X-ray emission} \label{Xdata}
Magnetically confined winds of early-type stars are expected to produce a bright and relatively hard X-ray emission \citep[see][and references therein]{udD16}. \citet{Naz14} found the X-ray luminosity of magnetic massive stars to be strongly correlated with the stellar wind mass-loss rate, with a power-law form that is slightly steeper than linear for the less luminous, lower mass-loss rate B-type stars and flattens for the more luminous, higher mass-loss rate O-type stars. As these winds are radiatively driven, these scalings can be equivalently expressed as relations with the bolometric luminosity. The observed X-ray luminosities, and their trend with mass-loss rates, were found to be well reproduced by magnetohydrodynamic (MHD) models, except for a a few overluminous B-type stars which were mostly rapidly rotating objects \citep{Naz14}. In this section, we revisit the X-ray properties of HD~108 to see how it fits into the global picture of magnetic massive stars. 

We fitted the archival {\it XMM-Newton}-EPIC and RGS spectra and the new {\it Chandra}-ACIS spectrum with {\tt xspec} version 12.9.0i. The data from both satellites were modeled using a model of the kind
{\tt phabs$_{\rm ISM}$ * (phabs$_1$ * apec(kT$_1$) + phabs$_2$ * apec(kT$_2$))}
where {\tt phabs$_{\rm ISM}$} stands for the photoelectric absorption by the interstellar medium \citep[for which we adopted a neutral hydrogen column density of $0.34\,10^{22}$\,cm$^{-2}$,][]{Dip94}, whilst the other {\tt phabs} components are meant to represent photoelectric absorption by the circumstellar environment (i.e.\ the stellar wind and/or the magnetosphere). The {\tt apec} models represent the X-ray emission of a collisionally-ionized optically thin thermal plasma \citep{Smi01} with solar abundances taken from \citet{Asp09}. The results of these fits are described in Table\,\ref{Xfit}. The normalization parameters of the {\tt apec} components correspond to $\frac{10^{-14}}{4\,\pi\,d^2}\,\int n_e\,n_H\,dV$ where $d$ is the distance to the source in cm, whilst $n_e$ and $n_H$ are the electron and hydrogen densities (in cm$^{-3}$).

\begin{table}
  \caption{Results of the fits of the X-ray spectra of HD~108}    \label{Xfit}
  \begin{center}
  \begin{tabular}{l c c}
    \hline
    & {\it XMM-Newton} & {\it Chandra} \\
    \hline
    Date (HJD - 2450000) & 2507.864 & 9553.178 \\
    Phase (rot.) & 0.41 & 0.77 \\
    Phase (bin.) & 0.03 & 0.23 \\
    N$_{\rm H, ISM}$ ($10^{22}$\,cm$^{-2}$) & 0.34 (fixed) & 0.34 (fixed) \\
    \hline
    & \multicolumn{2}{c}{Solar abundances} \\
    \hline
    N$_{\rm H, 1}$ ($10^{22}$\,cm$^{-2}$) & $0.46^{+.08}_{-.05}$ & $0.15^{+.25}_{-.15}$ \\
    kT$_1$ (keV) & $0.28^{+.01}_{-.03}$ & $0.77^{+.20}_{-.51}$ \\
    norm$_1$ ($10^{-3}$\,cm$^{-5}$) & $3.0^{+1.6}_{-0.6}$ & $0.45^{+.30}_{-.19}$ \\
    N$_{\rm H, 2}$ ($10^{22}$\,cm$^{-2}$) & $0.64^{+.22}_{-.21}$ & $0.33^{+.60}_{-.33}$ \\
    kT$_2$ (keV) & $1.86^{+.14}_{-.12}$ & $2.06^{+.93}_{-.40}$ \\
    norm$_2$ ($10^{-3}$\,cm$^{-5}$) & $0.56^{+.11}_{-.05}$ & $0.60^{+.16}_{-.22}$ \\
    $f_{\rm X}$ ($10^{-13}$\,erg\,cm$^{-2}$\,s$^{-1}$) & $6.46 \pm 0.04$ & $8.18 \pm 0.04$ \\
    $f_{\rm X}^{\rm un}$ ($10^{-13}$\,erg\,cm$^{-2}$\,s$^{-1}$) & $11.23$ & $12.54$ \\
    $\chi^2_n$ (d.o.f.) & 1.50 (569) & 0.99 (76) \\
    \hline
    & \multicolumn{2}{c}{CNO from \citet{Mar15}} \\
    \hline
    N$_{\rm H, 1}$ ($10^{22}$\,cm$^{-2}$) & $0.44^{+.04}_{-.04}$ & $0.15^{+.25}_{-.15}$ \\
    kT$_1$ (keV) & $0.24^{+.01}_{-.01}$ & $0.77^{+.20}_{-.46}$ \\
    norm$_1$ ($10^{-3}$\,cm$^{-5}$) & $4.6^{+1.0}_{-0.8}$ & $0.46^{+.30}_{-.19}$ \\
    N$_{\rm H, 2}$ ($10^{22}$\,cm$^{-2}$) & $0.76^{+.14}_{-.12}$ & $0.34^{+.59}_{-.34}$ \\
    kT$_2$ (keV) & $1.61^{+.07}_{-.07}$ & $2.06^{+.86}_{-.40}$ \\
    norm$_2$ ($10^{-3}$\,cm$^{-5}$) & $0.66^{+.05}_{-.04}$ & $0.61^{+.20}_{-.23}$ \\
    $f_{\rm X}$ ($10^{-13}$\,erg\,cm$^{-2}$\,s$^{-1}$) & $6.40 \pm 0.04$ & $8.18 \pm 0.04$ \\
    $f_{\rm X}^{\rm un}$ ($10^{-13}$\,erg\,cm$^{-2}$\,s$^{-1}$) & $11.43$ & $12.54$ \\
    $\chi^2_n$ (d.o.f.) & 1.35 (569) & 0.92 (76) \\
    \hline
  \end{tabular}
  \end{center}
  
      {\scriptsize The upper part of the table yields the best-fit results assuming solar abundances, whilst the lower part quotes the results adopting the CNO abundances of \citet{Mar15}. The $f_{\rm X}$ and $f_{\rm X}^{\rm un}$ values are respectively the observed fluxes and the fluxes corrected for absorption by the interstellar medium, both in the 0.5 -- 10\,keV energy band.}
\end{table}

\begin{figure}
\begin{center}
  \resizebox{8.5cm}{!}{\includegraphics{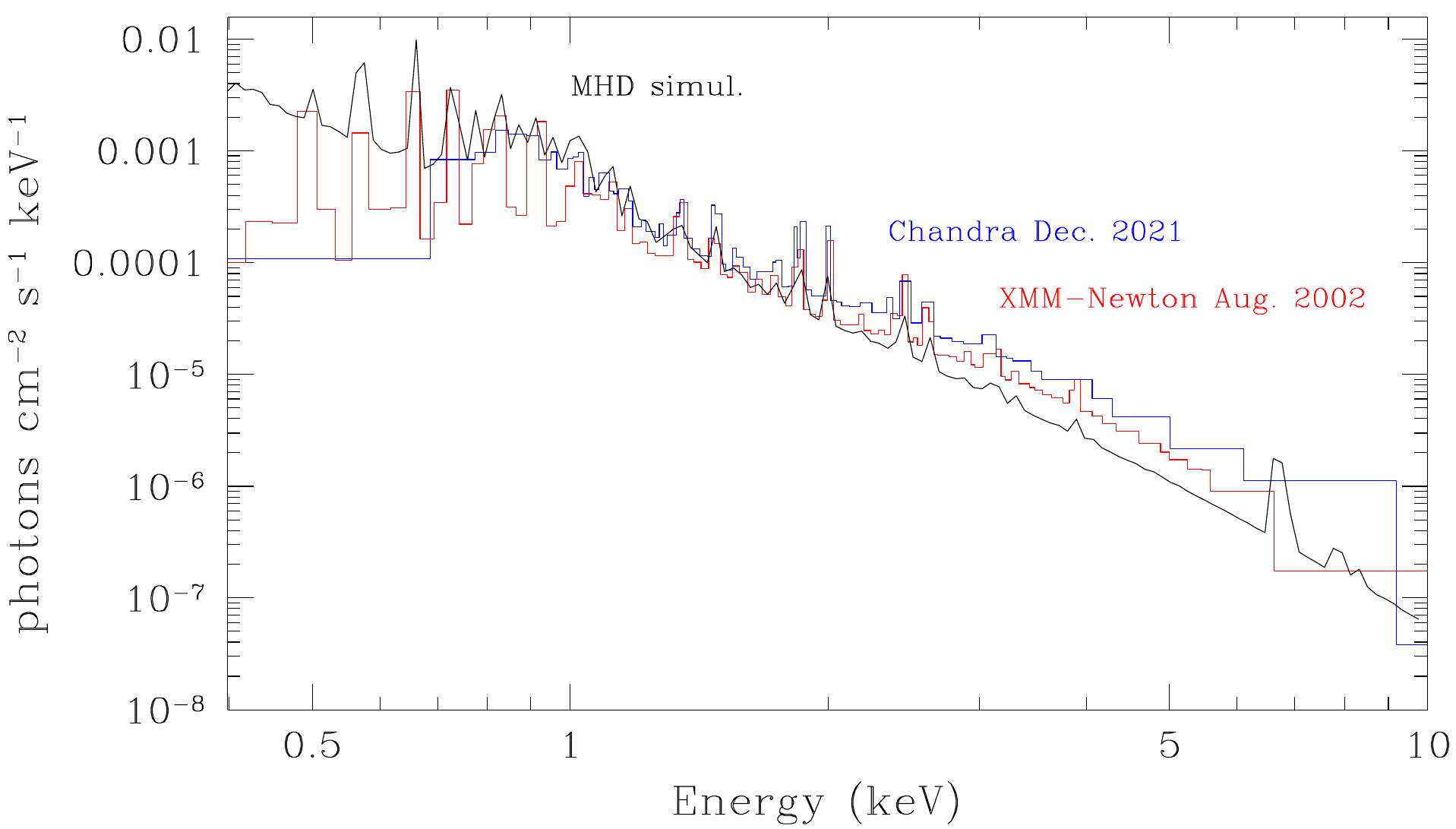}}
\end{center}  
\caption{Unfolded X-ray model spectra of HD~108 as adjusted to the {\it XMM-Newton} August 2002 (rotation phase $\simeq 0.41$) spectra shown in red and the {\it Chandra} December 2021 (rotation phase $\simeq 0.77$) data displayed in blue. The models were computed with the CNO abundances of \citet{Mar15}. The black line yields the spectral energy distribution predicted by our MHD calculations corresponding to Model II parameters (see the forthcoming Sect.\,\ref{MHD}).\label{ufX}}
\end{figure}

\citet{Mar12} and \citet{Mar15} analysed the nitrogen and CNO content of HD~108, concluding that the star displays a N enhancement by about a factor of 5 with respect to solar abundances of \citet{Asp09}, whilst C and O are depleted by factors of 0.6 and 0.5, respectively. Since lines of hydrogen-like and helium-like ions of C, N and O are present in the X-ray spectrum below 1\,keV, these non-solar abundances could impact the results of the spectral fits. We thus reanalysed the spectra with a model
{\tt phabs$_{\rm ISM}$ $\times$ (phabs$_1$ $\times$ vapec(kT$_1$) + phabs$_2$ $\times$ vapec(kT$_2$))}
where the {\tt vapec} plasma models indicate {\tt apec} models with non-solar abundances. We fixed the C, N and O abundances respectively to 0.6, 5.0, and 0.5 times solar, and kept all other elements at a solar abundance. The results of these fits are described in the lower half of Table\,\ref{Xfit}. These modified abundances slightly improve the quality of the fits, but overall do not change the global results significantly. 

Our parameters for the {\it XMM-Newton} observation slightly differ from those reported by \citet{Naz04,Naz14}. Several factors could explain these differences:  inclusion of the RGS data in our present fit, different model formulation (global or individual circumstellar absorptions, {\tt mekal} - \citealt{KaMe00} - versus {\tt apec} - \citealt{Smi01} - plasma models), different versions of the {\tt xspec} software (and thus of atomic parameters), different abundances (solar abundances taken from \citealt{AG} in the previous studies versus abundances from \citealt{Asp09}, or modified according to \citealt{Mar15} here), and evolving knowledge of the calibrations of the {\it XMM-Newton} instruments. Following some trials, the most important reason appears to be the evolving instrument calibration, with the other effects accounting for flux differences of a few percent at most.   

We also stress that one must be careful when interpreting the parameters in Table\,\ref{Xfit}. First of all, there exists a well-known degeneracy between the column density of the material responsible for the photoelectric absorption and the temperature of the emitting plasma. Especially in the case of low-resolution CCD spectra, a given spectrum can be equally well represented by a low-temperature plasma absorbed by a high column density or a higher temperature plasma seen through a lower column density. This degeneracy most strongly affects the kT$_1$ component and the associated column density N$_{\rm H, 1}$. It typically affects the results as seen in the {\it XMM} versus {\it Chandra} fittings (low N$_{\rm H}$ and high $kT$ versus high N$_{\rm H}$ and lower $kT$). However, since the spectral energy distribution is well fitted, this degeneracy does not alter much the derived X-ray fluxes. Second, there remain some cross-calibration uncertainties between {\it XMM-Newton} and {\it Chandra}. These can affect both the plasma parameters and the inferred fluxes. Using observations of the line-rich supernova remnant (SNR) 1E~0102.2-7219 by {\it Chandra}, {\it Suzaku}, {\it Swift} and {\it XMM-Newton}, \citet{Plu12} showed that the fluxes of the prominent lines in the 0.5 -- 1.5\,keV domain measured by these satellites agree within 10\%. At higher energies, \citet{Mad21} found differences of $\sim 15$\% between {\it Chandra} and {\it XMM-Newton} in the 4.5 -- 8.0\,keV domain for the SNR N~132D. Based on this, we consider that the fluxes derived from our X-ray observations of HD~108 could be affected by cross-calibration uncertainties of 10 -- 15\%. 

Nevertheless, the {\it Chandra} observation yields an observed flux in the 0.5 -- 10\,keV band that is 28\% higher than that recorded in the {\it XMM-Newton} observation. Hence, the flux very probably increased between the two observations (i.e.\ between rotation phase $\simeq 0.41$ and $\simeq 0.77$), but the increase is rather modest when looking at the full energy band. Figure\,\ref{ufX} illustrates the unfolded spectra of both observations. In general, accounting for the different widths of the energy bins, the overall shapes of the spectra are in good agreement. Yet, the {\it Chandra} observation yields a somewhat higher flux especially at higher energies. Whilst the {\it Chandra} flux is 18\% higher than the {\it XMM-Newton} flux at energies between 0.5 and 2.0\,keV, the increase amounts to 50\% at energies between 2.0 and 10\,keV. If we consider the X-ray flux corrected for the interstellar absorption (but not corrected for the circumstellar absorption), then the two observations yield the same flux in the soft (0.5 -- 2.0\,keV) band, but a 50\% increase is still seen in the hard (2.0 -- 10\,keV) domain.

The {\it GAIA} EDR3 \citep{EDR3} lists a parallax of $0.4980 \pm 0.0240$\,mas for HD~108, which according to \citet{Bai21} translates into a distance of ($1.91 \pm 0.11$)\,kpc. Adopting this distance, the {\it XMM-Newton} observation yields an X-ray luminosity corrected for interstellar absorption of $L_{\rm X} = 5.0\,10^{32}$\,erg\,s$^{-1}$ in the 0.5 -- 10\,keV band, whilst the {\it Chandra} observation yields $L_{\rm X} = 5.5\,10^{32}$\,erg\,s$^{-1}$. Scaling the bolometric luminosity determined by \citet{Mar12} to the Gaia distance, we finally infer $\log{L_{\rm X}/L_{\rm bol}} = -6.35$ and $\log{L_{\rm X}/L_{\rm bol}} = -6.31$ respectively for the {\it XMM-Newton} and {\it Chandra} data. These values clearly place HD~108 among the X-ray fainter Of?p stars, but are still fully compatible with other magnetic O-type stars \citep[$\log{L_{\rm X}/L_{\rm bol}}$ value of $-6.23 \pm 0.07$ on average,][]{Naz14}. 

Comparing the {\it XMM-Newton} X-ray flux to previous {\it Einstein} (1979) and {\it ROSAT} (1991) data, \citet{Naz04} noted the absence of strong variations (by more than a factor of two). Our results confirm the relatively low amplitude of the variation of the X-ray flux. We can further compare the behaviour of HD~108 to that of other Of?p stars. For HD~191\,612, a $\sim 40$\% variation of the X-ray fluxes (both the observed and the ISM-corrected values) between high and low state was measured \citep{Naz07}. Similar to HD~108, the X-ray spectrum of HD~191\,612 is also dominated by the cooler spectral component. CPD $-28^{\circ}$\,2561 was found to display bright and hard X-ray emission which varies in fluxes by $\sim 55$\% in phase with the optical variations \citep{Naz15a}. No significant hardness variations were found between minima and maxima for this star. Hence, from the data recorded so far, and accounting for the fact that the maximum emission has not yet been reached, the level of variability of HD~108's X-ray emission appears similar to that of other Of?p stars.

The detection of a binary signature in Sect.\,\ref{radvel} raises the question whether or not the secondary star or a wind interaction between the Of?p star and the secondary could contribute significantly to the X-ray emission of HD~108. Given the likely mass of the secondary and assuming it to be a non-degenerate star, its intrinsic X-ray emission should be several orders of magnitude lower than the emission of the Of?p star. Such a star would also have a very weak wind, implying that any putative wind interaction would most likely consist in a collision of the Of?p wind with the companion's photosphere. Given the wide orbital separation, such an interaction would only concern a very small part of the Of?p wind. Thus the associated X-ray emission should be negligible. Finally, if we assume the secondary to be a degenerate object, it would have to be a black hole given the minimum mass we have inferred. Accretion by such an object should lead to a very different spectral energy distribution than the one observed. Hence, we conclude that the companion is unlikely to play a role in the X-ray emission of HD~108. Beside the emission from the hot plasma of the magnetically confined wind, some soft X-ray emission could arise from intrinsic wind instabilities. The latter should typically account for $\log{L_{\rm X}/L_{\rm bol}} \simeq -7$, i.e.\ about 25\% of the observed X-ray emission.    

\begin{table*}
  \caption{Adopted parameters of HD~108}\label{param}
  \begin{tabular}{l c l c c c}
    \hline
    Parameter & Adopted value & Reference & Model I & Model II & Model III\\
    \hline
    $d$ (kpc) & $1.91 \pm 0.11$ & \citet{Bai21} &  &  & \\
    $T_{\rm eff}$ (kK) & $35 \pm 2$ & \citet{Mar12} &  &  & \\
    $\log{\frac{L_{\rm bol}}{\rm L_{\odot}}}$ & $5.46 \pm 0.10$ & This work & & &\\
    $R_*$ (R$_{\odot}$) & $14.7 \pm 1.2$ & This work & & & \\
    $M_{*, \rm evol}$ (M$_{\odot}$) & $38.5 \pm 4$ & This work & & &  \\
    $\log(\dot{M})$ (M$_{\odot}$\,yr$^{-1}$) & $-6.06 \pm 0.23$ & This work & & & \\
    $v_{\rm esc}$ (km\,s$^{-1}$) & $890 \pm 60$ & This work & & &\\
    $v_{\infty}$ (km\,s$^{-1}$) & $2300 \pm 300$ & This work & & &\\
    $B_{\rm d}$ (G) & $\geq 1150$ & \citet{Shu17} & 1150 & 2000 & 4000\\
    $\eta_*$ & & & 27.5& 83.1 & 332.4 \\
    $R_A$ ($R_*$) & & & 2.59 & 3.32 & 4.57\\
    \hline
  \end{tabular}
\end{table*}

\section{The magnetosphere of HD~108} \label{MHDADM}
\citet{Mar10} first reported the spectropolarimetric signature of the magnetic field of HD~108. The longitudinal field strength was found to increase between 2007, 2008 and 2009, as the star was emerging from its minimum state \citep{Mar10}. These measurements suggested a dipolar magnetic field of at least 0.5\,kG, but more likely 1 -- 2\,kG. In January 2010, \citet{Hub10} detected a magnetic field with a longitudinal strength marginally higher than that found by \citet{Mar10}. \citet{Shu17} presented a new spectropolarimetric observation of HD~108 taken in September 2015, that yielded $<B_z> = -325 \pm 45$\,G, about three times the value measured by \citet{Mar10}. This led to a revised minimum value of the dipolar field strength of 1150\,G \citep{Shu17}.

For a dipolar magnetic field tilted by an obliquity angle $\beta$ with respect to the stellar rotation axis, the strength of the longitudinal field should vary with time as a simple sine wave \citep{Pre67,Don02}:
\begin{equation}
  <B_z> = B_d\,\frac{15 + u}{20\,(3 - u)}\,(\cos{\beta}\,\cos{i} + \sin{\beta}\,\sin{i}\,\cos{(2\,\pi\,\phi)})
  \label{eqDonati}
\end{equation}
where $i$ is the inclination of the rotation axis with respect to the line of sight, $u$ is the continuum linear limb darkening coefficient, and $\phi = \frac{(t-t_0)}{P_{\rm rot}}$ is the phase of the stellar rotation with $\phi = 0$ at $t_0$ when the magnetic axis is closest to  our line of sight. For stars with effective temperatures and gravities close to those of HD~108, \citet{Cla19} quotes values of $u$ between 0.22 and 0.35. However, this range of values translates only into a small change of $<B_z>$ by about 5\%. Adopting $u = 0.3$, one finds that the minimum value of the longitudinal field amounts to
$$<B_{z, \rm min}> = 0.283\,B_d\,\cos{(\beta - i)}$$
whilst its maximum value is given by
$$<B_{z, \rm max}> = 0.283\,B_d\,\cos{(\beta + i)}$$ 
In principle, the values of $B_d$, $i$ and $\beta$ can be constrained by adjusting equation \ref{eqDonati} to the values of $<B_z>$ determined at different rotational phases. Yet, in the case of HD~108, the currently existing spectropolarimetric data do not provide a sufficient sampling of the 54\,yrs cycle to fit all parameters independently. Moreover, the uncertainties on the existing measurements allow for a rather wide range of $B_d$ values, even when we fix the $i$ and $\beta$ angles to the values we infer from the EW(H$\alpha$) light curve (see Sect.\,\ref{ADMfit}). Nevertheless, even though we cannot infer a precise $B_d$ value at this stage, the existing data suggest a value of $B_d$ which is likely somewhere between 2000\,G and 4000\,G. We thus tested models with $B_{\rm d}$ values of 1150\,G (Model I), 2000\,G (Model II), and 4000\,G (Model III, see Table\,\ref{param}) in our calculations hereafter.

Fundamental stellar and wind parameters of HD~108 were determined by \citet{Mar12} assuming a distance of 2.51\,kpc. Using instead the revised distance determined by \citet{Bai21}, we have scaled these parameters to infer $R_* = (14.7 \pm 1.2)$\,R$_{\odot}$, $\log{\frac{L_{\rm bol}}{\rm L_{\odot}}} = 5.46 \pm 0.1$, and $M_{*, \rm evol} = 38.5 \pm 4$\,M$_{\odot}$.

Observationally-determined values of the mass-loss rate \citep[e.g.][]{Mar10,Marc12} were derived by means of model atmosphere codes that assume a spherically symmetric and homogeneous wind and hence do not account for the actual wind geometry and the action of the magnetic field on the wind. To evaluate the impact of the magnetic field on the outflow, one must instead use the value of $\dot{M}_{B=0}$ that would be observed in the absence of a magnetic field \citep[][]{udD02}. Therefore, \citet{Shu17} estimated $\dot{M}_{B=0}$ via the formalism of \citet{Vin01} that predicts a mass-loss rate of $2.8\,10^{-6}$\,M$_{\odot}$\,yr$^{-1}$, about 1\,dex higher than the value inferred from spectroscopic analyses of the UV spectrum \citep{Mar10,Marc12}. Using instead the luminosity and stellar mass estimates that stem from the revised distance determination, the \citet{Vin01} recipe leads to a lower value of $\dot{M}_{B=0} = 0.9\,10^{-6}$\,M$_{\odot}$\,yr$^{-1}$ ($\log{\dot{M}_{B=0}} = -6.06 \pm 0.23$).

Similar remarks apply to the asymptotic wind velocity $v_{\infty}$. In the computation of the wind confinement parameter $\eta_* = \frac{B_d^2\,R_*^2}{4\,\dot{M}_{B=0}\,v_{\infty}}$ \citep[eg.][and referencese therein]{udD22} and in the MHD simulations, one needs the value of $v_{\infty}$ that would be observed in the absence of a magnetic field. Typical values of the $v_{\infty}/v_{\rm esc}$ ratio are expected to be around 2.6, which would lead to $v_{\infty} \simeq (2300 \pm 150)$\,km\,s$^{-1}$. 
The resulting model parameters are summarized in Table\,\ref{param}.

\subsection{Fitting the EW variations with an ADM-type model}\label{ADMfit}
In the oblique magnetic rotator model, the axis of the dipolar magnetic field is inclined by an obliquity angle $\beta$ with respect to the stellar rotational axis. The rotation axis itself is seen under an inclination $i$ by the external observer. The optical emission lines are strongest when the magnetically confined wind is seen as much as possible face-on, i.e.\ when the angle between the direction of the dipolar magnetic field and the line of sight is minimum and amounts to $\beta - i$. Conversely, the emission lines are weakest when the magnetosphere is seen near edge-on, i.e.\ when the angle between the magnetic axis and the line of sight is equal to $\beta + i$. 

\citet{Owo16} designed an analytic dynamical magnetosphere (ADM) model that provides analytical relations to express the temperature, density and flow velocities of the different ingredients of the stellar wind and of a dynamical magnetosphere. This model has been used successfully by \citet{Mun20} to analyse the optical photometric light curves of several Galactic and Magellanic Cloud Of?p stars as the result of occultations of the stellar photosphere by regions of different densities. In their description, \citet{Owo16} distinguish three components of the magnetosphere:\begin{itemize}
\item the wind upflow, driven by radiation pressure at the stellar surface and channeled along the magnetic field lines,
\item the hot (several MK) post-shock gas, which is produced when the wind upflowing along a closed magnetic loop encounters a hydrodynamic shock as it approaches the magnetic equator, and
\item the cooled downflow, made of post-shock gas that has undergone radiative cooling and falls back to the star.
\end{itemize}
In MHD simulations, these flows are intermingled, forming complex, variable, 3D structures \citep[e.g.][]{udD13}. The ADM model instead considers a time-averaged structure, where the stochastic variations are smoothed out, thereby allowing to infer analytic expressions for the velocities and densities of all three components \citep{Owo16}. As far as the optical line emission is concerned, it is mostly the cooled downflow that matters.  

To compare the observed line strengths with model calculations, we must correct the observed values for the strength of the underlying photospheric absorption line. \citet{Mar12} determined an effective temperature of T$_{\rm eff} = (35 \pm 2)$\,kK and a $\log{g}$ of $3.50 \pm 0.10$. Using a TLUSTY synthetic spectrum \citep{Lan03} with these parameters, we estimate the EW of the photospheric H$\alpha$ absorption to be 2.6\,\AA. Hence, the 2022-2023 spectra (with the strongest H$\alpha$ emission observed so far) correspond to a net EW of the H$\alpha$ emission of $-12.6$\,\AA.

\begin{figure*}
  \begin{center}
    \begin{minipage}{5.5cm}    
      \resizebox{5.5cm}{!}{\includegraphics{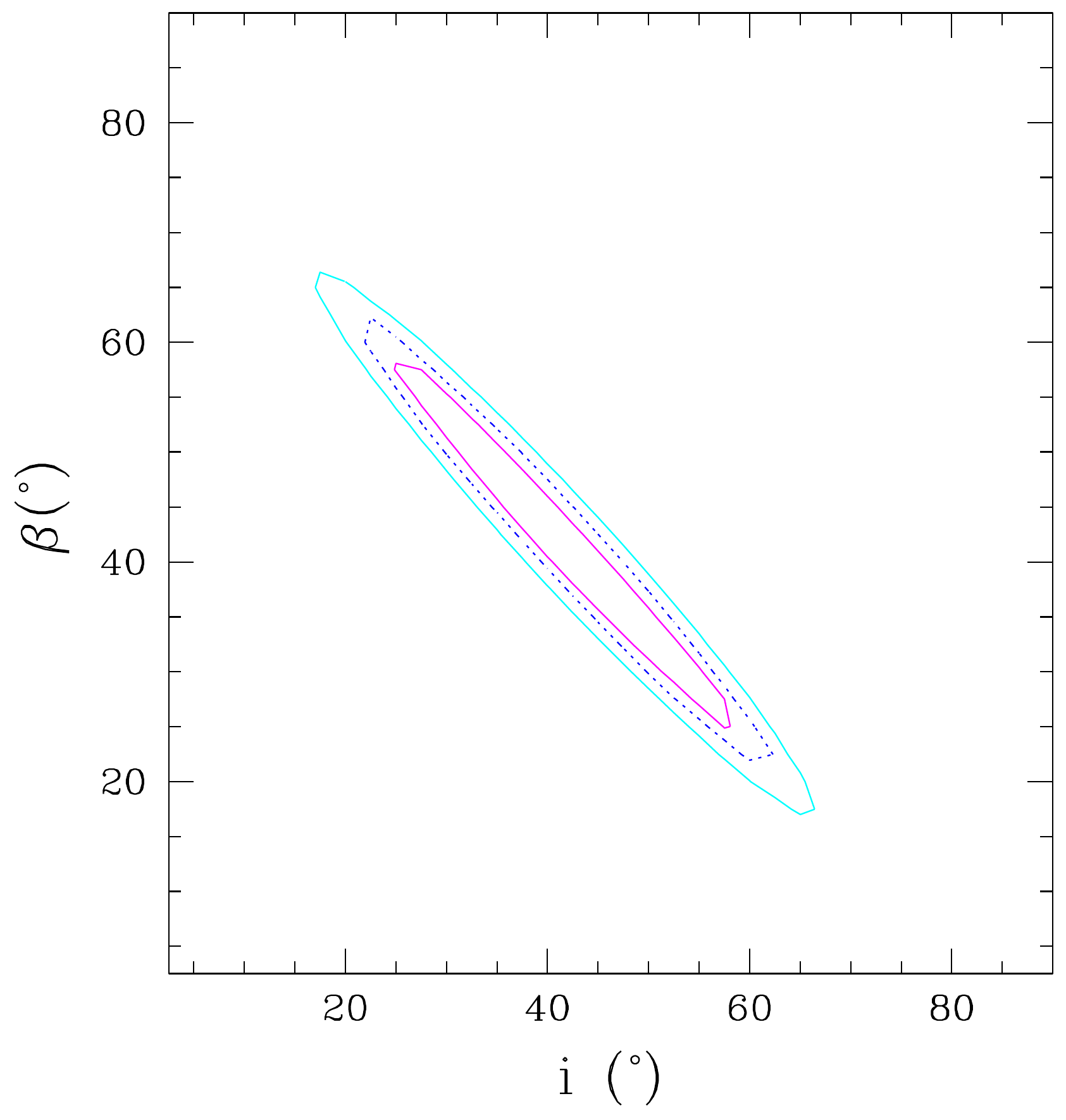}}
    \end{minipage}
    \hfill
    \begin{minipage}{5.5cm}    
      \resizebox{5.5cm}{!}{\includegraphics{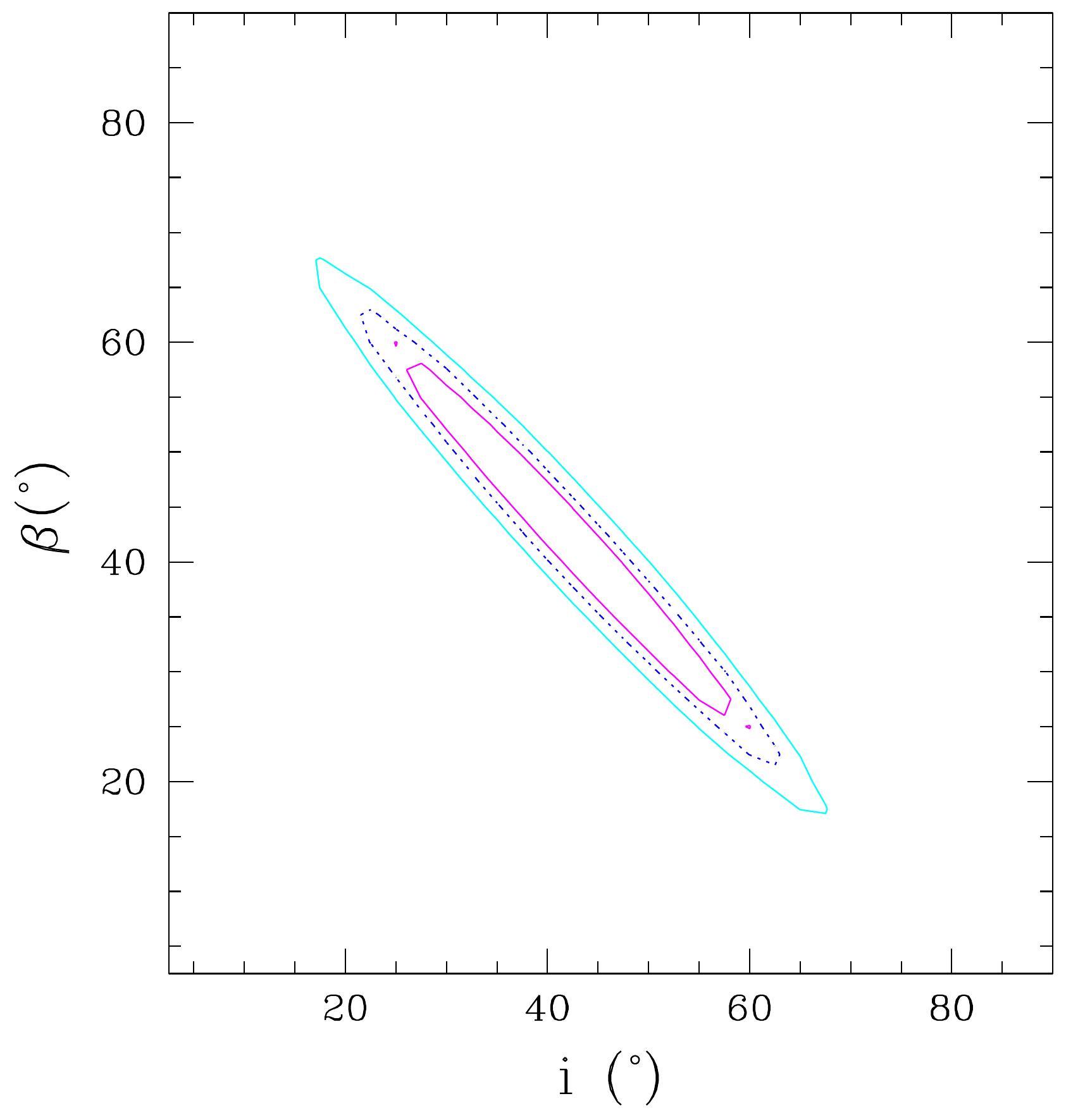}}
    \end{minipage}
    \begin{minipage}{5.5cm}    
      \resizebox{5.5cm}{!}{\includegraphics{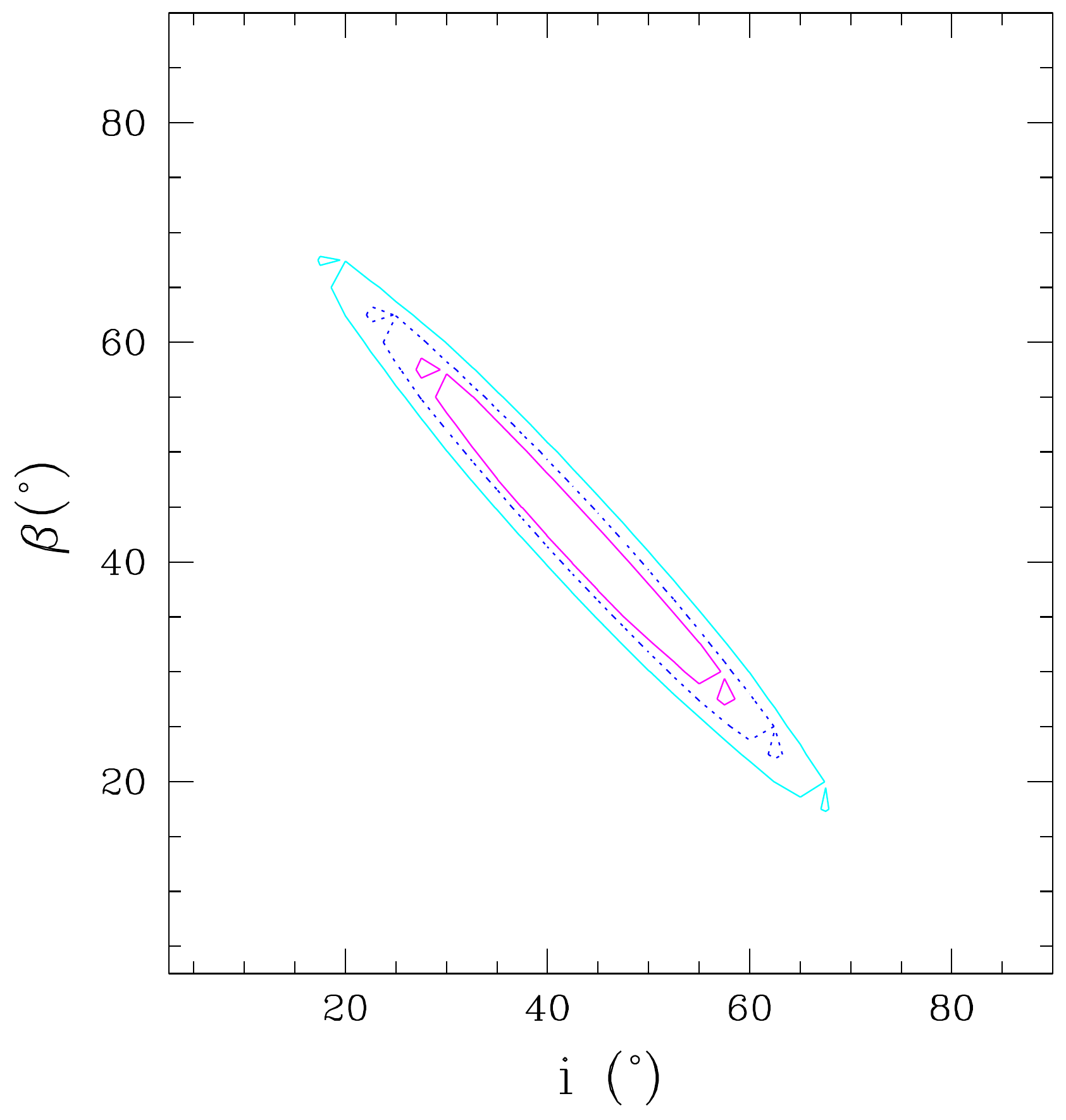}}
    \end{minipage}
  \end{center}  
\caption{$\chi^2$ contours of the ADM models for $\delta/R_* = 0.1$. The different contours correspond to values of the reduced $\chi^2$ of $\chi^2_{\min} + 2.30$ (magenta solid line), $\chi^2_{\min} + 4.61$ (blue dotted line), and $\chi^2_{\min} + 9.21$ (cyan dashed line). The left panel assumes parameters of Model I ($\chi^2_{\rm min} = 1.53$), the middle panel stands for Model II ($\chi^2_{\rm min} = 1.75$), whilst the right panel corresponds to Model III ($\chi^2_{\rm min} = 1.57$).\label{contADM}}
\end{figure*}
\begin{figure*}
  \begin{center}
    \begin{minipage}{5.5cm}    
      \resizebox{5.5cm}{!}{\includegraphics{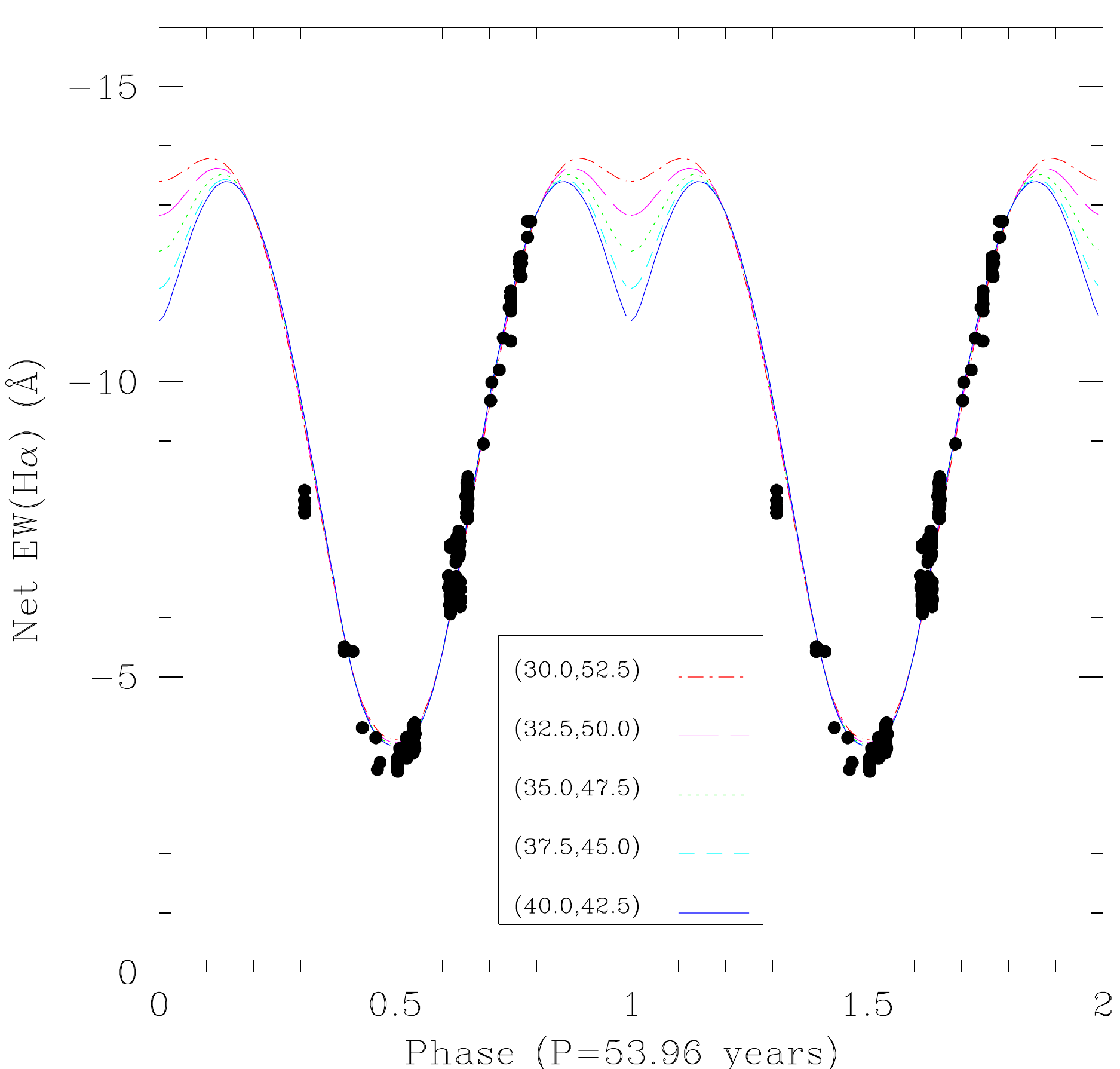}}
    \end{minipage}
    \hfill
    \begin{minipage}{5.5cm}    
      \resizebox{5.5cm}{!}{\includegraphics{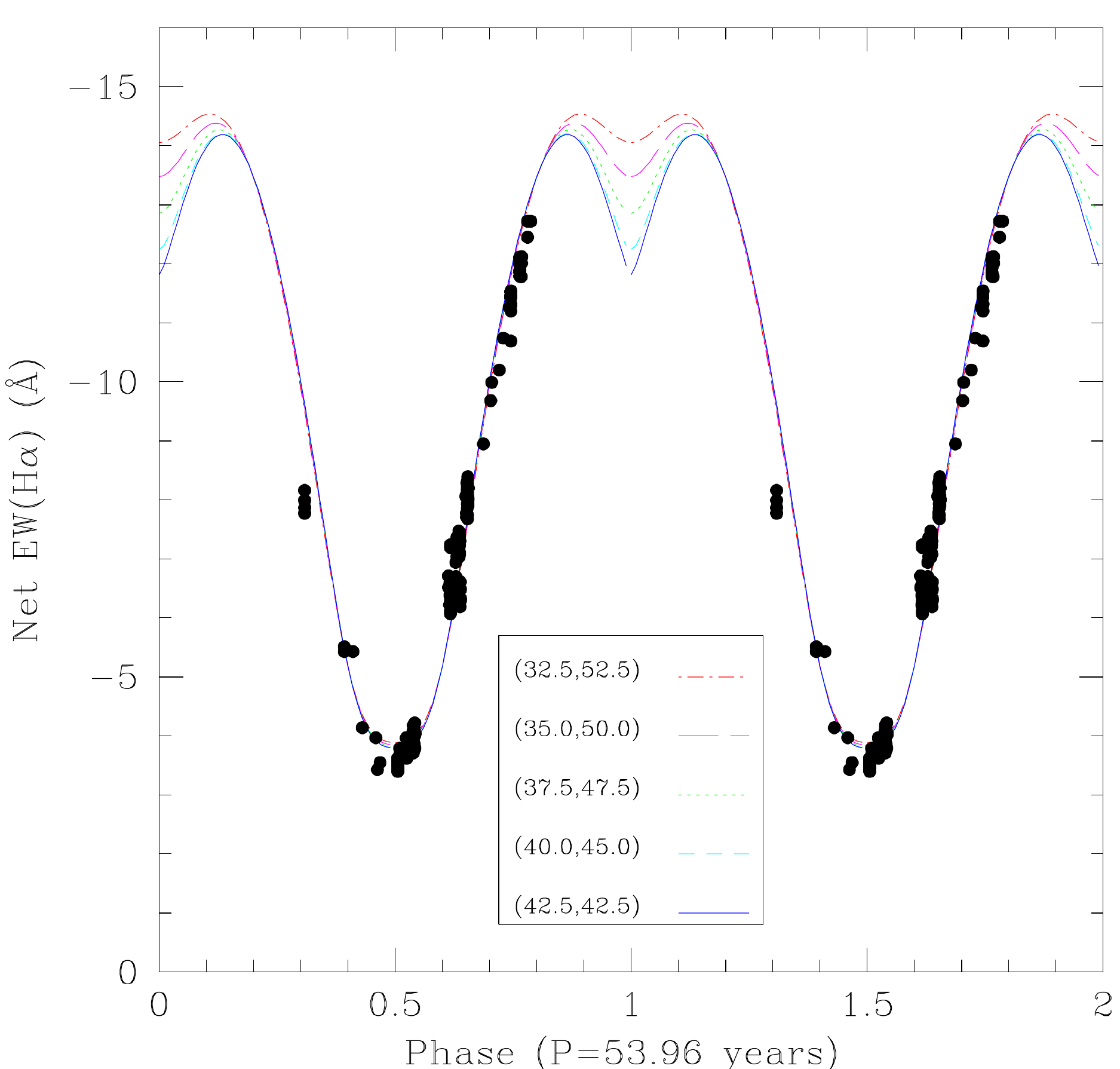}}
    \end{minipage}
    \begin{minipage}{5.5cm}    
      \resizebox{5.5cm}{!}{\includegraphics{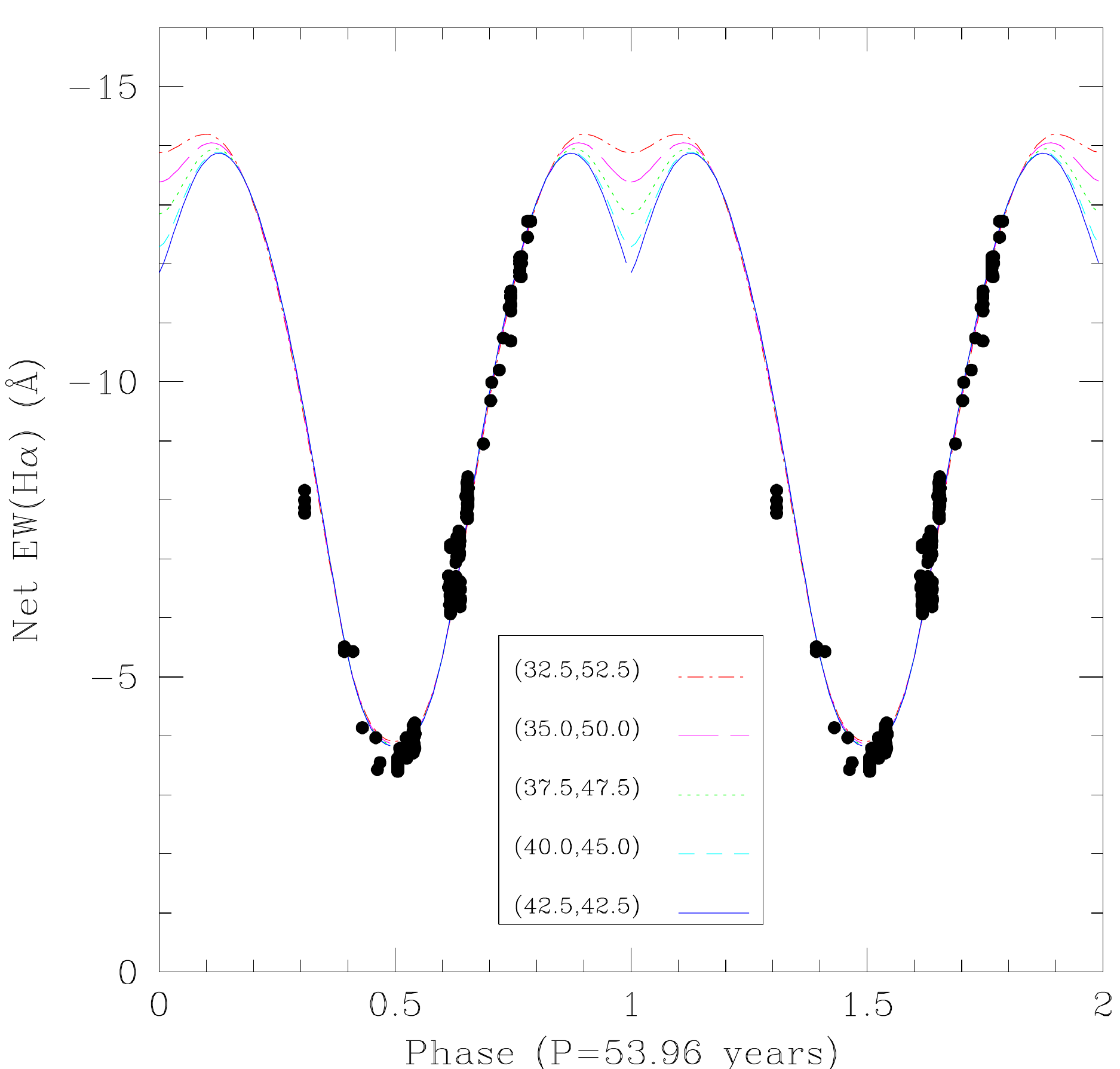}}
    \end{minipage}
  \end{center}
\caption{Fit to the net (absorption-corrected) EW(H$\alpha$) light curve of HD~108 with ADM models (from left to right Model I, II and III parameters are used). Different pairs of $(i, \beta)$ that yield comparable fits to the part of the curve that is sampled with the existing data are illustrated by different colours.\label{ADMfits}}
\end{figure*}

To formulate constraints on $i$ and $\beta$, we implemented the ADM formalism for EW(H$\alpha$) following the prescription for the cooled downflow \citep[equations 22 and 25 of][]{Owo16}. We assumed that the radiative transfer can be treated by means of the Sobolev escape probability approximation \citep[e.g.][]{Ryb78}. The optical depth in the H$\alpha$ line measured from a position $\vec{r}$ along the direction $\vec{n}$ towards the observer hence becomes 
\begin{eqnarray}
  \tau(\vec{r},\vec{n}) & = & \frac{6.6\,10^{20}\,T^{-1.5}}{|Q|}\,\frac{1 + y_{\rm He}\,I_{\rm He}}{(1 + 4\,y_{\rm He})^2}\,\rho^2 \nonumber \\
  & \times & \left[b_2\,\exp{\left(\frac{3.3944\,10^4}{T}\right)} - b_3\,\exp{\left(\frac{1.753\,10^4}{T}\right)}\right] 
\end{eqnarray}
\citep[see equation 1 of][]{PP96}. Here $T$ and $\rho$ are the gas temperature and gas density at the position $\vec{r}$, $y_{\rm He}$ and $I_{\rm He}$ are the number abundance of helium and the number of free electrons provided per helium atom, and $b_2$ and $b_3$ are the NLTE departure coefficients of the $n=2$ and $n=3$ energy levels of the hydrogen atom computed following the parametrization of \citet{Pul96}.
The quantity $Q$ stands for the projection of the gradient of the line-of-sight component of the gas velocity onto the line of sight $\vec{n}$ \citep{Ryb78}, and is given by
\begin{equation}
  Q = \vec{n} \cdot \vec{\nabla}(\vec{v} \cdot \vec{n})
\end{equation}

Using the stellar and wind parameters of Models I, II and III from Table\,\ref{param}, we computed grids of ADM models for $i$ and $\beta$ between $2.5^{\circ}$ and $90^{\circ}$ with steps of $2.5^{\circ}$. For each pair of $i$ and $\beta$, we computed a full synthetic EW(H$\alpha$) curve as a function of rotational phase. A re-scaled version of this curve, to match the mean level of the observed EW(H$\alpha$), was then compared to the observed data points. The re-scaling is needed to account for the effects of clumping. Indeed, the fact that the emissivity scales as $\rho^2$ implies that the clumpiness of the downflow affects the overall emission strength. Such a clumping is seen in the MHD simulations \citep[e.g.][]{udD13} and thus probably exists also in the genuine Of?p magnetosphere, but it is not included in the ADM calculations. The comparison between the re-scaled synthetic light curve and the observed data then allowed us to compute $\chi^2$ maps that unveil those values of $i$ and $\beta$ that yield the best formal match to the observations. \citet{Owo16} introduced a parameter $\delta/R_*$ that expresses the smoothing of the density near the magnetic equator in the ADM. In our calculations, we considered three different values of this parameter (0.0, 0.1 and 0.3). Figure\,\ref{contADM} illustrates the $\chi^2$ contour plot in the $(i,\beta)$ parameter space for $\delta/R_* = 0.1$ which, among the three values tested here, yields the best matches between the synthetic curve and our data.

\citet{Mun20} showed that photometric light curves computed with the ADM model were subject to a degeneracy between $i$ and $\beta$. A very similar degeneracy applies to our analysis. Indeed, our existing EW(H$\alpha$) light curve samples HD~108's minimum emission state well, but does not yet cover the phases of maximum emission state. Thus, we expect the value $i + \beta$ to be better constrained than the value of $i - \beta$. Our contour plots indeed unveil the ensuing degeneracy between $i$ and $\beta$. For Model I, the best formal fits are obtained for $i = 40.0^{\circ}$  and $\beta = 42.5^{\circ}$ but $(i,\beta)$ pairs with $i + \beta = 82.5^{\circ}$ and $i \in [27.5^{\circ},55^{\circ}]$ yield essentially the same fitting quality (see Fig.\,\ref{ADMfits}). For Models II and III, the same degeneracy is observed with a slightly higher value of $i + \beta = 85^{\circ}$, a formally best fit at $i = \beta = 42.5^{\circ}$, and an acceptable range of inclinations of $i \in [30^{\circ},55^{\circ}]$. Therefore, the fits of the current EW(H$\alpha$) light curve are relatively independent of the assumed value of $B_d$. As expected in Sect.\,\ref{longtermspec}, our best-fit $(i, \beta)$ pairs differ from those of HD~191\,612 for which \citet{Owo16} rather inferred $(i, \beta) = (23^{\circ}, 73^{\circ})$ or $(73^{\circ}, 23^{\circ})$.  

\begin{figure*}
  \begin{center}
    \begin{minipage}{5.5cm}
      \resizebox{5.5cm}{!}{\includegraphics{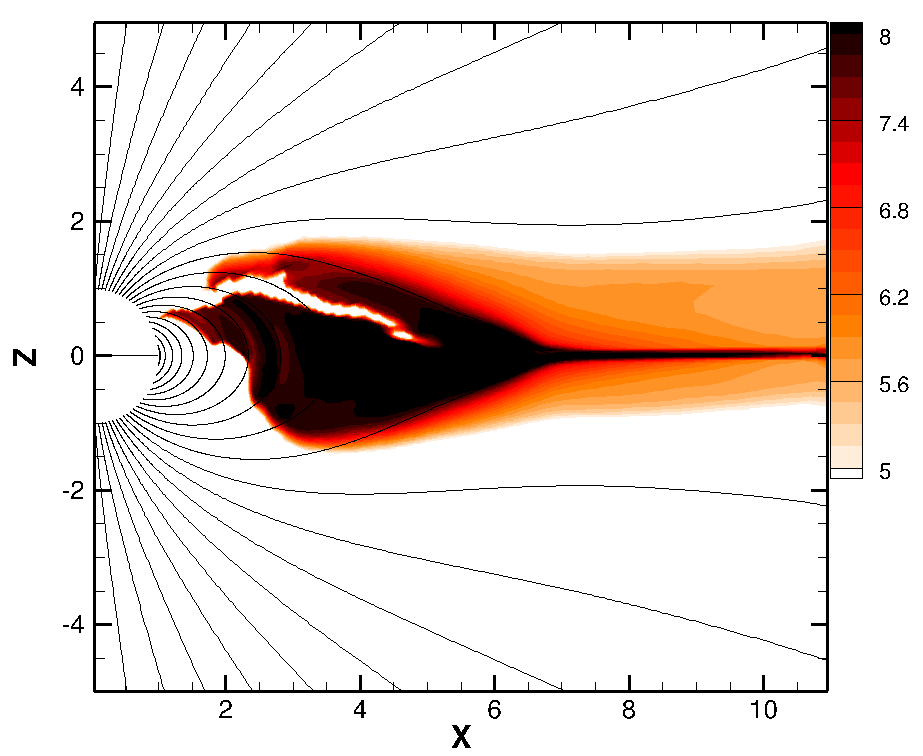}}
    \end{minipage}
    \hfill
    \begin{minipage}{5.5cm}
      \resizebox{5.5cm}{!}{\includegraphics{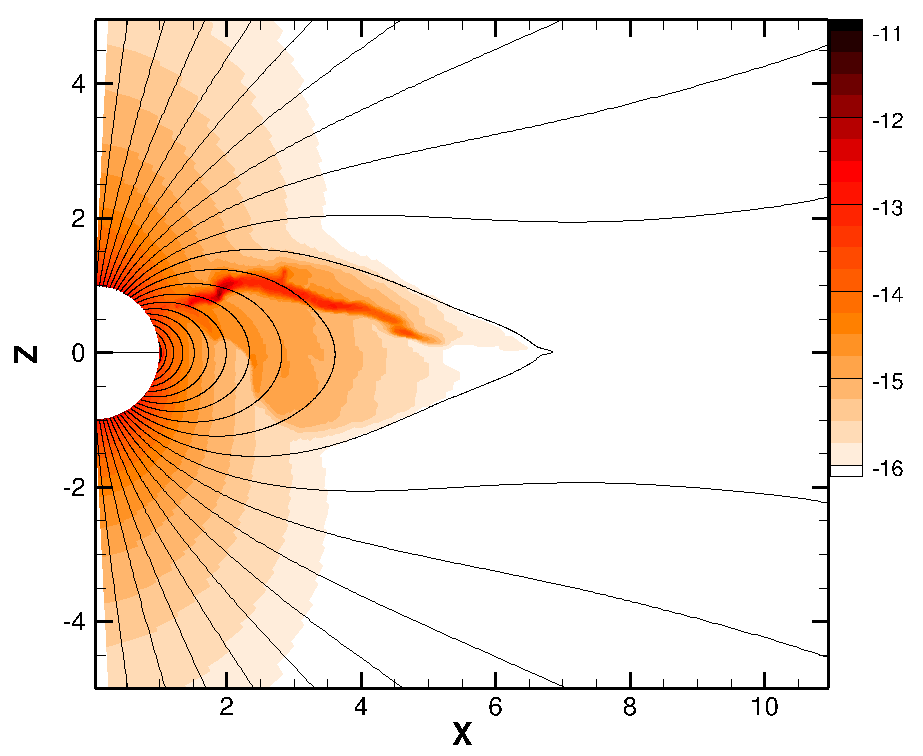}}
    \end{minipage}
    \hfill
    \begin{minipage}{5.5cm}
      \resizebox{5.5cm}{!}{\includegraphics{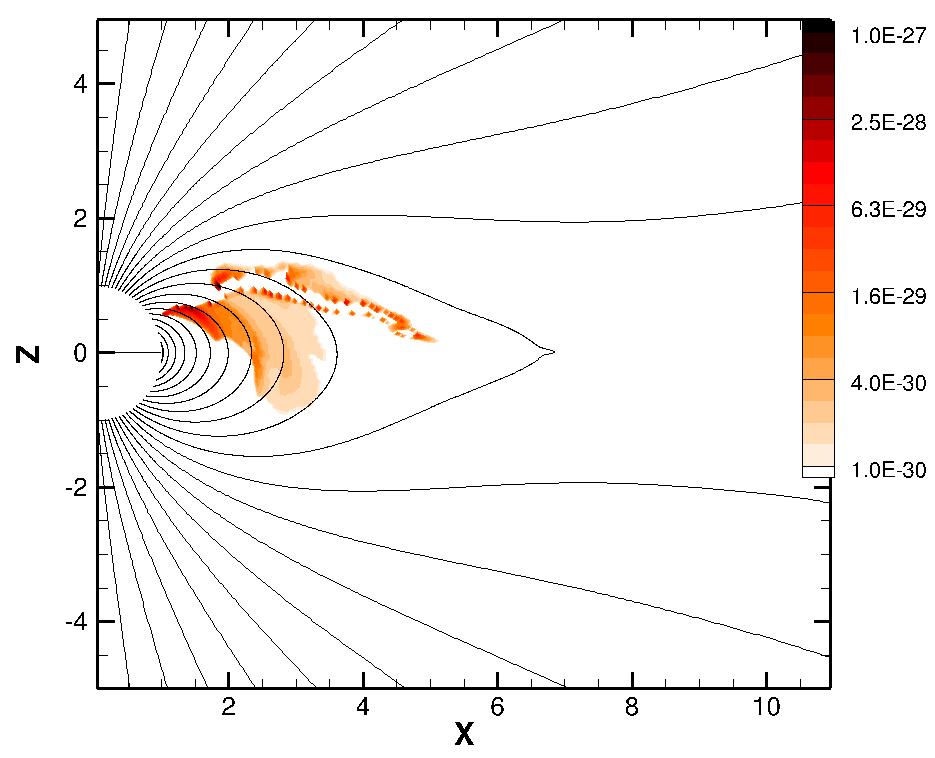}}
    \end{minipage}
  \end{center}  
  \caption{Snapshot of the 2D MHD simulation of the magnetically confined wind of HD~108 (Model II) at 6\,Ms after the onset of the simulation. The $z$ axis indicates the direction of the magnetic dipole axis. From left to right, the panels illustrate the logarithm of the plasma temperature (in K), the logarithm of the density (in g\,cm$^{-3}$), and the emission measure of the plasma (in g$^2$\,cm$^{-6}$) with $T \geq T_{\rm thresh} = 1.5$\,MK (equivalent to $kT_{\rm threshold} = 0.13$\,keV). The emission measure is defined as $\rho^2\,\exp{(-T/T_{\rm thresh})}$ \citep{udD14}.\label{MHD1}}
\end{figure*}

\subsection{Magnetohydrodynamic simulations}\label{MHD}       
We performed a fully self-consistent MHD simulation of HD~108 using the stellar parameters listed in Table\,\ref{param}. For this purpose, we used the publicly available MHD code Zeus-3D adapted to our own needs.

The magnetospheres of massive stars are divided into two broad categories depending on the comparison between the Keplerian corotation radius $R_{\rm K}$ and the Alfv\'en radius $R_{\rm A}$ \citep{Pet13}. Situations where $R_{\rm K} < R_{\rm A}$, corresponding to rapid rotators, are called centrifugal magnetospheres. In those stars, the trapped wind material accumulates in a stable long-lived rigidly rotating disk-like structure. For slow rotators, one has instead $R_{\rm K} > R_{\rm A}$. Material from these so-called dynamical magnetospheres falls back onto the star. Owing to its very slow rotation, HD~108 clearly belongs to the class of stars with dynamical magnetospheres.

The extremely slow rotation of HD~108 simplifies the MHD simulations of the star's magnetosphere. Indeed, in such a case, rotation has no significant dynamical effects on the magnetospheric structure. As such, it can thus be modeled in two dimensions (2D) assuming a field aligned with the rotation axis \citep{Sun12}. We follow the basic methods and formalism presented for 2D simulations by \citet{udD02}, and  extended by \citet{Gag05} to include a full energy equation with optically thin radiative cooling \citep{Mac81}. Following \citet{udD02}, we use radiative driving by line-scattering based on the standard CAK \citep{Cas75} formalism, corrected for the finite cone angle of the star, using a spherical expansion approximation for the local flow gradients \citep{Fri86,Pau86}.

The computational grid and boundary conditions are adopted from \citet{udD02}, and \citet{udD08} to include the non-zero azimuthal velocity at the lower boundary arising from rotation.  We use 300 radial grid points logarithmically stretched from stellar surface to 20 stellar radii with a 2\% increase in length, i.e. the ratio between two subsequent zones is 1.02. For the mesh in co-latitude, again following \citet{udD02}, we use 100 grid points with a higher number of points around the magnetic equator to capture the compressed disk structure more efficiently. We ran our model for 6 Ms which is long enough to be free from transients resulting from any initial condition.
   
A snapshot of the structure of the magnetosphere (taken at 6\,Ms) corresponding to Model II is shown in Fig.\,\ref{MHD1}. The X-ray emission of each grid cell is computed assuming an optically thin plasma and taking $n_e = n_{\rm H}$. The synthesized X-ray luminosity integrated between 0.1 and 10\,keV of this simulation is $2.48\,10^{33}$\,erg\,s$^{-1}$, whilst it amounts to $7.3\,10^{32}$\,erg\,s$^{-1}$ between 0.5 and 10.0\,keV. These synthetic luminosities assume no stellar occultation or absorption by the stellar wind, and thus represent an upper limit to the observable luminosity. Stellar wind absorption would mostly affect the softer part of the X-ray spectrum (below 2\,keV). However, significant absorption by the circumstellar material has been securely detected only in the extreme case of NGC~1624-2 \citep{Pet15}. On the other hand, given that the X-ray emission arises at a distance of about 3.5\,$R_*$ (Fig.\,\ref{MHD1}), occultation effects will only play a role at rotational phases close to the minimum state. With these caveats in mind, the 0.5 -- 10\,keV luminosity predicted by the model is in very good agreement with the observationally-determined values (see Sect.\,\ref{Xdata}). Figure\,\ref{ufX} further shows that the spectral energy distribution predicted by the Model II MHD calculations nicely matches the observed one.

Very similar values of the X-ray luminosities are obtained for Model I ($2.52\,10^{33}$ and $7.7\,10^{32}$\,erg\,s$^{-1}$), whilst Model III yields lower values ($9.8\,10^{32}$ and $2.2\,10^{32}$\,erg\,s$^{-1}$). At first sight, it is surprising to see Model III produce less X-ray emission than the two other models with lower magnetic field strengths as this is contrary to the predictions of the semi-analytic XADM paradigm \citep{udD14}. However, HD~108's wind properties result in a wind cooling parameter $\chi_{\infty}$ at the border between the adiabatic and radiative regimes. Any small deviation in the wind parameters can thus lead to significant changes in the ensuing X-ray emission. In particular, the much higher magnetic confinement parameter in Model III leads to much larger closed loops wherein wind speed can substantially exceed the presumed $v_{\infty}$ of the non-magnetic wind, thus making cooling much less efficient as the cooling parameter is very sensitive to speed ($\chi_{\infty} \propto v^4$). This in turn enhances the shock retreat effects wherein less efficient cooling pushes the shock back along the loops towards the star, hence to lower pre-shock wind speeds. As a result, the X-ray emission becomes softer and weaker. Further studies will be required to investigate this issue in more details.

The emission measure plot (right panel of Fig.\,\ref{MHD1}) indicates that the bulk of the X-ray emission in Model II is generated between about 1.5 and 3.5\,$R_*$ (i.e., out to the Alfv\'en radius). Though the simulation also predicts hot material with a $kT$ above 0.13\,keV beyond this region (left panel of Fig.\,\ref{MHD1}), this plasma is too tenuous to contribute significantly to the overall X-ray emission as revealed by comparing the left and right panels of Fig.\,\ref{MHD1}.

\begin{figure}
  \begin{center}
    \resizebox{8cm}{!}{\includegraphics{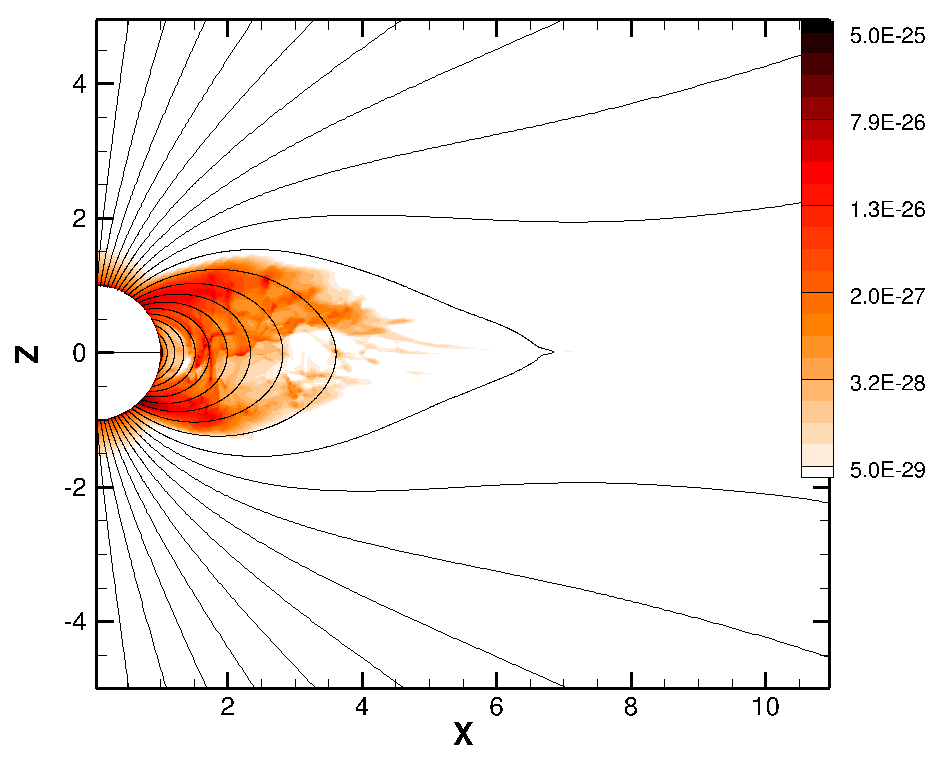}}
  \end{center}  
\caption{Time-averaged $\rho^2$ distribution as inferred from the 2D MHD simulation of Model II. \label{MHD2}}
\end{figure}

The variable optical emission lines are expected to arise from the radiatively cooled magnetospheric material that falls back to the stellar surface \citep{Sun12,Owo16}. Since the H$\alpha$ line emission is formed via the recombination process, the best indicator of the line forming region is $\rho^2$. Figure\,\ref{MHD2} indicates that most of the optical line emission is expected to arise from material located between the stellar surface and about $3.5\,R_*$, i.e.\ from within the Alfv\'en radius ($R_A = 3.3\,R_*$ for Model II).  

\section{Discussion and conclusions}\label{discuss}
In this study, we analysed optical and X-ray spectra of the Of?p star HD~108 to constrain the properties of this star and its magnetosphere. Beside the $54 \pm 3$\,yrs cycle that modulates the strength of most emission lines, our data unveil an 8.5\,yr periodic variation of the radial velocities with a peak-to-peak amplitude of 10 -- 11\,km\,s$^{-1}$. This is strong evidence for binarity with an unseen companion of at least 4\,M$_{\odot}$. Alike two other Of?p stars, HD~108 thus also is a binary system, although the binary properties (orbital period, eccentricity, etc.) differ strongly among these three objects. 

The variable emission lines in the optical spectrum of HD~108 have now recovered from the minimum emission state that was reached in the years 2007--2008. Whilst our spectroscopic monitoring of HD~108 with modern instrumentations does not yet cover the full 54\,yr cycle, our fitting of the EW(H$\alpha$) light curve with an analytic dynamical magnetosphere model yields constraints on the inclination angle of the rotation axis ($i$) and the obliquity of the magnetic axis ($\beta$). We find that pairs of ($i$, $\beta$) that satisfy $i + \beta \simeq 85^{\circ}$ with $i \in [30^{\circ},55^{\circ}]$ yield results of similar quality. This implies that near minimum state, the magnetosphere was seen under an angle of $85^{\circ}$, i.e.\ almost edge-on. The detection of inverse P-Cygni-type profiles in the Si\,{\sc iii} $\lambda\lambda$\,4552, 4568, 4575 lines around the 2007--2008  minimum state (see the left panel of Fig.\,\ref{montage}) then clearly indicates that these lines arise in material that was observed to flow back to the stellar surface. This line morphology is fully consistent with our configuration of the magnetically confined wind: around minimum state, the magnetosphere is seen nearly edge-on and the backfalling material where the Si\,{\sc iii} emissions arise is seen projected against the stellar disk. 

{\it TESS} photometry of HD~108 unveiled a strong red noise variability which likely reflects the same phenomenon as the macroturbulence that broadens the spectral lines of this very slow rotator. Beside this red noise component, the {\it TESS} data also display a transient cyclic modulation on a timescale of about six days. We found similar, though not identical, timescales in the short-term variations of the H$\alpha$ emission. This suggests a possible link between the transient photometric cycle and the spectroscopic variability. Although we cannot rule out a purely photospheric origin, this variability could also arise from episodes of enhanced downflow of magnetospheric material. Indeed, \citet{Bar07} reported on a slow photometric fading of HD~108 as the star was going to the minimum emission state. This suggests that reflection of photospheric light by cool material in the magnetosphere contributes to the optical brightness of this star. The same cool magnetospheric material is responsible for the H$\alpha$ emission. The existence of modulations on timescales of $\sim$ 4 -- 8 days in photometry and line emission strength could thus reflect a roughly cyclic behaviour of the magnetosphere, where material first piles up locally before it cools and falls back to the star.

\begin{figure}
  \begin{center}
    \resizebox{8cm}{!}{\includegraphics{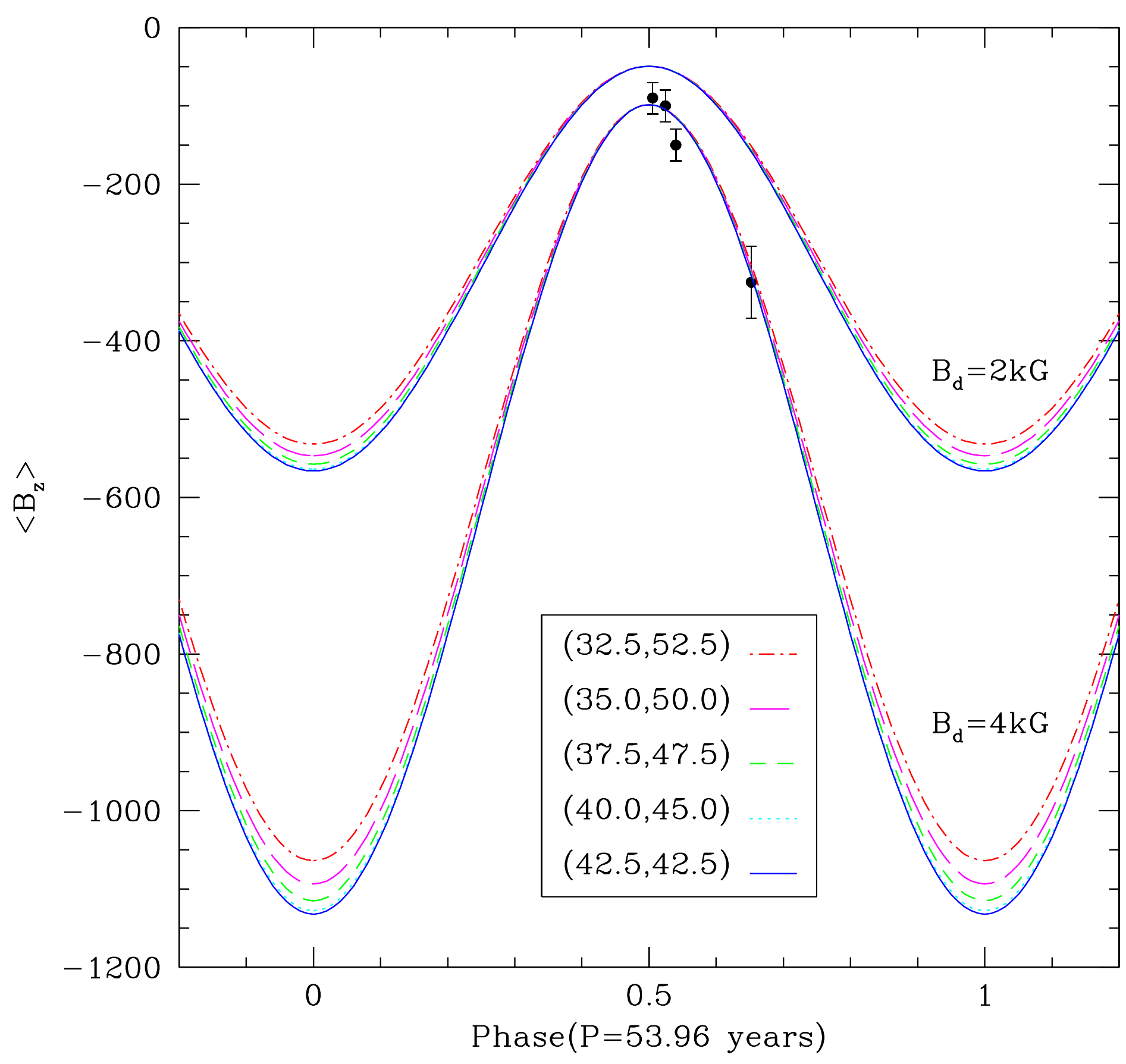}}
  \end{center}  
\caption{Predicted values of $<B_z>$ computed with equation\,\ref{eqDonati} for the $(i, \beta)$ pairs found in Sect.\,\ref{ADMfit} (see Fig.\,\ref{ADMfits}). Two different values of $B_d$ have been tested. The points with their error bars stand for the data from \citet{Mar10} and \citet{Shu17}. \label{Bz}}
\end{figure}
The best-fit ADM models predict that the emission strength should further increase by about 1.4\,\AA\ in the coming decade. Spectroscopic monitoring of the star in the coming decade will show whether the light curve near phase 0.5 is flat or presents a secondary dip, thus leading to even more stringent constraints on $i$ and $\beta$. Meanwhile, our results indicate that the dipolar magnetic field strength $B_d$ of HD~108 is probably significantly stronger than the value previously estimated by \citet{Mar10} or \citet{Shu17}. Indeed, in an oblique magnetic rotator, the emission lines are strongest when the magnetically confined wind is seen near face-on, i.e.\ when the angle between the direction of the dipolar magnetic field and the line of sight is minimum. Figure\,\ref{Bz} illustrates the value of $<B_Z>$ computed with equation \ref{eqDonati} \citep{Don02} for the different pairs of $i$ and $\beta$ allowed by our ADM models, and assuming values of $B_d$ of $-2000$ or $-4000$\,G. Whilst the existing spectropolarimetric determinations of $<B_z>$ are not sufficient to achieve a precise evaluation of $B_d$, the figure clearly shows that they are in much better agreement with values of $B_d$ of $-4000$\,G than with values of $-2000$\,G or lower. Forthcoming spectropolarimetric observations of HD~108 should thus indicate significantly larger values of $<B_z>$, hence allowing to further constrain the value of HD~108's dipolar magnetic field strength.

The {\it GAIA} distance determination of HD~108 led to a revision of the star's mass-loss rate. This could have important implications on the spin-down time. Indeed, the value of the magnetic field originally determined by \citet{Mar10} and the low value of $\dot{M}$ inferred by \citet{Marc12} implied an upper limit on the spin-down time more than twice longer than the estimated age of HD~108 \citep{Pet13}. \citet{Shu17} obtained a higher value of $<B_z>$ which solved this issue, provided the stellar wind mass-loss rate was taken to be $2.8\,10^{-6}$\,M$_{\odot}$\,yr$^{-1}$. In this paper, we have shown that scaling the stellar parameters to the {\it GAIA} distance yields a lower $\dot{M}_{B=0}$ of $0.9\,10^{-6}$\,M$_{\odot}$\,yr$^{-1}$. The spin-down timescale $\tau_J$ depends on the stellar mass-loss timescale ($M_*/\dot{M}_{B=0}$) and the wind confinement parameter $\eta_*$ \citep{udD08,udD22}. Using equation (8) of \citet{udD22} with the parameters of Table\,\ref{param}, we infer $\tau_J$ values of $9.9\,10^5$, $6.0\,10^5$ and $3.2\,10^5$\,yr respectively for Models I, II and III. As shown above, Models II and especially III are in better agreement with the existing $<B_z>$ measurements. We thus conclude that $\tau_J$ is most probably in the range $3.2\,10^5$ to $6.0\,10^5$\,yr. The maximum spin-down age, assuming the star was at critical rotation initially, is then given by $t_{\rm s,max} = -\tau_J\,\ln{W}$ where $W = V_{\rm rot}/{\sqrt{G\,M_*/R_*}}$ is the present-day critical rotation fraction. Again using the parameters of Table\,\ref{param}, we obtain $W \simeq 5.3\,10^{-5}$ and thus $t_{\rm s,max}$ between 3.1 and 5.9\,Myr respectively for $B_d = -4000$\,G and $B_d = -2000$\,G. With the revised distance determination, HD~108 would have an evolutionary age of about 3.5\,Myr \citep{Mar12}. We thus conclude that the spin-down age is consistent with the evolutionary age provided that the dipole field strength is indeed near 4\,kG. 

Our observations of HD~108 indicate that our sightline is now nearly aligned with the magnetic axis, allowing us to gain precious insight into the properties of its magnetosphere. Continuing the monitoring of this star in optical spectroscopy, spectropolarimetry and X-ray spectroscopy over at least another decade will allow to finally lift the remaining ambiguities.  

\section*{Acknowledgements}
GR and YN would like to express their gratitude to Jean-Marie Vreux for stimulating their interest in HD~108 and to the technical staff of the Haute Provence Observatory, especially to the telescope operators J.-P.\ Bretagne, R.\ Giraud, D.\ Gravallon, J.-C.\ Mevolhon, and J.-P.\ Troncin for their efficient help during many observing nights. We thank the referee, Gregg Wade, for his report that helped us improve our manuscript. YN and GR acknowledge support from the Fonds National de la Recherche Scientifique (Belgium), the Communaut\'e Fran\c caise de Belgique (including notably support for the observing runs at OHP), and the Belgian Federal Science Policy Office (BELSPO) in the framework of the PRODEX Programme (contract HERMeS). AuD acknowledges NASA ATP grant number 80NSSC22K0628 and support by NASA through Chandra Award number TM1-22001B and GO2-23003X issued by the Chandra X-ray Observatory 27 Center, which is operated by the Smithsonian Astrophysical Observatory for and on behalf of NASA under contract NAS8-03060. This project makes use of data collected by the TESS mission, whose funding is provided by the NASA Explorer Program. ADS and CDS were used for this research. This research uses optical spectra collected with the TIGRE telescope (La Luz, Mexico). TIGRE is a collaboration of Hamburger Sternwarte, the Universities of Hamburg, Guanajuato and Li\`ege. 

\section*{Data availability}
The {\it TESS} data underlying this article are available from the MAST archives, while the OHP and TIGRE spectra can be made available upon reasonable request. {\it Chandra} and {\it XMM-Newton} data are available from their respective archives.

\appendix
\section{Table of observations}
Table\,\ref{Journal} below provides the list of the optical spectra of HD~108 used in this study along with the measured RVs and EWs. The spectra are ordered by date.\\

\begin{table*}
  \caption{Journal of our optical spectroscopic observations \label{Journal}}
  \tiny
  \begin{tabular}{l c | r r r r r r | r r r r r r r r}
    \hline
    HJD-2400000 & Obs. & & \multicolumn{4}{c}{RVs (km\,s$^{-1}$)} & & & \multicolumn{6}{c}{EWs (\AA)} & \\
    & & He\,{\sc ii} & He\,{\sc ii} & He\,{\sc ii} & N\,{\sc iii} & O\,{\sc iii} &  C\,{\sc iii} & He\,{\sc ii} & H$\gamma$ & He\,{\sc i} & H$\beta$ & C\,{\sc iii} & He\,{\sc i} & H$\alpha$ & He\,{\sc i} \\
    & & $\lambda$\,4200 & $\lambda$\,4542 & $\lambda$\,4686 & & $\lambda$\,5592 &  $\lambda$\,5696 & $\lambda$\,4200 & & $\lambda$\,4471 & & $\lambda$\,5696 & $\lambda$\,5876 & & $\lambda$\,6678\\
    \hline
46627.5820 & OHP &$-70.1$ &$ $ &$ $ &$ $ &$ $ &$ $ &$ 0.55$ &$ $ &$ $ &$ $ &$ $ &$ $ &$ $ &$ $ \\
46628.5480 & OHP &$-59.5$ &$ $ &$ $ &$ $ &$ $ &$ $ &$ 0.55$ &$ $ &$ $ &$ $ &$ $ &$ $ &$ $ &$ $ \\
47007.5360 & OHP &$-77.2$ &$ $ &$ $ &$ $ &$ $ &$ $ &$ 0.51$ &$ 0.18$ &$ $ &$ $ &$ $ &$ $ &$ $ &$ $ \\
47007.5440 & OHP &$-78.9$ &$ $ &$ $ &$ $ &$ $ &$ $ &$ 0.51$ &$ 0.15$ &$ $ &$ $ &$ $ &$ $ &$ $ &$ $ \\
47007.5490 & OHP &$-82.1$ &$ $ &$ $ &$ $ &$ $ &$ $ &$ 0.53$ &$ 0.15$ &$ $ &$ $ &$ $ &$ $ &$ $ &$ $ \\
47007.5540 & OHP &$-80.5$ &$ $ &$ $ &$ $ &$ $ &$ $ &$ 0.52$ &$ 0.16$ &$ $ &$ $ &$ $ &$ $ &$ $ &$ $ \\
47008.5020 & OHP &$-85.1$ &$ $ &$ $ &$ $ &$ $ &$ $ &$ 0.53$ &$ 0.12$ &$ $ &$ $ &$ $ &$ $ &$ $ &$ $ \\
47008.5190 & OHP &$-82.1$ &$ $ &$ $ &$ $ &$ $ &$ $ &$ 0.51$ &$ 0.11$ &$ $ &$ $ &$ $ &$ $ &$ $ &$ $ \\
47008.5240 & OHP &$-86.1$ &$ $ &$ $ &$ $ &$ $ &$ $ &$ 0.51$ &$ 0.13$ &$ $ &$ $ &$ $ &$ $ &$ $ &$ $ \\
47009.5260 & OHP &$-83.6$ &$ $ &$ $ &$ $ &$ $ &$ $ &$ 0.50$ &$ 0.18$ &$ $ &$ $ &$ $ &$ $ &$ $ &$ $ \\
47009.5380 & OHP &$-79.7$ &$ $ &$ $ &$ $ &$ $ &$ $ &$ 0.53$ &$ 0.17$ &$ $ &$ $ &$ $ &$ $ &$ $ &$ $ \\
47010.5490 & OHP &$-79.8$ &$ $ &$ $ &$ $ &$ $ &$ $ &$ 0.53$ &$ 0.15$ &$ $ &$ $ &$ $ &$ $ &$ $ &$ $ \\
47010.5580 & OHP &$-78.7$ &$ $ &$ $ &$ $ &$ $ &$ $ &$ 0.53$ &$ 0.12$ &$ $ &$ $ &$ $ &$ $ &$ $ &$ $ \\
47011.5540 & OHP &$-77.2$ &$ $ &$ $ &$ $ &$ $ &$ $ &$ 0.49$ &$ 0.12$ &$ $ &$ $ &$ $ &$ $ &$ $ &$ $ \\
47011.5610 & OHP &$-81.1$ &$ $ &$ $ &$ $ &$ $ &$ $ &$ 0.50$ &$ 0.11$ &$ $ &$ $ &$ $ &$ $ &$ $ &$ $ \\
47011.5680 & OHP &$-82.7$ &$ $ &$ $ &$ $ &$ $ &$ $ &$ 0.49$ &$ 0.14$ &$ $ &$ $ &$ $ &$ $ &$ $ &$ $ \\
47011.5720 & OHP &$-79.5$ &$ $ &$ $ &$ $ &$ $ &$ $ &$ 0.49$ &$ 0.15$ &$ $ &$ $ &$ $ &$ $ &$ $ &$ $ \\
47012.5660 & OHP &$-75.0$ &$ $ &$ $ &$ $ &$ $ &$ $ &$ 0.53$ &$ 0.19$ &$ $ &$ $ &$ $ &$ $ &$ $ &$ $ \\
47012.5730 & OHP &$-76.7$ &$ $ &$ $ &$ $ &$ $ &$ $ &$ 0.49$ &$ 0.18$ &$ $ &$ $ &$ $ &$ $ &$ $ &$ $ \\
47013.5400 & OHP &$-76.8$ &$ $ &$ $ &$ $ &$ $ &$ $ &$ 0.53$ &$ 0.27$ &$ $ &$ $ &$ $ &$ $ &$ $ &$ $ \\
47013.5450 & OHP &$-70.4$ &$ $ &$ $ &$ $ &$ $ &$ $ &$ 0.52$ &$ 0.17$ &$ $ &$ $ &$ $ &$ $ &$ $ &$ $ \\
47013.5490 & OHP &$-76.8$ &$ $ &$ $ &$ $ &$ $ &$ $ &$ 0.52$ &$ 0.18$ &$ $ &$ $ &$ $ &$ $ &$ $ &$ $ \\
47016.6010 & OHP &$ $ &$ -72.0$ &$ -68.6$ &$ $ &$ $ &$ $ &$ $ &$ $ &$ 0.13$ &$ -1.32$ &$ $ &$ $ &$ $ &$ $ \\
48490.4780 & OHP &$ $ &$ -95.1$ &$ $ &$ $ &$ $ &$ $ &$ $ &$ 0.43$ &$ 0.27$ &$ $ &$ $ &$ $ &$ $ &$ $ \\
48490.4890 & OHP &$ $ &$ -88.8$ &$ $ &$ $ &$ $ &$ $ &$ $ &$ 0.38$ &$ 0.24$ &$ $ &$ $ &$ $ &$ $ &$ $ \\
48490.5050 & OHP &$ $ &$ -88.2$ &$ $ &$ $ &$ $ &$ $ &$ $ &$ 0.34$ &$ 0.25$ &$ $ &$ $ &$ $ &$ $ &$ $ \\
48492.6390 & OHP &$ $ &$ -95.5$ &$ $ &$ $ &$ $ &$ $ &$ $ &$ 0.31$ &$ 0.22$ &$ $ &$ $ &$ $ &$ $ &$ $ \\
48496.5900 & OHP &$ $ &$ -85.7$ &$ $ &$ $ &$ $ &$ $ &$ $ &$ 0.34$ &$ 0.22$ &$ $ &$ $ &$ $ &$ $ &$ $ \\
48497.5810 & OHP &$ $ &$ -78.7$ &$ $ &$ $ &$ $ &$ $ &$ $ &$ 0.40$ &$ 0.23$ &$ $ &$ $ &$ $ &$ $ &$ $ \\
49264.4470 & OHP &$-62.8$&$ $&$ $&$ $&$ $&$ $&$0.51$&$0.51$&$ $&$ $&$ $&$ $&$ $&$ $\\
49264.4520 & OHP &$-68.1$&$ $&$ $&$ $&$ $&$ $&$0.57$&$0.64$&$ $&$ $&$ $&$ $&$ $&$ $\\
49576.5530 & OHP &$-49.0$&$ $&$ $&$ $&$ $&$ $&$0.51$&$0.61$&$ $&$ $&$ $&$ $&$ $&$ $\\
49577.5420 & OHP &$-68.8$&$ $&$ $&$ $&$ $&$ $&$0.51$&$0.62$&$ $&$ $&$ $&$ $&$ $&$ $\\
49579.5220 & OHP &$-74.8$&$ $&$ $&$ $&$ $&$ $&$0.48$&$0.62$&$ $&$ $&$ $&$ $&$ $&$ $\\
49580.4810 & OHP &$-64.7$&$ $&$ $&$ $&$ $&$ $&$0.51$&$0.63$&$ $&$ $&$ $&$ $&$ $&$ $\\
49581.5350 & OHP &$-62.4$&$ $&$ $&$ $&$ $&$ $&$0.50$&$0.62$&$ $&$ $&$ $&$ $&$ $&$ $\\
49582.5110 & OHP &$-63.8$&$ $&$ $&$ $&$ $&$ $&$0.51$&$0.65$&$ $&$ $&$ $&$ $&$ $&$ $\\
49583.5330 & OHP &$-63.4$&$ $&$ $&$ $&$ $&$ $&$0.52$&$0.65$&$ $&$ $&$ $&$ $&$ $&$ $\\
50316.6500 & OHP &$-61.6$&$-66.6$&$-55.7$&$-82.8$&$ $&$ $&$0.53$&$0.78$&$0.40$&$0.19$&$ $&$ $&$ $&$ $\\
50316.6580 & OHP &$-68.2$&$-63.2$&$-56.9$&$-78.2$&$ $&$ $&$0.58$&$0.77$&$0.41$&$0.14$&$ $&$ $&$ $&$ $\\
50318.6440 & OHP &$-58.2$&$-64.0$&$-54.2$&$-80.5$&$ $&$ $&$0.52$&$0.79$&$0.43$&$0.23$&$ $&$ $&$ $&$ $\\
50506.2780 & OHP &$ $&$ $&$ $&$ $&$ $&$ $&$ $&$ $&$ $&$ $&$ $&$ $&$-5.39$&$-1.16$\\
50507.2850 & OHP &$ $&$ $&$ $&$ $&$ $&$ $&$ $&$ $&$ $&$ $&$ $&$ $&$-5.56$&$-0.99$\\
50508.2730 & OHP &$ $&$ $&$ $&$ $&$ $&$ $&$ $&$ $&$ $&$ $&$ $&$ $&$-5.17$&$-0.95$\\
50510.2720 & OHP &$ $&$ $&$ $&$ $&$ $&$ $&$ $&$ $&$ $&$ $&$ $&$ $&$-5.27$&$-1.07$\\
50638.5740 & OHP &$-68.1$&$-61.1$&$-55.6$&$-78.0$&$ $&$ $&$0.52$&$0.87$&$0.54$&$0.38$&$ $&$ $&$ $&$ $\\
50639.5570 & OHP &$-70.6$&$-58.7$&$-54.5$&$-74.4$&$ $&$ $&$0.55$&$0.77$&$0.53$&$0.39$&$ $&$ $&$ $&$ $\\
50642.5730 & OHP &$-66.3$&$-60.5$&$-53.1$&$-72.6$&$ $&$ $&$0.58$&$0.76$&$0.53$&$0.37$&$ $&$ $&$ $&$ $\\
51066.6350 & OHP &$ $&$-70.6$&$-67.1$&$-85.7$&$ $&$ $&$ $&$ $&$0.64$&$ $&$ $&$ $&$ $&$ $\\
51070.6290 & OHP &$ $&$-71.1$&$-66.5$&$-86.5$&$ $&$ $&$ $&$ $&$0.62$&$ $&$ $&$ $&$ $&$ $\\
51072.6380 & OHP &$ $&$-69.8$&$-63.1$&$-85.0$&$ $&$ $&$ $&$ $&$0.56$&$ $&$ $&$ $&$ $&$ $\\
51132.3770 & OHP &$ $&$-76.3$&$ $&$-90.8$&$ $&$ $&$ $&$ $&$0.75$&$ $&$ $&$ $&$ $&$ $\\
51132.4210 & OHP &$ $&$-72.2$&$ $&$-88.9$&$ $&$ $&$ $&$ $&$0.77$&$ $&$ $&$ $&$ $&$ $\\
51133.3300 & OHP &$ $&$-73.6$&$ $&$-87.8$&$ $&$ $&$ $&$ $&$0.74$&$ $&$ $&$ $&$ $&$ $\\
51133.3800 & OHP &$ $&$-73.6$&$ $&$-83.2$&$ $&$ $&$ $&$ $&$0.74$&$ $&$ $&$ $&$ $&$ $\\
51133.4320 & OHP &$ $&$-74.1$&$ $&$-90.2$&$ $&$ $&$ $&$ $&$0.74$&$ $&$ $&$ $&$ $&$ $\\
51133.5890 & OHP &$ $&$-73.5$&$ $&$-89.0$&$ $&$ $&$ $&$ $&$0.72$&$ $&$ $&$ $&$ $&$ $\\
51134.3780 & OHP &$ $&$-74.1$&$ $&$-90.3$&$ $&$ $&$ $&$ $&$0.68$&$ $&$ $&$ $&$ $&$ $\\
51134.4480 & OHP &$ $&$-71.9$&$ $&$-87.5$&$ $&$ $&$ $&$ $&$0.68$&$ $&$ $&$ $&$ $&$ $\\
51135.3390 & OHP &$ $&$-72.1$&$ $&$-84.7$&$ $&$ $&$ $&$ $&$0.72$&$ $&$ $&$ $&$ $&$ $\\
51135.3940 & OHP &$ $&$-74.2$&$ $&$-90.1$&$ $&$ $&$ $&$ $&$0.74$&$ $&$ $&$ $&$ $&$ $\\
51135.4580 & OHP &$ $&$-74.7$&$ $&$-93.4$&$ $&$ $&$ $&$ $&$0.70$&$ $&$ $&$ $&$ $&$ $\\
51136.3730 & OHP &$ $&$-72.0$&$ $&$-89.9$&$ $&$ $&$ $&$ $&$0.70$&$ $&$ $&$ $&$ $&$ $\\
51136.4390 & OHP &$ $&$-74.1$&$ $&$-90.4$&$ $&$ $&$ $&$ $&$0.65$&$ $&$ $&$ $&$ $&$ $\\
51137.4010 & OHP &$ $&$-73.2$&$ $&$-93.7$&$ $&$ $&$ $&$ $&$0.70$&$ $&$ $&$ $&$ $&$ $\\
51137.4670 & OHP &$ $&$-71.4$&$ $&$-94.8$&$ $&$ $&$ $&$ $&$0.73$&$ $&$ $&$ $&$ $&$ $\\
51374.5966 & OHP &$-78.8$&$-71.7$&$-69.0$&$-88.5$&$ $&$ $&$0.57$&$1.05$&$0.69$&$0.57$&$ $&$ $&$ $&$ $\\
51376.5322 & OHP &$-79.8$&$-76.4$&$-69.9$&$-90.7$&$ $&$ $&$0.62$&$1.14$&$0.76$&$0.66$&$ $&$ $&$ $&$ $\\
51378.5299 & OHP &$-78.0$&$-71.7$&$-67.0$&$-89.3$&$ $&$ $&$0.50$&$0.98$&$0.73$&$0.64$&$ $&$ $&$ $&$ $\\
51396.5996 & OHP &$-75.0$&$-72.7$&$-68.1$&$-86.8$&$ $&$ $&$0.47$&$1.11$&$0.80$&$0.65$&$ $&$ $&$ $&$ $\\
51403.6353 & OHP &$-76.0$&$-76.1$&$-70.4$&$-91.6$&$ $&$ $&$0.53$&$1.01$&$0.78$&$0.64$&$ $&$ $&$ $&$ $\\
51406.6338 & OHP &$-75.8$&$-75.1$&$-69.0$&$-90.9$&$ $&$ $&$0.57$&$1.07$&$0.77$&$0.60$&$ $&$ $&$ $&$ $\\
51407.6390 & OHP &$-76.4$&$-75.7$&$-69.9$&$-91.7$&$ $&$ $&$0.57$&$1.32$&$0.80$&$0.66$&$ $&$ $&$ $&$ $\\
51810.5860 & OHP &$ $&$-75.3$&$-68.2$&$-91.1$&$ $&$ $&$ $&$ $&$0.87$&$0.95$&$ $&$ $&$ $&$ $\\
51810.5950 & OHP &$ $&$-75.2$&$-70.0$&$-94.5$&$ $&$ $&$ $&$ $&$0.88$&$0.98$&$ $&$ $&$ $&$ $\\
51811.5670 & OHP &$ $&$-74.1$&$-68.2$&$-92.8$&$ $&$ $&$ $&$ $&$0.82$&$0.93$&$ $&$ $&$ $&$ $\\
51811.5750 & OHP &$ $&$-73.3$&$-68.2$&$-91.3$&$ $&$ $&$ $&$ $&$0.86$&$0.93$&$ $&$ $&$ $&$ $\\
51812.5940 & OHP &$ $&$-76.3$&$-71.4$&$-95.1$&$ $&$ $&$ $&$ $&$0.82$&$0.87$&$ $&$ $&$ $&$ $\\
51813.6030 & OHP &$ $&$-72.8$&$-65.8$&$-91.0$&$ $&$ $&$ $&$ $&$0.87$&$0.96$&$ $&$ $&$ $&$ $\\
51814.5950 & OHP &$ $&$-75.1$&$-67.3$&$-91.8$&$ $&$ $&$ $&$ $&$0.91$&$1.00$&$ $&$ $&$ $&$ $\\
51815.5960 & OHP &$ $&$-74.9$&$-70.2$&$-92.2$&$ $&$ $&$ $&$ $&$0.86$&$0.95$&$ $&$ $&$ $&$ $\\
51818.5310 & OHP &$ $&$-71.9$&$-65.8$&$-86.0$&$ $&$ $&$ $&$ $&$0.85$&$0.91$&$ $&$ $&$ $&$ $\\
51819.5600 & OHP &$ $&$-74.3$&$-68.0$&$-90.9$&$ $&$ $&$ $&$ $&$0.92$&$0.96$&$ $&$ $&$ $&$ $\\
51821.5610 & OHP &$ $&$-76.9$&$-71.2$&$-90.7$&$ $&$ $&$ $&$ $&$0.89$&$0.94$&$ $&$ $&$ $&$ $\\
52163.5010 & OHP &$ $&$-70.7$&$-68.4$&$-90.2$&$ $&$ $&$ $&$ $&$0.96$&$1.11$&$ $&$ $&$ $&$ $\\
52163.6530 & OHP &$ $&$ $&$ $&$ $&$ $&$ $&$ $&$ $&$ $&$ $&$ $&$ $&$-2.83$&$-0.32$\\
52164.4950 & OHP &$ $&$-72.2$&$-68.2$&$-89.0$&$ $&$ $&$ $&$ $&$1.00$&$1.05$&$ $&$ $&$ $&$ $\\
52165.6280 & OHP &$ $&$-70.8$&$-67.2$&$-88.7$&$ $&$ $&$ $&$ $&$1.01$&$1.19$&$ $&$ $&$ $&$ $\\
52167.6240 & OHP &$ $&$-72.6$&$-66.4$&$-88.9$&$ $&$ $&$ $&$ $&$1.01$&$1.15$&$ $&$ $&$ $&$ $\\
52167.6530 & OHP &$-70.0$&$ $&$ $&$ $&$ $&$ $&$0.67$&$1.37$&$ $&$ $&$ $&$ $&$ $&$ $\\
52167.6670 & OHP &$-69.1$&$ $&$ $&$ $&$ $&$ $&$0.64$&$1.32$&$ $&$ $&$ $&$ $&$ $&$ $\\
52170.6510 & OHP &$ $&$-70.5$&$-67.3$&$-89.1$&$ $&$ $&$ $&$ $&$1.03$&$1.28$&$ $&$ $&$ $&$ $\\
52170.6650 & OHP &$ $&$-70.6$&$-66.7$&$-89.1$&$ $&$ $&$ $&$ $&$1.04$&$1.29$&$ $&$ $&$ $&$ $\\
52518.4763 & OHP &$ $&$-67.8$&$-63.2$&$-86.2$&$ $&$ $&$ $&$ $&$1.12$&$1.41$&$ $&$ $&$ $&$ $\\
52524.4224 & OHP &$ $&$-68.8$&$-66.1$&$-84.7$&$ $&$ $&$ $&$ $&$1.13$&$1.35$&$ $&$ $&$ $&$ $\\
52524.6509 & OHP &$ $&$ $&$ $&$ $&$ $&$ $&$ $&$ $&$ $&$ $&$ $&$ $&$-2.83$&$-0.12$\\
52613.6360 & OHP &$ $&$ $&$ $&$ $&$ $&$ $&$ $&$ $&$ $&$ $&$ $&$ $&$-2.91$&$-0.35$\\
52917.5310 & OHP &$ $&$-65.0$&$-64.5$&$-84.1$&$ $&$ $&$ $&$ $&$1.19$&$1.62$&$ $&$ $&$ $&$ $\\
52917.5410 & OHP &$ $&$-63.9$&$-62.2$&$ $&$ $&$ $&$ $&$ $&$1.32$&$1.50$&$ $&$ $&$ $&$ $\\
52918.5330 & OHP &$ $&$-66.7$&$-61.3$&$-82.3$&$ $&$ $&$ $&$ $&$1.14$&$1.59$&$ $&$ $&$ $&$ $\\
52918.5470 & OHP &$ $&$-63.9$&$-60.5$&$-78.1$&$ $&$ $&$ $&$ $&$1.17$&$1.68$&$ $&$ $&$ $&$ $\\
52919.3380 & OHP &$ $&$ $&$ $&$ $&$ $&$ $&$ $&$ $&$ $&$ $&$ $&$ $&$-1.54$&$0.06$\\
52923.4040 & OHP &$ $&$-62.9$&$-63.3$&$-79.9$&$ $&$ $&$ $&$ $&$1.16$&$1.73$&$ $&$ $&$ $&$ $\\
53289.4940 & OHP &$ $&$-65.6$&$-59.6$&$-80.3$&$ $&$ $&$ $&$ $&$1.24$&$1.78$&$ $&$ $&$ $&$ $\\
53480.6160 & OHP &$ $&$ $&$ $&$ $&$ $&$ $&$ $&$ $&$ $&$ $&$ $&$ $&$-1.37$&$0.07$\\
53548.5720 & OHP &$ $&$-61.8$&$-60.9$&$-77.0$&$ $&$ $&$ $&$ $&$1.26$&$1.89$&$ $&$ $&$ $&$ $\\
\hline
  \end{tabular}
\end{table*}
\addtocounter{table}{-1}
\begin{table*}
  \caption{Continued}
  \tiny
  \begin{tabular}{l c | r r r r r r | r r r r r r r r}
    \hline
    HJD-2400000 & Obs. & & \multicolumn{4}{c}{RVs (km\,s$^{-1}$)} & & & \multicolumn{6}{c}{EWs (\AA)} & \\
    & & He\,{\sc ii} & He\,{\sc ii} & He\,{\sc ii} & N\,{\sc iii} & O\,{\sc iii} &  C\,{\sc iii} & He\,{\sc ii} & H$\gamma$ & He\,{\sc i} & H$\beta$ & C\,{\sc iii} & He\,{\sc i} & H$\alpha$ & He\,{\sc i} \\
    & & $\lambda$\,4200 & $\lambda$\,4542 & $\lambda$\,4686 & & $\lambda$\,5592 &  $\lambda$\,5696 & $\lambda$\,4200 & & $\lambda$\,4471 & & $\lambda$\,5696 & $\lambda$\,5876 & & $\lambda$\,6678\\
    \hline
    53549.5850 & OHP &$ $&$ $&$ $&$ $&$ $&$ $&$ $&$ $&$ $&$ $&$ $&$ $&$-0.83$&$0.21$\\
53650.6420 & OHP &$ $&$-62.6$&$-57.2$&$-84.0$&$ $&$ $&$ $&$ $&$1.27$&$1.96$&$ $&$ $&$ $&$ $\\
53654.5800 & OHP &$ $&$ $&$ $&$ $&$ $&$ $&$ $&$ $&$ $&$ $&$ $&$ $&$-0.95$&$0.24$\\
53775.2820 & OHP &$ $&$ $&$ $&$ $&$-67.9$&$-55.6$&$ $&$ $&$ $&$ $&$-0.73$&$2.06$&$ $&$ $\\
53776.2560 & OHP &$ $&$ $&$ $&$ $&$-73.2$&$-55.4$&$ $&$ $&$ $&$ $&$-0.56$&$1.91$&$ $&$ $\\
53778.2630 & OHP &$ $&$ $&$ $&$ $&$-71.0$&$-56.7$&$ $&$ $&$ $&$ $&$-0.66$&$2.13$&$ $&$ $\\
54389.3850 & TBL &$-72.6$&$-73.1$&$-66.1$&$ $&$-80.5$&$-66.1$&$ $&$1.77$&$1.33$&$1.82$&$-0.63$&$2.01$&$-0.88$&$0.14$\\
54389.4290 & TBL &$-73.3$&$-73.8$&$-67.2$&$ $&$-81.2$&$-66.5$&$ $&$1.73$&$1.34$&$1.84$&$-0.64$&$1.99$&$-0.99$&$0.14$\\
54389.4731 & TBL &$-73.6$&$-73.7$&$-67.3$&$ $&$-80.4$&$-67.1$&$ $&$1.80$&$1.34$&$1.76$&$-0.62$&$2.02$&$-1.02$&$0.15$\\
54390.4814 & TBL &$-73.5$&$-73.7$&$-66.9$&$ $&$-80.1$&$-66.0$&$ $&$1.77$&$1.36$&$1.79$&$-0.65$&$1.92$&$-1.01$&$0.21$\\
54390.5254 & TBL &$-72.9$&$-73.6$&$-67.0$&$ $&$-79.2$&$-66.3$&$ $&$1.80$&$1.35$&$1.78$&$-0.65$&$1.92$&$-0.98$&$0.18$\\
54390.5695 & TBL &$-73.6$&$-73.1$&$-66.1$&$ $&$-80.0$&$-65.9$&$ $&$1.75$&$1.36$&$1.79$&$-0.65$&$1.92$&$-1.01$&$0.19$\\
54392.3271 & TBL &$-73.0$&$-73.2$&$-68.0$&$ $&$-79.8$&$-67.3$&$ $&$1.61$&$1.36$&$1.90$&$-0.61$&$2.03$&$-0.90$&$0.13$\\
54392.3711 & TBL &$-72.6$&$-73.6$&$-68.5$&$ $&$-80.0$&$-66.8$&$ $&$1.64$&$1.37$&$1.92$&$-0.62$&$2.04$&$-0.95$&$0.13$\\
54392.4152 & TBL &$-72.5$&$-73.1$&$-67.6$&$ $&$-80.0$&$-67.7$&$ $&$1.49$&$1.37$&$1.92$&$-0.61$&$1.99$&$-0.92$&$0.16$\\
54393.3589 & TBL &$-72.8$&$-73.6$&$-67.2$&$ $&$-80.8$&$-67.1$&$ $&$1.82$&$1.35$&$1.82$&$-0.65$&$2.06$&$-0.88$&$0.15$\\
54393.4030 & TBL &$-73.1$&$-74.1$&$-67.2$&$ $&$-81.0$&$-66.9$&$ $&$1.81$&$1.34$&$1.82$&$-0.66$&$2.09$&$-0.92$&$0.15$\\
54393.4470 & TBL &$-73.2$&$-73.3$&$-67.3$&$ $&$-80.5$&$-66.9$&$ $&$1.81$&$1.34$&$1.80$&$-0.64$&$2.08$&$-0.87$&$0.15$\\
54394.3417 & TBL &$-73.3$&$-72.8$&$-66.5$&$ $&$-80.2$&$-66.1$&$ $&$1.80$&$1.35$&$1.84$&$-0.65$&$2.10$&$-0.83$&$0.15$\\
54394.3858 & TBL &$-73.3$&$-72.9$&$-66.4$&$ $&$-80.3$&$-66.3$&$ $&$1.82$&$1.35$&$1.84$&$-0.63$&$2.10$&$-0.83$&$0.14$\\
54394.4299 & TBL &$-73.0$&$-73.4$&$-66.5$&$ $&$-80.4$&$-66.0$&$ $&$1.85$&$1.34$&$1.82$&$-0.64$&$2.10$&$-0.80$&$0.15$\\
54395.3698 & TBL &$-73.5$&$-73.6$&$-67.4$&$ $&$-80.8$&$-67.2$&$ $&$ $&$ $&$ $&$-0.64$&$2.10$&$ $&$0.15$\\
54395.4138 & TBL &$-73.4$&$-73.8$&$-67.0$&$ $&$-80.4$&$-66.9$&$ $&$1.70$&$1.36$&$1.87$&$-0.64$&$2.06$&$-0.95$&$0.15$\\
54395.4579 & TBL &$-73.4$&$-74.2$&$ $&$ $&$-80.7$&$-67.1$&$ $&$ $&$ $&$ $&$ $&$ $&$ $&$ $\\
54397.3470 & TBL &$-72.0$&$-73.2$&$-67.9$&$ $&$-80.2$&$-67.2$&$ $&$1.79$&$1.35$&$1.87$&$-0.65$&$2.01$&$-0.84$&$0.21$\\
54397.3911 & TBL &$-72.6$&$-73.4$&$-68.7$&$ $&$-79.8$&$-66.8$&$ $&$1.83$&$1.37$&$1.88$&$-0.64$&$2.04$&$-0.80$&$0.22$\\
54397.4352 & TBL &$-72.4$&$-72.9$&$-68.1$&$ $&$-80.0$&$-66.7$&$ $&$1.86$&$1.35$&$1.87$&$-0.65$&$2.03$&$-0.81$&$0.25$\\
54398.3486 & TBL &$-72.4$&$-73.2$&$-67.7$&$ $&$-80.0$&$-66.8$&$ $&$1.82$&$1.35$&$1.84$&$-0.62$&$2.05$&$-0.85$&$0.19$\\
54398.3927 & TBL &$-73.1$&$-73.3$&$-67.3$&$ $&$-80.1$&$-66.7$&$ $&$1.83$&$1.35$&$1.83$&$-0.62$&$2.03$&$-0.88$&$0.18$\\
54398.4368 & TBL &$-73.2$&$-73.0$&$-67.1$&$ $&$-79.9$&$-67.1$&$ $&$1.79$&$1.35$&$1.78$&$-0.63$&$2.03$&$-0.91$&$0.17$\\
54416.3270 & OHP &$ $&$-70.0$&$-67.7$&$-86.6$&$ $&$ $&$ $&$ $&$1.28$&$1.85$&$ $&$ $&$ $&$ $\\
54416.3970 & OHP &$ $&$-72.7$&$-68.2$&$-89.6$&$ $&$ $&$ $&$ $&$1.27$&$1.85$&$ $&$ $&$ $&$ $\\
54416.4760 & OHP &$ $&$-70.9$&$-66.5$&$-87.0$&$ $&$ $&$ $&$ $&$1.27$&$1.86$&$ $&$ $&$ $&$ $\\
54416.6270 & OHP &$ $&$-70.8$&$-68.1$&$-85.2$&$ $&$ $&$ $&$ $&$1.33$&$1.84$&$ $&$ $&$ $&$ $\\
54417.3600 & OHP &$ $&$-67.2$&$-62.5$&$-81.2$&$ $&$ $&$ $&$ $&$1.32$&$1.89$&$ $&$ $&$ $&$ $\\
54418.3080 & OHP &$ $&$-72.2$&$-66.7$&$-89.3$&$ $&$ $&$ $&$ $&$1.27$&$1.89$&$ $&$ $&$ $&$ $\\
54419.3800 & OHP &$ $&$-72.1$&$-67.7$&$-90.1$&$ $&$ $&$ $&$ $&$1.27$&$1.91$&$ $&$ $&$ $&$ $\\
54472.2798 & OHP &$ $&$ $&$ $&$ $&$-80.1$&$-66.5$&$ $&$ $&$ $&$ $&$-0.66$&$2.00$&$ $&$ $\\
54473.2353 & OHP &$ $&$ $&$ $&$ $&$ $&$ $&$ $&$ $&$ $&$ $&$ $&$ $&$-1.07$&$0.15$\\
54473.2513 & OHP &$ $&$ $&$ $&$ $&$ $&$ $&$ $&$ $&$ $&$ $&$ $&$ $&$-1.19$&$0.16$\\
54474.2838 & OHP &$ $&$ $&$ $&$ $&$-81.7$&$-68.7$&$ $&$ $&$ $&$ $&$-0.62$&$2.05$&$ $&$ $\\
54475.2470 & OHP &$ $&$ $&$ $&$ $&$-83.7$&$-68.0$&$ $&$ $&$ $&$ $&$-0.66$&$1.96$&$ $&$ $\\
54475.2679 & OHP &$ $&$ $&$ $&$ $&$-82.3$&$-67.6$&$ $&$ $&$ $&$ $&$-0.65$&$1.94$&$ $&$ $\\
54711.6371 & OHP &$ $&$-73.1$&$-69.6$&$-89.7$&$ $&$ $&$ $&$ $&$1.31$&$1.74$&$ $&$ $&$ $&$ $\\
54717.6173 & OHP &$ $&$-72.6$&$-69.8$&$-88.9$&$ $&$ $&$ $&$ $&$1.27$&$1.82$&$ $&$ $&$ $&$ $\\
54718.6083 & OHP &$ $&$-75.0$&$-73.1$&$-94.4$&$ $&$ $&$ $&$ $&$1.27$&$1.64$&$ $&$ $&$ $&$ $\\
54718.6228 & OHP &$ $&$-77.0$&$-74.0$&$-90.8$&$ $&$ $&$ $&$ $&$1.25$&$1.71$&$ $&$ $&$ $&$ $\\
54740.5681 & OHP &$ $&$-70.1$&$-65.8$&$-90.3$&$ $&$ $&$ $&$ $&$1.30$&$1.78$&$ $&$ $&$ $&$ $\\
54754.3051 & TBL &$-73.9$&$-74.5$&$-69.5$&$ $&$-80.7$&$-68.0$&$ $&$1.72$&$1.32$&$1.73$&$-0.66$&$1.88$&$-1.07$&$0.09$\\
54754.3492 & TBL &$-74.7$&$-74.7$&$-68.7$&$ $&$-81.3$&$-67.6$&$ $&$1.76$&$1.32$&$1.70$&$-0.66$&$1.88$&$-1.03$&$0.09$\\
54754.3932 & TBL &$-74.5$&$-74.5$&$-69.0$&$ $&$-81.8$&$-67.8$&$ $&$1.78$&$1.32$&$1.70$&$-0.67$&$1.89$&$-1.06$&$0.09$\\
54754.4372 & TBL &$-73.6$&$-73.9$&$-70.3$&$ $&$-80.9$&$-67.5$&$ $&$1.73$&$1.32$&$1.67$&$-0.67$&$1.88$&$-1.11$&$0.10$\\
54754.4813 & TBL &$-73.4$&$-73.9$&$-69.4$&$ $&$-81.2$&$-67.7$&$ $&$1.74$&$1.32$&$1.68$&$-0.67$&$1.88$&$-1.16$&$0.08$\\
54755.8528 & CFHT &$-75.1$&$-75.1$&$-69.1$&$ $&$-82.0$&$-69.0$&$0.72$&$1.76$&$1.30$&$1.72$&$-0.64$&$1.92$&$-1.37$&$0.09$\\
54755.8850 & CFHT &$-74.6$&$-74.6$&$-69.7$&$ $&$-82.5$&$-68.8$&$0.69$&$1.78$&$1.30$&$1.71$&$-0.65$&$1.91$&$-1.36$&$0.12$\\
54763.2869 & TBL &$-74.4$&$-74.9$&$-68.9$&$ $&$-82.3$&$-67.9$&$ $&$1.70$&$1.31$&$1.75$&$-0.64$&$1.91$&$-1.14$&$0.10$\\
54763.3309 & TBL &$-75.3$&$-75.8$&$-69.6$&$ $&$-82.1$&$-67.9$&$ $&$1.72$&$1.31$&$1.71$&$-0.63$&$1.87$&$-1.16$&$0.09$\\
54763.3750 & TBL &$-74.5$&$-75.4$&$-69.1$&$ $&$-82.3$&$-68.1$&$ $&$1.73$&$1.31$&$1.75$&$-0.66$&$1.89$&$-1.19$&$0.10$\\
54763.4191 & TBL &$-74.8$&$-75.2$&$-68.9$&$ $&$-82.7$&$-68.2$&$ $&$1.68$&$1.32$&$1.69$&$-0.64$&$1.86$&$-1.21$&$0.12$\\
54763.4631 & TBL &$-74.4$&$-75.5$&$-69.0$&$ $&$-81.6$&$-67.9$&$ $&$1.67$&$1.33$&$1.78$&$-0.63$&$1.85$&$-1.19$&$0.12$\\
54764.2673 & TBL &$-74.7$&$-74.2$&$-69.3$&$ $&$-82.2$&$-68.1$&$ $&$1.76$&$1.34$&$1.74$&$-0.67$&$1.89$&$-1.12$&$0.11$\\
54764.3113 & TBL &$-73.8$&$-74.7$&$-68.5$&$ $&$-81.9$&$-67.9$&$ $&$1.76$&$1.34$&$1.73$&$-0.67$&$1.88$&$-1.13$&$0.11$\\
54764.3554 & TBL &$-74.2$&$-74.4$&$-68.5$&$ $&$-82.1$&$-67.9$&$ $&$1.82$&$1.33$&$1.72$&$-0.67$&$1.90$&$-1.13$&$0.11$\\
54764.3994 & TBL &$-74.7$&$-75.1$&$-69.1$&$ $&$-82.2$&$-67.9$&$ $&$1.80$&$1.34$&$1.73$&$-0.67$&$1.91$&$-1.15$&$0.10$\\
54764.4435 & TBL &$-75.1$&$-75.0$&$-69.0$&$ $&$-82.3$&$-67.7$&$ $&$1.81$&$1.33$&$1.72$&$-0.68$&$1.91$&$-1.13$&$0.10$\\
54765.2791 & TBL &$-74.6$&$-75.3$&$-68.4$&$ $&$-82.7$&$-68.4$&$ $&$1.75$&$1.35$&$1.76$&$-0.65$&$2.01$&$-1.14$&$0.08$\\
54765.3231 & TBL &$-74.2$&$-74.8$&$-69.4$&$ $&$-82.5$&$-68.4$&$ $&$1.76$&$1.34$&$1.76$&$-0.65$&$2.01$&$-1.10$&$0.06$\\
54765.3672 & TBL &$-74.5$&$-74.2$&$-69.5$&$ $&$-81.0$&$-68.4$&$ $&$1.70$&$1.35$&$1.81$&$-0.64$&$1.99$&$-1.10$&$0.04$\\
54765.4113 & TBL &$-74.2$&$-74.4$&$-68.6$&$ $&$-81.8$&$-68.1$&$ $&$1.61$&$1.34$&$1.82$&$-0.64$&$1.98$&$-1.02$&$0.06$\\
54765.4553 & TBL &$-74.7$&$-75.6$&$-69.4$&$ $&$-81.6$&$-68.7$&$ $&$1.62$&$1.34$&$1.83$&$-0.65$&$1.98$&$-1.15$&$0.05$\\
54766.2661 & TBL &$-74.6$&$-74.9$&$-69.1$&$ $&$-81.7$&$-68.9$&$ $&$1.62$&$1.31$&$1.76$&$-0.64$&$1.94$&$-1.14$&$0.08$\\
54766.3101 & TBL &$-74.4$&$-75.0$&$-69.8$&$ $&$-82.0$&$-68.8$&$ $&$1.61$&$1.29$&$1.78$&$-0.65$&$1.90$&$-1.23$&$0.03$\\
54766.3542 & TBL &$-74.5$&$-74.9$&$-68.9$&$ $&$-81.9$&$-68.5$&$ $&$1.63$&$1.30$&$1.79$&$-0.65$&$1.90$&$-1.20$&$0.02$\\
54766.3983 & TBL &$-73.8$&$-74.4$&$-68.3$&$ $&$-81.5$&$-68.5$&$ $&$1.70$&$1.29$&$1.74$&$-0.64$&$1.92$&$-1.16$&$0.04$\\
54766.4423 & TBL &$-74.1$&$-74.4$&$-69.1$&$ $&$-81.9$&$-68.9$&$ $&$1.70$&$1.30$&$1.75$&$-0.65$&$1.92$&$-1.17$&$0.06$\\
54989.6730 & OHP &$ $&$-77.6$&$-72.3$&$ $&$-91.4$&$-74.9$&$ $&$ $&$1.39$&$1.71$&$-0.66$&$1.91$&$-1.29$&$ $\\
55018.0422 & CFHT &$-74.2$&$-73.3$&$ $&$ $&$-81.9$&$-67.1$&$0.70$&$1.72$&$1.27$&$1.73$&$-0.62$&$1.82$&$-1.33$&$0.07$\\
55018.1049 & CFHT &$-73.4$&$-72.9$&$ $&$ $&$-81.8$&$-67.1$&$0.71$&$1.72$&$1.26$&$ $&$-0.62$&$1.83$&$-1.35$&$0.04$\\
55020.0391 & CFHT &$-73.9$&$-74.1$&$-67.6$&$ $&$-82.4$&$-67.4$&$0.67$&$1.69$&$1.24$&$ $&$-0.64$&$1.86$&$-1.16$&$0.05$\\
55020.1030 & CFHT &$-74.2$&$-73.7$&$-67.5$&$ $&$-81.3$&$-67.6$&$0.71$&$1.70$&$1.24$&$ $&$-0.63$&$1.88$&$-1.13$&$0.06$\\
55022.0417 & CFHT &$-74.0$&$-73.5$&$-66.7$&$ $&$-80.6$&$-67.3$&$0.69$&$1.75$&$1.27$&$ $&$-0.64$&$1.82$&$-1.36$&$0.10$\\
55022.1042 & CFHT &$-73.7$&$-73.2$&$-67.1$&$ $&$-81.7$&$-67.2$&$0.72$&$1.75$&$1.27$&$1.73$&$-0.64$&$1.83$&$-1.33$&$0.11$\\
55026.0366 & CFHT &$-74.8$&$-73.9$&$-67.1$&$ $&$-82.8$&$-66.9$&$0.68$&$1.72$&$1.24$&$ $&$-0.64$&$1.91$&$-1.19$&$0.03$\\
55026.1000 & CFHT &$-74.9$&$-74.1$&$-67.0$&$ $&$-82.6$&$-66.4$&$0.68$&$1.73$&$1.25$&$1.75$&$-0.65$&$1.93$&$-1.20$&$0.04$\\
55027.0498 & CFHT &$-74.5$&$-74.2$&$ $&$ $&$-82.4$&$-67.2$&$0.69$&$1.71$&$1.24$&$ $&$-0.62$&$1.85$&$-1.31$&$0.11$\\
55027.1125 & CFHT &$-74.2$&$-73.4$&$ $&$ $&$-82.1$&$-67.1$&$0.69$&$1.71$&$1.26$&$ $&$-0.62$&$1.84$&$-1.35$&$0.11$\\
55033.5365 & TBL &$-74.4$&$-73.9$&$-70.1$&$ $&$-81.6$&$-66.9$&$ $&$1.44$&$1.32$&$1.70$&$-0.66$&$1.68$&$-1.45$&$0.12$\\
55033.5961 & TBL &$-72.8$&$-73.0$&$-69.5$&$ $&$-81.4$&$-67.0$&$ $&$1.63$&$1.31$&$1.66$&$-0.66$&$1.74$&$-1.37$&$0.10$\\
55033.6332 & TBL &$-73.0$&$-73.2$&$-69.2$&$ $&$-81.0$&$-67.5$&$ $&$1.69$&$1.30$&$1.63$&$-0.66$&$1.75$&$-1.28$&$0.11$\\
55037.5291 & TBL &$-73.4$&$-73.4$&$-69.8$&$ $&$-81.2$&$-66.9$&$ $&$1.69$&$1.29$&$1.64$&$-0.63$&$1.85$&$-1.26$&$0.10$\\
55037.5662 & TBL &$-73.8$&$-74.2$&$-69.8$&$ $&$-81.2$&$-67.0$&$ $&$1.74$&$1.29$&$1.64$&$-0.63$&$1.89$&$-1.24$&$0.10$\\
55037.6033 & TBL &$-74.2$&$-74.1$&$-70.0$&$ $&$-81.6$&$-66.8$&$ $&$1.75$&$1.29$&$1.62$&$-0.62$&$1.89$&$-1.29$&$0.10$\\
55038.5750 & TBL &$-72.8$&$-74.3$&$-71.8$&$ $&$-82.3$&$-67.3$&$ $&$ $&$1.30$&$1.77$&$-0.67$&$1.76$&$-1.29$&$0.07$\\
55038.6098 & TBL &$-74.0$&$-74.2$&$-68.2$&$ $&$-81.1$&$-67.6$&$ $&$ $&$1.30$&$1.71$&$-0.66$&$1.80$&$-1.29$&$0.07$\\
55038.6446 & TBL &$-73.2$&$-75.1$&$-67.9$&$ $&$-82.2$&$-67.8$&$ $&$ $&$1.28$&$1.66$&$-0.65$&$1.78$&$-1.30$&$0.06$\\
55039.5371 & TBL &$-74.6$&$-74.3$&$-66.8$&$ $&$-81.8$&$-66.8$&$ $&$ $&$1.25$&$1.85$&$-0.68$&$1.68$&$-1.35$&$-0.01$\\
55039.5768 & TBL &$-74.2$&$-73.7$&$-70.3$&$ $&$-81.3$&$-67.1$&$ $&$ $&$1.24$&$1.71$&$-0.67$&$1.71$&$-1.39$&$-0.01$\\
55039.6165 & TBL &$-74.0$&$-74.4$&$-66.9$&$ $&$-81.2$&$-67.2$&$ $&$ $&$1.24$&$1.65$&$-0.65$&$1.75$&$-1.41$&$-0.01$\\
55039.6562 & TBL &$-73.5$&$-74.0$&$-66.9$&$ $&$-81.9$&$-67.0$&$ $&$ $&$1.20$&$1.62$&$-0.65$&$1.75$&$-1.40$&$-0.03$\\
55040.5356 & TBL &$-73.7$&$-74.8$&$-68.5$&$ $&$-82.6$&$-66.8$&$ $&$ $&$1.22$&$1.66$&$-0.65$&$1.64$&$-1.48$&$0.01$\\
55040.5797 & TBL &$-73.9$&$-74.3$&$-67.7$&$ $&$-82.6$&$-66.7$&$ $&$1.45$&$1.21$&$1.64$&$-0.63$&$1.66$&$-1.41$&$-0.01$\\
55040.6237 & TBL &$-74.0$&$-74.7$&$-68.5$&$ $&$-82.3$&$-66.7$&$ $&$1.49$&$1.20$&$1.54$&$-0.64$&$1.66$&$-1.46$&$0.01$\\
55041.5277 & TBL &$-73.4$&$-73.6$&$-67.9$&$ $&$-82.1$&$-66.7$&$ $&$1.68$&$1.24$&$1.55$&$-0.65$&$1.78$&$-1.44$&$0.05$\\
55041.5718 & TBL &$-73.8$&$-73.7$&$-68.0$&$ $&$-81.5$&$-66.4$&$ $&$1.71$&$1.25$&$1.55$&$-0.63$&$1.77$&$-1.44$&$0.06$\\
55041.6158 & TBL &$-73.5$&$-73.7$&$-69.0$&$ $&$-81.6$&$-66.3$&$ $&$1.70$&$1.26$&$1.58$&$-0.64$&$1.75$&$-1.39$&$0.04$\\
55042.5392 & TBL &$-73.7$&$-73.9$&$-68.9$&$ $&$-81.6$&$-67.0$&$ $&$1.77$&$1.28$&$1.61$&$-0.64$&$1.87$&$-1.16$&$0.10$\\
\hline
  \end{tabular}
\end{table*}
\addtocounter{table}{-1}
\begin{table*}
  \caption{Continued}
  \tiny
  \begin{tabular}{l c | r r r r r r | r r r r r r r r}
    \hline
    HJD-2400000 & Obs. & & \multicolumn{4}{c}{RVs (km\,s$^{-1}$)} & & & \multicolumn{6}{c}{EWs (\AA)} & \\
    & & He\,{\sc ii} & He\,{\sc ii} & He\,{\sc ii} & N\,{\sc iii} & O\,{\sc iii} &  C\,{\sc iii} & He\,{\sc ii} & H$\gamma$ & He\,{\sc i} & H$\beta$ & C\,{\sc iii} & He\,{\sc i} & H$\alpha$ & He\,{\sc i} \\
    & & $\lambda$\,4200 & $\lambda$\,4542 & $\lambda$\,4686 & & $\lambda$\,5592 &  $\lambda$\,5696 & $\lambda$\,4200 & & $\lambda$\,4471 & & $\lambda$\,5696 & $\lambda$\,5876 & & $\lambda$\,6678\\
    \hline
55042.5832 & TBL &$-73.1$&$-73.8$&$-70.1$&$ $&$-81.2$&$-67.0$&$ $&$1.78$&$1.28$&$1.63$&$-0.64$&$1.88$&$-1.15$&$0.09$\\
55042.6273 & TBL &$-73.4$&$-74.0$&$-69.8$&$ $&$-81.3$&$-67.0$&$0.81$&$1.74$&$1.28$&$1.66$&$-0.64$&$1.88$&$-1.19$&$0.09$\\
55043.5212 & TBL &$-74.2$&$-74.1$&$-70.0$&$ $&$-82.1$&$-67.2$&$ $&$1.69$&$1.26$&$1.64$&$-0.63$&$1.81$&$-1.15$&$0.05$\\
55043.5611 & TBL &$-74.0$&$-73.6$&$-70.3$&$ $&$-82.2$&$-67.4$&$ $&$1.69$&$1.26$&$1.64$&$-0.64$&$ $&$-1.11$&$0.04$\\
55043.6011 & TBL &$-74.0$&$-74.0$&$-69.8$&$ $&$-82.1$&$-67.5$&$ $&$1.71$&$1.25$&$1.61$&$-0.63$&$1.84$&$-1.13$&$0.04$\\
55043.6411 & TBL &$-73.7$&$-74.1$&$-70.6$&$ $&$-81.4$&$-67.1$&$ $&$1.69$&$1.25$&$1.64$&$-0.64$&$1.82$&$-1.13$&$0.05$\\
55044.5389 & TBL &$-74.0$&$-74.0$&$-70.2$&$ $&$-80.9$&$-66.5$&$ $&$1.64$&$1.24$&$1.58$&$-0.67$&$1.69$&$-1.32$&$0.00$\\
55044.5813 & TBL &$-73.7$&$-74.0$&$-68.9$&$ $&$-81.5$&$-66.5$&$ $&$1.68$&$1.23$&$1.55$&$-0.66$&$1.69$&$-1.29$&$-0.01$\\
55044.6213 & TBL &$-73.7$&$-74.2$&$-68.5$&$ $&$-81.7$&$-66.5$&$ $&$1.70$&$1.23$&$1.57$&$-0.65$&$1.70$&$-1.28$&$0.00$\\
55045.5349 & TBL &$-73.1$&$-74.2$&$-68.5$&$ $&$-81.9$&$-66.3$&$ $&$ $&$1.26$&$1.58$&$-0.64$&$1.70$&$-1.58$&$-0.02$\\
55045.5790 & TBL &$-74.1$&$-74.6$&$-69.0$&$ $&$-82.3$&$-66.3$&$ $&$ $&$1.24$&$1.62$&$-0.65$&$1.69$&$-1.43$&$-0.01$\\
55045.6232 & TBL &$-73.7$&$-74.0$&$-69.0$&$ $&$-82.5$&$-66.3$&$ $&$ $&$1.25$&$1.67$&$-0.65$&$1.68$&$-1.49$&$-0.02$\\
55047.5158 & TBL &$-73.3$&$-74.0$&$-67.8$&$ $&$-81.7$&$-67.0$&$ $&$ $&$1.28$&$1.61$&$-0.66$&$1.67$&$-1.45$&$0.11$\\
55047.5588 & TBL &$-73.4$&$-73.7$&$-69.5$&$ $&$-81.1$&$-67.0$&$ $&$ $&$1.28$&$1.60$&$-0.66$&$1.68$&$-1.49$&$0.11$\\
55047.6029 & TBL &$-73.4$&$-73.9$&$-67.3$&$ $&$-81.3$&$-66.6$&$ $&$ $&$1.28$&$1.62$&$-0.66$&$1.68$&$-1.53$&$0.11$\\
55047.6469 & TBL &$-73.4$&$-74.3$&$-67.9$&$ $&$-81.5$&$-67.1$&$ $&$ $&$1.29$&$1.64$&$-0.66$&$1.82$&$-1.25$&$0.06$\\
55048.5598 & TBL &$-73.7$&$-74.2$&$-67.4$&$ $&$-82.1$&$-66.7$&$ $&$1.71$&$1.29$&$1.62$&$-0.67$&$1.81$&$-1.20$&$0.07$\\
55048.6039 & TBL &$-73.0$&$-74.2$&$-68.5$&$ $&$-82.1$&$-66.7$&$ $&$1.70$&$1.29$&$1.66$&$-0.67$&$1.83$&$-1.17$&$0.06$\\
55078.0702 & CFHT &$-73.5$&$-73.1$&$ $&$ $&$-81.7$&$-67.3$&$0.70$&$1.71$&$1.23$&$1.64$&$-0.63$&$1.80$&$-1.51$&$0.06$\\
55082.9461 & CFHT &$-73.9$&$-73.7$&$ $&$ $&$-81.6$&$-67.3$&$0.74$&$1.82$&$1.28$&$1.78$&$-0.62$&$1.90$&$-1.23$&$0.13$\\
55083.0085 & CFHT &$-73.6$&$-73.7$&$ $&$ $&$-81.9$&$-67.4$&$0.72$&$1.80$&$1.29$&$1.80$&$-0.62$&$1.91$&$-1.26$&$0.13$\\
55099.8774 & CFHT &$-73.7$&$-73.7$&$-66.5$&$ $&$-81.1$&$-66.9$&$0.72$&$1.78$&$1.27$&$1.74$&$-0.65$&$1.87$&$-1.20$&$0.08$\\
55099.9415 & CFHT &$-73.4$&$-73.2$&$-67.3$&$ $&$-80.8$&$-67.0$&$0.71$&$1.77$&$1.27$&$ $&$-0.65$&$1.86$&$-1.18$&$0.09$\\
55105.8579 & CFHT &$-73.8$&$-73.0$&$ $&$ $&$-80.9$&$-67.0$&$0.72$&$ $&$ $&$1.61$&$-0.62$&$1.78$&$-1.62$&$0.09$\\
55105.9207 & CFHT &$-73.5$&$-73.4$&$ $&$ $&$-80.9$&$-67.2$&$0.71$&$ $&$ $&$ $&$-0.62$&$1.76$&$-1.62$&$0.09$\\
55109.9580 & CFHT &$-73.0$&$-72.7$&$ $&$ $&$-80.0$&$-66.7$&$0.72$&$1.70$&$1.25$&$1.67$&$-0.68$&$1.67$&$-1.45$&$0.05$\\
55110.0211 & CFHT &$-73.0$&$-73.2$&$ $&$ $&$-80.6$&$-66.7$&$0.71$&$1.69$&$1.25$&$1.68$&$-0.69$&$1.62$&$-1.41$&$0.08$\\
55114.8689 & CFHT &$-73.4$&$-73.1$&$ $&$ $&$-81.3$&$-67.0$&$0.74$&$ $&$ $&$ $&$-0.66$&$1.72$&$-1.44$&$0.03$\\
55114.9320 & CFHT &$-73.5$&$-73.0$&$ $&$ $&$-81.1$&$-67.2$&$0.71$&$ $&$1.25$&$ $&$-0.66$&$1.73$&$-1.44$&$0.04$\\
55175.2201 & OHP &$ $&$-71.9$&$ $&$-89.5$&$ $&$ $&$ $&$ $&$1.20$&$ $&$ $&$ $&$ $&$ $\\
55175.2410 & OHP &$ $&$-70.9$&$ $&$-87.5$&$ $&$ $&$ $&$ $&$1.19$&$ $&$ $&$ $&$ $&$ $\\
55177.4167 & OHP &$ $&$-72.0$&$ $&$-91.3$&$ $&$ $&$ $&$ $&$1.19$&$ $&$ $&$ $&$ $&$ $\\
55355.5903 & OHP &$ $&$-69.8$&$-66.1$&$-90.8$&$ $&$ $&$ $&$ $&$1.17$&$1.58$&$ $&$ $&$ $&$ $\\
55361.5958 & OHP &$ $&$-70.8$&$-66.6$&$-86.8$&$ $&$ $&$ $&$ $&$1.14$&$1.42$&$ $&$ $&$ $&$ $\\
55370.5972 & OHP &$ $&$-66.4$&$ $&$-87.0$&$ $&$ $&$ $&$ $&$1.13$&$ $&$ $&$ $&$ $&$ $\\
55373.6042 & OHP &$ $&$-66.6$&$ $&$-86.1$&$ $&$ $&$ $&$ $&$1.12$&$ $&$ $&$ $&$ $&$ $\\
55540.4047 & OHP &$ $&$-67.7$&$ $&$-85.2$&$ $&$ $&$ $&$ $&$1.05$&$ $&$ $&$ $&$ $&$ $\\
55541.4268 & OHP &$ $&$-68.6$&$ $&$-85.8$&$ $&$ $&$ $&$ $&$1.10$&$ $&$ $&$ $&$ $&$ $\\
55825.5705 & OHP &$ $&$-67.7$&$ $&$-84.9$&$ $&$ $&$ $&$ $&$1.00$&$ $&$ $&$ $&$ $&$ $\\
55825.5921 & OHP &$ $&$-68.2$&$ $&$-84.1$&$ $&$ $&$ $&$ $&$1.02$&$ $&$ $&$ $&$ $&$ $\\
55828.5075 & OHP &$ $&$-66.5$&$ $&$-82.8$&$ $&$ $&$ $&$ $&$1.02$&$ $&$ $&$ $&$ $&$ $\\
55828.5290 & OHP &$ $&$-66.6$&$ $&$-83.4$&$ $&$ $&$ $&$ $&$1.03$&$ $&$ $&$ $&$ $&$ $\\
55830.5096 & OHP &$ $&$-66.8$&$ $&$-84.1$&$ $&$ $&$ $&$ $&$1.03$&$ $&$ $&$ $&$ $&$ $\\
55830.5311 & OHP &$ $&$-67.9$&$ $&$-85.0$&$ $&$ $&$ $&$ $&$1.03$&$ $&$ $&$ $&$ $&$ $\\
56091.5878 & OHP &$ $&$-66.4$&$-60.2$&$-80.8$&$ $&$ $&$ $&$ $&$0.85$&$0.91$&$ $&$ $&$ $&$ $\\
56093.5747 & OHP &$ $&$-62.3$&$-59.6$&$-79.8$&$ $&$ $&$ $&$ $&$0.90$&$1.03$&$ $&$ $&$ $&$ $\\
56093.5893 & OHP &$ $&$-64.9$&$-61.0$&$-80.1$&$ $&$ $&$ $&$ $&$0.91$&$1.01$&$ $&$ $&$ $&$ $\\
56094.5664 & OHP &$ $&$-66.4$&$-61.7$&$-81.0$&$ $&$ $&$ $&$ $&$0.88$&$0.95$&$ $&$ $&$ $&$ $\\
56094.5851 & OHP &$ $&$-65.2$&$-61.5$&$-83.7$&$ $&$ $&$ $&$ $&$0.86$&$0.95$&$ $&$ $&$ $&$ $\\
56095.5664 & OHP &$ $&$-64.3$&$-62.4$&$-80.8$&$ $&$ $&$ $&$ $&$0.87$&$0.97$&$ $&$ $&$ $&$ $\\
56095.5880 & OHP &$ $&$-64.0$&$-62.8$&$-80.8$&$ $&$ $&$ $&$ $&$0.88$&$0.96$&$ $&$ $&$ $&$ $\\
56096.5706 & OHP &$ $&$-63.1$&$-61.3$&$-78.5$&$ $&$ $&$ $&$ $&$0.92$&$0.91$&$ $&$ $&$ $&$ $\\
56096.5880 & OHP &$ $&$-63.4$&$-61.2$&$-78.3$&$ $&$ $&$ $&$ $&$0.88$&$0.98$&$ $&$ $&$ $&$ $\\
56456.5760 & OHP &$ $&$-60.9$&$-58.4$&$-82.7$&$ $&$ $&$ $&$ $&$0.85$&$0.87$&$ $&$ $&$ $&$ $\\
56456.5864 & OHP &$ $&$-61.7$&$-56.5$&$-73.3$&$ $&$ $&$ $&$ $&$0.84$&$0.84$&$ $&$ $&$ $&$ $\\
56457.5489 & OHP &$ $&$-61.1$&$-57.6$&$-78.6$&$ $&$ $&$ $&$ $&$0.81$&$0.77$&$ $&$ $&$ $&$ $\\
56457.5697 & OHP &$ $&$-63.0$&$-56.5$&$-80.9$&$ $&$ $&$ $&$ $&$0.77$&$0.76$&$ $&$ $&$ $&$ $\\
56458.5601 & OHP &$ $&$-60.5$&$-56.8$&$-80.2$&$ $&$ $&$ $&$ $&$0.79$&$0.77$&$ $&$ $&$ $&$ $\\
56458.5809 & OHP &$ $&$-59.4$&$-56.0$&$-77.3$&$ $&$ $&$ $&$ $&$0.81$&$0.76$&$ $&$ $&$ $&$ $\\
56459.5573 & OHP &$ $&$-59.9$&$-55.5$&$-82.0$&$ $&$ $&$ $&$ $&$0.85$&$0.82$&$ $&$ $&$ $&$ $\\
56459.5790 & OHP &$ $&$-60.3$&$-55.1$&$-77.5$&$ $&$ $&$ $&$ $&$0.86$&$0.80$&$ $&$ $&$ $&$ $\\
56460.5574 & OHP &$ $&$-59.0$&$-56.3$&$-75.8$&$ $&$ $&$ $&$ $&$0.86$&$0.93$&$ $&$ $&$ $&$ $\\
56460.5790 & OHP &$ $&$-59.2$&$-55.0$&$-77.4$&$ $&$ $&$ $&$ $&$0.86$&$0.88$&$ $&$ $&$ $&$ $\\
56514.8382 & TIGRE &$-66.8$&$-61.1$&$-62.9$&$-80.5$&$-80.0$&$-50.9$&$0.78$&$1.54$&$0.76$&$0.88$&$ $&$0.85$&$-4.11$&$-0.30$\\
56519.7645 & TIGRE &$-61.4$&$-67.3$&$-61.8$&$-87.7$&$-78.5$&$-60.3$&$0.76$&$1.50$&$0.78$&$0.62$&$ $&$0.74$&$-3.92$&$-0.34$\\
56533.7688 & TIGRE &$-62.9$&$-65.2$&$-60.9$&$-80.4$&$-78.1$&$-47.0$&$0.69$&$1.20$&$0.81$&$0.72$&$ $&$0.81$&$-3.91$&$-0.37$\\
56561.7402 & TIGRE &$-61.9$&$-66.6$&$-62.8$&$-87.2$&$-65.0$&$-59.3$&$0.58$&$1.02$&$0.75$&$0.44$&$ $&$0.74$&$-3.62$&$-0.64$\\
56568.7055 & TIGRE &$-62.2$&$-66.9$&$-59.0$&$-86.1$&$-69.5$&$-61.6$&$0.54$&$1.14$&$0.87$&$0.65$&$ $&$0.81$&$-3.78$&$-0.53$\\
56580.6415 & TIGRE &$-64.9$&$-65.6$&$-61.3$&$-81.7$&$-69.7$&$-63.6$&$0.61$&$0.89$&$0.76$&$0.48$&$ $&$0.66$&$-4.06$&$-0.67$\\
56581.7194 & TIGRE &$-62.1$&$-63.2$&$-59.4$&$-79.3$&$-78.9$&$-55.6$&$0.53$&$1.06$&$0.79$&$0.64$&$ $&$0.79$&$-4.00$&$-0.44$\\
56584.7444 & TIGRE &$-60.7$&$-62.9$&$-62.3$&$-83.1$&$-67.1$&$-58.3$&$0.62$&$0.75$&$0.67$&$0.35$&$ $&$0.61$&$-4.10$&$-0.50$\\
56590.7162 & TIGRE &$-67.5$&$-63.9$&$-62.4$&$-86.5$&$ $&$-55.4$&$0.64$&$1.07$&$0.80$&$0.60$&$ $&$0.78$&$-3.69$&$-0.55$\\
56591.6544 & TIGRE &$-76.2$&$-73.8$&$-60.3$&$-85.6$&$ $&$ $&$0.52$&$0.89$&$0.70$&$1.00$&$ $&$0.88$&$-3.52$&$-0.57$\\
56592.6282 & TIGRE &$-68.0$&$-66.2$&$-62.5$&$-86.7$&$-74.9$&$-60.3$&$0.75$&$1.05$&$0.63$&$0.20$&$ $&$0.37$&$-4.64$&$-0.85$\\
56592.6378 & TIGRE &$-67.8$&$-63.9$&$-65.0$&$-77.9$&$-75.8$&$ $&$0.47$&$0.99$&$0.82$&$0.71$&$ $&$0.65$&$-3.47$&$-0.56$\\
56592.7294 & TIGRE &$-67.4$&$-67.1$&$-62.5$&$-91.2$&$-83.7$&$-53.2$&$0.79$&$1.20$&$0.65$&$0.20$&$ $&$0.43$&$-4.59$&$-0.85$\\
56604.6883 & TIGRE &$-67.2$&$-69.9$&$-62.3$&$-81.6$&$-69.7$&$-61.3$&$0.52$&$0.84$&$0.68$&$0.72$&$ $&$0.73$&$-3.78$&$-0.48$\\
56640.5975 & TIGRE &$-69.9$&$-69.2$&$-62.3$&$-87.4$&$-77.9$&$-61.4$&$0.52$&$1.02$&$0.73$&$0.49$&$ $&$0.71$&$-3.93$&$-0.56$\\
56643.5737 & TIGRE &$-69.6$&$-68.7$&$-64.3$&$-86.8$&$-77.7$&$-63.6$&$0.81$&$1.08$&$0.76$&$0.52$&$ $&$0.67$&$-4.03$&$-0.68$\\
56644.5574 & TIGRE &$-73.1$&$-69.4$&$-67.1$&$-89.2$&$-74.8$&$-63.8$&$0.84$&$1.13$&$0.75$&$0.41$&$ $&$0.73$&$-3.97$&$-0.61$\\
56645.5550 & TIGRE &$-64.1$&$-65.4$&$-60.5$&$-86.7$&$-77.1$&$-54.9$&$0.57$&$0.93$&$0.72$&$0.29$&$ $&$0.66$&$-4.07$&$-0.61$\\
56646.5620 & TIGRE &$-67.1$&$-65.9$&$-59.5$&$-86.2$&$-75.3$&$-58.0$&$0.58$&$0.96$&$0.74$&$0.39$&$ $&$0.63$&$-4.07$&$-0.55$\\
56813.9293 & TIGRE &$-65.8$&$-62.1$&$-60.9$&$-82.4$&$ $&$ $&$0.87$&$1.40$&$0.67$&$0.30$&$ $&$0.60$&$-4.34$&$-0.63$\\
56818.9307 & TIGRE &$ $&$ $&$ $&$ $&$ $&$ $&$ $&$ $&$ $&$ $&$ $&$0.49$&$-4.10$&$-0.54$\\
56819.9135 & TIGRE &$-65.3$&$-67.2$&$-66.8$&$-87.3$&$ $&$ $&$0.68$&$1.37$&$0.75$&$0.64$&$ $&$0.61$&$-3.72$&$-0.53$\\
56860.9128 & TIGRE &$-65.4$&$-68.8$&$-59.9$&$-84.4$&$ $&$ $&$0.71$&$1.36$&$0.78$&$0.41$&$ $&$0.80$&$-4.44$&$-0.73$\\
56861.8841 & TIGRE &$-72.6$&$-67.4$&$-59.9$&$-77.8$&$ $&$ $&$0.83$&$1.26$&$0.71$&$0.46$&$ $&$0.36$&$-4.76$&$-0.63$\\
56863.9736 & TIGRE &$-66.5$&$-67.1$&$-58.6$&$-82.3$&$-78.0$&$ $&$0.64$&$1.40$&$0.66$&$0.33$&$ $&$0.57$&$-4.56$&$-0.45$\\
56867.8779 & TIGRE &$-67.8$&$-66.6$&$-59.8$&$-82.0$&$-80.4$&$-54.8$&$0.69$&$1.30$&$0.75$&$0.43$&$ $&$0.45$&$-4.54$&$-0.74$\\
56895.9119 & TIGRE &$-65.8$&$-64.5$&$-60.2$&$-82.2$&$-76.0$&$-56.4$&$0.72$&$1.34$&$0.77$&$0.40$&$ $&$0.54$&$-4.77$&$-0.74$\\
56897.9207 & TIGRE &$-65.3$&$-63.6$&$-62.6$&$-84.1$&$-72.5$&$-55.4$&$0.63$&$1.16$&$0.71$&$0.54$&$ $&$0.48$&$-4.57$&$-0.79$\\
56907.8048 & TIGRE &$-68.7$&$-63.3$&$-60.3$&$-83.7$&$-77.7$&$-57.0$&$0.64$&$1.38$&$0.67$&$0.43$&$ $&$0.58$&$-4.42$&$-0.75$\\
56910.8340 & TIGRE &$-67.6$&$-66.9$&$-61.6$&$-85.6$&$-75.4$&$-54.4$&$0.67$&$1.37$&$0.73$&$0.29$&$ $&$0.43$&$-4.71$&$-0.80$\\
56920.7862 & TIGRE &$-66.3$&$-66.7$&$-62.9$&$-89.9$&$-70.6$&$-57.6$&$0.71$&$1.34$&$0.73$&$0.36$&$ $&$0.38$&$-4.57$&$-0.74$\\
56939.7407 & TIGRE &$-69.0$&$-65.3$&$-62.5$&$-87.9$&$-74.1$&$-60.0$&$0.62$&$1.40$&$0.77$&$0.38$&$ $&$0.44$&$-4.87$&$-0.83$\\
56940.7051 & TIGRE &$-68.3$&$-64.9$&$-63.4$&$-83.6$&$-73.9$&$-61.7$&$0.70$&$1.27$&$0.73$&$0.54$&$ $&$0.47$&$-4.58$&$-0.83$\\
56941.6536 & TIGRE &$-68.0$&$-66.1$&$-61.8$&$-83.2$&$-71.7$&$-54.8$&$0.66$&$1.28$&$0.77$&$0.50$&$ $&$0.39$&$-4.66$&$-0.87$\\
56953.6629 & TIGRE &$-65.3$&$-68.0$&$-64.1$&$-85.3$&$-63.0$&$-53.0$&$0.70$&$1.20$&$0.60$&$0.43$&$ $&$0.49$&$-4.43$&$-0.84$\\
56954.5536 & TIGRE &$-73.6$&$-62.8$&$-62.8$&$-84.4$&$ $&$ $&$0.73$&$1.00$&$0.71$&$0.27$&$ $&$0.31$&$-4.79$&$-0.87$\\
56956.6882 & TIGRE &$-67.8$&$-65.7$&$-62.9$&$-82.8$&$-69.4$&$-66.7$&$0.77$&$1.08$&$0.68$&$0.57$&$ $&$0.49$&$-4.59$&$-0.78$\\
56957.6525 & TIGRE &$-68.0$&$-66.1$&$-62.3$&$-87.3$&$-71.5$&$-55.9$&$0.75$&$1.19$&$0.72$&$0.58$&$ $&$0.50$&$-4.71$&$-0.83$\\
56959.6429 & TIGRE &$-66.4$&$-67.5$&$-63.1$&$-89.1$&$-73.1$&$-63.7$&$0.61$&$1.18$&$0.62$&$0.35$&$ $&$0.39$&$-4.64$&$-0.92$\\
56960.6178 & TIGRE &$-66.4$&$-66.1$&$-63.2$&$-88.6$&$-73.4$&$-63.6$&$0.72$&$1.11$&$0.69$&$0.54$&$ $&$0.51$&$-4.70$&$-0.89$\\
56960.7054 & TIGRE &$-65.4$&$-65.6$&$-61.9$&$-86.6$&$-72.1$&$-58.8$&$0.71$&$1.21$&$0.68$&$0.47$&$ $&$0.51$&$-4.63$&$-0.78$\\
56961.8111 & TIGRE &$-67.4$&$-62.4$&$-61.8$&$-88.1$&$-66.1$&$-59.1$&$0.65$&$1.29$&$0.64$&$0.56$&$ $&$0.44$&$-4.60$&$-0.93$\\
56965.6312 & TIGRE &$-65.0$&$-67.2$&$-63.0$&$-81.4$&$ $&$ $&$0.66$&$1.46$&$0.73$&$0.65$&$ $&$0.52$&$-4.79$&$-0.87$\\
\hline
  \end{tabular}
\end{table*}
\addtocounter{table}{-1}
\begin{table*}
  \caption{Continued}
  \tiny
  \begin{tabular}{l c | r r r r r r | r r r r r r r r}
    \hline
    HJD-2400000 & Obs. & & \multicolumn{4}{c}{RVs (km\,s$^{-1}$)} & & & \multicolumn{6}{c}{EWs (\AA)} & \\
    & & He\,{\sc ii} & He\,{\sc ii} & He\,{\sc ii} & N\,{\sc iii} & O\,{\sc iii} &  C\,{\sc iii} & He\,{\sc ii} & H$\gamma$ & He\,{\sc i} & H$\beta$ & C\,{\sc iii} & He\,{\sc i} & H$\alpha$ & He\,{\sc i} \\
    & & $\lambda$\,4200 & $\lambda$\,4542 & $\lambda$\,4686 & & $\lambda$\,5592 &  $\lambda$\,5696 & $\lambda$\,4200 & & $\lambda$\,4471 & & $\lambda$\,5696 & $\lambda$\,5876 & & $\lambda$\,6678\\
    \hline
56971.7209 & TIGRE &$-66.2$&$-64.8$&$-60.6$&$-85.7$&$-74.6$&$-55.7$&$0.66$&$1.27$&$0.58$&$0.25$&$ $&$0.28$&$-4.85$&$-0.86$\\
56974.6493 & TIGRE &$-68.6$&$-67.8$&$-61.9$&$-94.7$&$ $&$-60.5$&$0.66$&$1.29$&$0.69$&$0.32$&$ $&$0.44$&$-4.51$&$-0.87$\\
56976.6454 & TIGRE &$-67.0$&$-66.0$&$-62.4$&$-87.9$&$-70.5$&$-61.2$&$0.63$&$1.19$&$0.64$&$0.41$&$ $&$0.39$&$-4.63$&$-0.88$\\
56978.6015 & TIGRE &$-70.7$&$-67.2$&$-64.3$&$-91.2$&$-74.8$&$-63.1$&$0.64$&$1.32$&$0.70$&$0.42$&$ $&$0.42$&$-4.74$&$-0.76$\\
56980.6470 & TIGRE &$-65.0$&$-65.4$&$-62.2$&$-89.6$&$-67.7$&$-58.6$&$0.61$&$1.21$&$0.70$&$0.49$&$ $&$0.54$&$-4.47$&$-0.87$\\
56984.6739 & TIGRE &$-67.7$&$-66.1$&$-61.6$&$-84.1$&$-73.6$&$-58.1$&$0.68$&$1.36$&$0.66$&$0.34$&$ $&$0.36$&$-4.70$&$-0.78$\\
56999.6410 & TIGRE &$-66.1$&$-63.9$&$-61.6$&$-84.8$&$-70.9$&$-60.9$&$0.60$&$1.40$&$0.70$&$0.52$&$ $&$0.55$&$-3.59$&$-0.58$\\
57007.5755 & TIGRE &$-65.2$&$-64.3$&$-59.8$&$-85.9$&$-74.8$&$-59.1$&$0.64$&$1.35$&$0.72$&$0.45$&$ $&$0.46$&$-3.71$&$-0.59$\\
57009.6294 & TIGRE &$-67.3$&$-64.6$&$-60.2$&$-85.6$&$-72.0$&$-61.7$&$0.66$&$1.39$&$0.76$&$0.56$&$ $&$0.44$&$-4.01$&$-0.62$\\
57011.5726 & TIGRE &$ $&$-65.3$&$-60.5$&$-84.7$&$-75.7$&$-53.6$&$ $&$0.95$&$0.59$&$0.29$&$ $&$0.40$&$-3.73$&$-0.34$\\
57013.6315 & TIGRE &$-67.0$&$-65.1$&$-63.8$&$-85.5$&$-73.1$&$-56.2$&$0.70$&$1.30$&$0.66$&$0.35$&$ $&$0.52$&$-3.70$&$-0.50$\\
57015.5998 & TIGRE &$-68.7$&$-65.3$&$-62.2$&$-85.6$&$-69.1$&$-61.8$&$0.65$&$1.26$&$0.71$&$0.43$&$ $&$0.48$&$-3.88$&$-0.42$\\
57231.9445 & TIGRE &$-73.4$&$-69.2$&$-66.8$&$-87.8$&$-74.7$&$-61.1$&$0.57$&$1.34$&$0.68$&$0.12$&$ $&$0.27$&$-5.46$&$-0.92$\\
57268.9059 & CFHT &$ $&$-68.5$&$-64.1$&$ $&$-76.0$&$-62.0$&$ $&$ $&$ $&$0.24$&$-0.70$&$-0.05$&$-5.11$&$-1.00$\\
57268.9681 & CFHT &$ $&$-68.6$&$-63.9$&$ $&$-76.8$&$-62.1$&$ $&$ $&$ $&$0.18$&$-0.69$&$-0.02$&$-5.17$&$-0.99$\\
57286.7480 & TIGRE &$-70.6$&$-70.7$&$-66.1$&$-89.4$&$-79.7$&$-61.6$&$0.57$&$1.18$&$0.63$&$0.04$&$ $&$0.15$&$-5.69$&$-1.03$\\
57293.8560 & TIGRE &$-70.5$&$-72.6$&$-66.7$&$-91.0$&$-78.8$&$-62.6$&$0.71$&$1.26$&$0.61$&$0.20$&$ $&$0.26$&$-5.25$&$-0.94$\\
57294.8060 & TIGRE &$-71.9$&$-71.9$&$-65.9$&$-93.4$&$-78.3$&$-68.0$&$0.55$&$1.12$&$0.61$&$0.10$&$ $&$0.13$&$-5.71$&$-1.00$\\
57297.8928 & TIGRE &$-70.0$&$-72.2$&$-66.7$&$-90.1$&$-80.3$&$-66.8$&$0.70$&$1.16$&$0.63$&$0.18$&$ $&$0.22$&$-5.57$&$-0.95$\\
57298.7138 & TIGRE &$-72.6$&$-70.0$&$-66.6$&$-90.0$&$-79.6$&$-74.3$&$0.54$&$1.16$&$0.62$&$0.20$&$ $&$0.21$&$-5.53$&$-1.01$\\
57298.8219 & TIGRE &$-71.8$&$-72.1$&$-66.3$&$-88.7$&$-79.0$&$-67.5$&$0.69$&$1.08$&$0.63$&$0.10$&$ $&$0.18$&$-5.62$&$-0.96$\\
57299.7684 & TIGRE &$-71.1$&$-70.9$&$-64.8$&$-90.6$&$-78.7$&$-68.5$&$0.56$&$1.29$&$0.61$&$0.01$&$ $&$0.10$&$-5.66$&$-1.00$\\
57300.7475 & TIGRE &$-69.5$&$-72.5$&$-66.4$&$-95.5$&$-76.5$&$-69.4$&$0.64$&$1.32$&$0.62$&$0.17$&$ $&$0.13$&$-5.33$&$-1.10$\\
57304.6782 & TIGRE &$-70.1$&$-73.4$&$-67.1$&$-93.5$&$-76.9$&$ $&$0.63$&$1.08$&$0.61$&$0.21$&$ $&$0.13$&$-5.48$&$-0.90$\\
57304.7608 & TIGRE &$-69.9$&$-73.1$&$-65.7$&$-90.1$&$-79.2$&$-65.4$&$0.71$&$1.26$&$0.63$&$0.08$&$ $&$0.13$&$-5.57$&$-0.91$\\
57307.7435 & TIGRE &$-68.8$&$-69.3$&$-66.1$&$-89.5$&$-80.1$&$-62.3$&$0.74$&$1.27$&$0.55$&$0.27$&$ $&$0.23$&$-5.52$&$-0.98$\\
57310.6171 & TIGRE &$-73.9$&$-70.0$&$-66.4$&$-91.3$&$-81.6$&$-63.9$&$0.64$&$1.37$&$0.71$&$0.23$&$ $&$0.20$&$-5.08$&$-0.96$\\
57310.7595 & TIGRE &$-71.9$&$-70.9$&$-67.0$&$-93.8$&$-76.2$&$-65.0$&$0.63$&$1.13$&$0.64$&$0.30$&$ $&$0.27$&$-5.14$&$-1.02$\\
57315.6104 & TIGRE &$-72.6$&$-72.1$&$-67.8$&$-90.5$&$-79.6$&$-67.6$&$0.69$&$1.15$&$0.69$&$0.09$&$ $&$0.12$&$-5.79$&$-0.96$\\
57315.7346 & TIGRE &$-72.4$&$-70.0$&$-66.4$&$-91.6$&$-79.6$&$-63.6$&$0.71$&$1.15$&$0.62$&$0.05$&$ $&$0.13$&$-5.77$&$-0.95$\\
57316.6673 & TIGRE &$-74.8$&$-70.3$&$-66.0$&$-91.2$&$-80.9$&$-67.1$&$0.66$&$1.08$&$0.62$&$0.09$&$ $&$0.21$&$-5.47$&$-0.98$\\
57317.6686 & TIGRE &$-70.1$&$-69.8$&$-67.4$&$-92.2$&$ $&$ $&$0.68$&$1.27$&$0.70$&$0.10$&$ $&$0.11$&$-5.29$&$-1.11$\\
57320.7903 & TIGRE &$-70.4$&$-70.5$&$-66.3$&$-89.9$&$-76.5$&$-61.3$&$0.68$&$1.32$&$0.59$&$0.06$&$ $&$0.22$&$-5.43$&$-0.94$\\
57330.7925 & TIGRE &$-72.0$&$-71.0$&$-67.7$&$-91.1$&$ $&$-65.8$&$0.53$&$1.25$&$0.58$&$0.10$&$ $&$0.15$&$-5.43$&$-1.05$\\
57332.6559 & TIGRE &$-72.9$&$-71.2$&$-68.5$&$-93.3$&$-81.4$&$-64.4$&$0.60$&$1.18$&$0.60$&$0.15$&$ $&$0.12$&$-5.33$&$-1.11$\\
57334.6326 & TIGRE &$-73.9$&$-72.6$&$-67.2$&$-91.1$&$ $&$ $&$0.60$&$1.23$&$0.56$&$0.33$&$ $&$0.17$&$-5.41$&$-1.08$\\
57336.6533 & TIGRE &$-71.8$&$-72.4$&$-68.9$&$-90.9$&$-82.4$&$-66.7$&$0.54$&$1.09$&$0.56$&$0.18$&$ $&$0.16$&$-5.39$&$-1.15$\\
57338.5743 & TIGRE &$-71.3$&$-71.6$&$-67.4$&$-92.3$&$-79.6$&$-65.8$&$0.63$&$1.09$&$0.60$&$0.19$&$ $&$0.16$&$-5.60$&$-1.10$\\
57973.8502 & TIGRE &$-75.7$&$-76.9$&$-75.0$&$-95.1$&$-85.2$&$-67.6$&$0.69$&$0.80$&$0.45$&$-0.44$&$ $&$-0.24$&$-6.35$&$-1.20$\\
58002.6230 & OHP &$ $&$-77.8$&$-73.2$&$-93.8$&$ $&$ $&$ $&$ $&$0.44$&$-0.31$&$ $&$ $&$ $&$ $\\
58002.6342 & OHP &$ $&$-77.1$&$-72.4$&$-93.0$&$ $&$ $&$ $&$ $&$0.48$&$-0.21$&$ $&$ $&$ $&$ $\\
58003.6318 & OHP &$ $&$-73.0$&$-68.2$&$-90.6$&$ $&$ $&$ $&$ $&$0.44$&$-0.21$&$ $&$ $&$ $&$ $\\
58006.6250 & OHP &$ $&$-75.1$&$-69.6$&$-91.9$&$ $&$ $&$ $&$ $&$0.49$&$-0.27$&$ $&$ $&$ $&$ $\\
58275.9330 & TIGRE &$-74.7$&$-74.4$&$-70.0$&$-93.6$&$-84.8$&$-65.2$&$0.55$&$0.41$&$0.38$&$-0.61$&$ $&$-0.42$&$-7.08$&$-1.45$\\
58318.9006 & TIGRE &$-78.7$&$-76.8$&$-72.5$&$-91.6$&$-81.4$&$-71.5$&$0.70$&$0.83$&$0.46$&$-0.70$&$ $&$-0.41$&$-7.39$&$-1.29$\\
58352.6205 & OHP &$ $&$-73.5$&$-66.3$&$-91.9$&$ $&$ $&$ $&$ $&$0.36$&$-0.53$&$ $&$ $&$ $&$ $\\
58353.6255 & OHP &$ $&$-73.6$&$-68.5$&$-91.8$&$ $&$ $&$ $&$ $&$0.39$&$-0.36$&$ $&$ $&$ $&$ $\\
58357.6083 & OHP &$ $&$-76.4$&$-70.7$&$-92.3$&$ $&$ $&$ $&$ $&$0.40$&$-0.42$&$ $&$ $&$ $&$ $\\
58357.6222 & OHP &$ $&$-74.3$&$-72.4$&$-90.7$&$ $&$ $&$ $&$ $&$0.39$&$-0.47$&$ $&$ $&$ $&$ $\\
58636.9404 & TIGRE &$-74.6$&$-72.9$&$-69.2$&$-90.6$&$-80.8$&$-63.4$&$0.56$&$0.37$&$0.32$&$-0.82$&$ $&$-0.58$&$-7.60$&$-1.42$\\
58721.5006 & OHP &$ $&$-71.0$&$-66.0$&$-83.2$&$ $&$ $&$ $&$ $&$0.35$&$-0.64$&$ $&$ $&$ $&$ $\\
58721.5219 & OHP &$ $&$-68.8$&$-65.2$&$-87.9$&$ $&$ $&$ $&$ $&$0.34$&$-0.65$&$ $&$ $&$ $&$ $\\
58722.4960 & OHP &$ $&$-69.3$&$-63.5$&$-85.5$&$ $&$ $&$ $&$ $&$0.35$&$-0.62$&$ $&$ $&$ $&$ $\\
58806.8216 & CFHT &$-70.5$&$-70.2$&$ $&$ $&$-78.4$&$-64.6$&$ $&$ $&$0.32$&$ $&$-0.63$&$-0.85$&$-8.14$&$-1.55$\\
59048.9064 & TIGRE &$-71.3$&$-70.4$&$-63.9$&$-89.0$&$-81.6$&$-60.8$&$0.62$&$0.51$&$0.33$&$-1.31$&$ $&$-0.88$&$-8.66$&$-1.53$\\
59118.5307 & OHP &$ $&$ $&$ $&$ $&$ $&$ $&$ $&$ $&$ $&$ $&$ $&$ $&$-8.69$&$-1.61$\\
59118.5522 & OHP &$ $&$ $&$ $&$ $&$ $&$ $&$ $&$ $&$ $&$ $&$ $&$ $&$-8.86$&$-1.64$\\
59119.5186 & OHP &$ $&$ $&$ $&$ $&$ $&$ $&$ $&$ $&$ $&$ $&$ $&$ $&$-8.83$&$-1.63$\\
59119.5333 & OHP &$ $&$ $&$ $&$ $&$ $&$ $&$ $&$ $&$ $&$ $&$ $&$ $&$-8.93$&$-1.54$\\
59121.5190 & OHP &$ $&$ $&$ $&$ $&$ $&$ $&$ $&$ $&$ $&$ $&$ $&$ $&$-8.60$&$-1.54$\\
59121.5330 & OHP &$ $&$ $&$ $&$ $&$ $&$ $&$ $&$ $&$ $&$ $&$ $&$ $&$-8.94$&$-1.52$\\
59122.5132 & OHP &$ $&$ $&$ $&$ $&$ $&$ $&$ $&$ $&$ $&$ $&$ $&$ $&$-8.72$&$-1.52$\\
59122.5272 & OHP &$ $&$ $&$ $&$ $&$ $&$ $&$ $&$ $&$ $&$ $&$ $&$ $&$-8.09$&$-1.47$\\
59493.5495 & OHP &$ $&$ $&$ $&$ $&$ $&$ $&$ $&$ $&$ $&$ $&$ $&$ $&$-9.40$&$-1.69$\\
59493.5672 & OHP &$ $&$ $&$ $&$ $&$ $&$ $&$ $&$ $&$ $&$ $&$ $&$ $&$-9.45$&$-1.65$\\
59493.5890 & OHP &$ $&$ $&$ $&$ $&$ $&$ $&$ $&$ $&$ $&$ $&$ $&$ $&$-9.51$&$-1.75$\\
59495.5233 & OHP &$ $&$ $&$ $&$ $&$ $&$ $&$ $&$ $&$ $&$ $&$ $&$ $&$-9.28$&$-1.73$\\
59495.5341 & OHP &$ $&$ $&$ $&$ $&$ $&$ $&$ $&$ $&$ $&$ $&$ $&$ $&$-9.26$&$-1.68$\\
59495.5446 & OHP &$ $&$ $&$ $&$ $&$ $&$ $&$ $&$ $&$ $&$ $&$ $&$ $&$-9.19$&$-1.70$\\
59496.4551 & OHP &$ $&$ $&$ $&$ $&$ $&$ $&$ $&$ $&$ $&$ $&$ $&$ $&$-9.50$&$-1.70$\\
59496.4656 & OHP &$ $&$ $&$ $&$ $&$ $&$ $&$ $&$ $&$ $&$ $&$ $&$ $&$-9.43$&$-1.71$\\
59496.4762 & OHP &$ $&$ $&$ $&$ $&$ $&$ $&$ $&$ $&$ $&$ $&$ $&$ $&$-9.39$&$-1.73$\\
59496.4867 & OHP &$ $&$ $&$ $&$ $&$ $&$ $&$ $&$ $&$ $&$ $&$ $&$ $&$-9.46$&$-1.68$\\
59496.4974 & OHP &$ $&$ $&$ $&$ $&$ $&$ $&$ $&$ $&$ $&$ $&$ $&$ $&$-9.42$&$-1.74$\\
59498.4893 & OHP &$ $&$ $&$ $&$ $&$ $&$ $&$ $&$ $&$ $&$ $&$ $&$ $&$-9.33$&$-1.72$\\
59498.5035 & OHP &$ $&$ $&$ $&$ $&$ $&$ $&$ $&$ $&$ $&$ $&$ $&$ $&$-9.25$&$-1.65$\\
59498.5175 & OHP &$ $&$ $&$ $&$ $&$ $&$ $&$ $&$ $&$ $&$ $&$ $&$ $&$-9.46$&$-1.67$\\
59498.5318 & OHP &$ $&$ $&$ $&$ $&$ $&$ $&$ $&$ $&$ $&$ $&$ $&$ $&$-9.40$&$-1.68$\\
59498.7516 & TIGRE &$-67.8$&$-67.2$&$-62.4$&$-85.2$&$-74.4$&$-58.5$&$0.69$&$0.47$&$0.28$&$-1.36$&$ $&$-1.15$&$-9.39$&$-1.86$\\
59552.5563 & TIGRE &$-67.9$&$-65.6$&$-60.0$&$-85.1$&$-72.5$&$-59.6$&$0.69$&$0.25$&$0.20$&$-1.35$&$ $&$-1.02$&$-9.18$&$-1.77$\\
59564.6448 & TIGRE &$-65.3$&$-65.3$&$-62.3$&$-88.1$&$-71.4$&$-61.3$&$0.57$&$0.29$&$0.20$&$-1.26$&$ $&$-1.13$&$-9.41$&$-1.87$\\
59570.5869 & TIGRE &$-66.7$&$-65.8$&$-59.7$&$-83.4$&$-73.7$&$-67.6$&$0.72$&$0.34$&$0.25$&$-1.41$&$ $&$-1.08$&$-9.52$&$-1.84$\\
59811.8731 & TIGRE &$-65.2$&$-62.1$&$-56.3$&$-78.3$&$-70.3$&$-54.7$&$0.75$&$0.43$&$0.18$&$-1.64$&$ $&$-1.21$&$-10.12$&$-1.65$\\
59814.8401 & TIGRE &$-65.8$&$-62.5$&$-59.8$&$-83.2$&$-69.3$&$-56.3$&$0.82$&$0.46$&$0.22$&$-1.53$&$ $&$-1.16$&$-9.85$&$-1.60$\\
59851.5520 & OHP &$ $&$-61.5$&$-60.3$&$-81.3$&$ $&$ $&$ $&$ $&$0.16$&$-1.18$&$ $&$ $&$ $&$ $\\
59851.5690 & OHP &$ $&$-59.3$&$-54.3$&$-78.7$&$ $&$ $&$ $&$ $&$0.22$&$-1.14$&$ $&$ $&$ $&$ $\\
59852.5420 & OHP &$ $&$-56.9$&$-53.6$&$-76.0$&$ $&$ $&$ $&$ $&$0.21$&$-1.22$&$ $&$ $&$ $&$ $\\
59852.5630 & OHP &$ $&$-56.4$&$-54.5$&$-75.3$&$ $&$ $&$ $&$ $&$0.18$&$-1.20$&$ $&$ $&$ $&$ $\\
59855.5060 & OHP &$ $&$-60.4$&$-57.3$&$-74.5$&$ $&$ $&$ $&$ $&$0.17$&$-1.19$&$ $&$ $&$ $&$ $\\
59855.5220 & OHP &$ $&$-61.0$&$-57.3$&$-78.6$&$ $&$ $&$ $&$ $&$0.18$&$-1.18$&$ $&$ $&$ $&$ $\\
59855.5500 & OHP &$ $&$-59.7$&$-58.4$&$-79.2$&$ $&$ $&$ $&$ $&$0.20$&$-1.15$&$ $&$ $&$ $&$ $\\
59855.5790 & OHP &$ $&$-58.5$&$-56.3$&$-80.2$&$ $&$ $&$ $&$ $&$0.18$&$-1.27$&$ $&$ $&$ $&$ $\\
59855.6070 & OHP &$ $&$-56.7$&$-55.0$&$-78.3$&$ $&$ $&$ $&$ $&$0.17$&$-1.13$&$ $&$ $&$ $&$ $\\
59855.6360 & OHP &$ $&$-61.0$&$-54.6$&$-79.1$&$ $&$ $&$ $&$ $&$0.19$&$-1.18$&$ $&$ $&$ $&$ $\\
59855.6610 & OHP &$ $&$-59.7$&$-55.2$&$-83.7$&$ $&$ $&$ $&$ $&$0.15$&$-1.18$&$ $&$ $&$ $&$ $\\
59951.6041 & TIGRE &$-60.7$&$-62.5$&$-59.7$&$-83.3$&$-69.6$&$-58.1$&$0.52$&$0.01$&$0.15$&$-1.54$&$-0.33$&$-1.34$&$-10.12$&$-1.80$\\
\hline
\end{tabular}
\end{table*}
\bsp
\label{lastpage}
\end{document}